\documentclass[aip,pof,onecolumn,floatfix]{revtex4-2}
\usepackage{amsmath,amssymb,amsfonts}
\usepackage{graphicx}
\usepackage{dcolumn}
\usepackage{siunitx}
\usepackage{booktabs}
\usepackage{bm}
\usepackage{multirow}
\usepackage{longtable}
\usepackage{rotating}

\makeatletter
\providecommand{\@bibitemShut}{\@empty}
\providecommand{\bibitemNoStop}{\@empty}
\providecommand{\bibitemStop}{\@empty}
\def\selectlanguage#1{}
\makeatother

\usepackage{CJKutf8}
\newcommand{\cn}[1]{\begin{CJK*}{UTF8}{gbsn}#1\end{CJK*}}

\providecommand{\alttext}[1]{}

\begin{document}
\raggedbottom

\title{Tandem jets in crossflow: the enclosed-recirculation momentum exchange and the unsteady shear-layer dynamics}

\author{Xunchen Liu\,(\cn{刘训臣})}
\email{liuxunchen@sdust.edu.cn}
\affiliation{\vspace{-3.82pt}College of Mechanical and Electronic Engineering, Shandong University of Science and Technology, No.~579 Qianwan'gang Road, Qingdao 266590, Shandong, China}

\begin{abstract}
Twin tandem jets in crossflow are examined by 40\,kHz time-resolved particle image velocimetry (PIV) over nine spacing--velocity-ratio conditions ($L=2d$--$4d$, $r\approx3$--7; 68 twin cases, 11 single-jet records). The front jet is diluted: its trajectory lies below the single-jet reference, with a near-field segment up to 40\% shorter, and its velocity ratio falls to an effective value $r_{\mathrm{eff}}<r$, while the shielded rear jet rises above the reference. This paper identifies the carrier of that dilution and the unsteady dynamics that sustains it. The rear-jet blockage closes a gap recirculation whose reversed flow puts the two gap-side layers in the counter-current shear-layer configuration, the classical route to enhanced entrainment. The measured Reynolds stress and Prandtl mixing length show that the absolute exchange is nearly independent of the velocity ratio and exceeds the single-jet leeward zone like-for-like by up to $1.5\times$ (zone peak, across conditions) and $2.8\times$ (area-integrated) at $r\approx7$. The gain comes from an enlarged mixing scale rather than from a steeper shear, and the front-jet near-field shortening follows the absolute exchange transport rather than the twin/single enhancement ratio. The turbulent term exceeds the pressure term in eight of the nine conditions, and the flux budget reads the dilution as a momentum redistribution (a centerline deficit, a widened profile, a momentum loss to the crossflow) rather than a net mass addition; the evidence remains observational. The gap-side layers remain convectively unstable---DMD and SPOD find no growing mode within the resolvable band and record length---so the exchange is carried by convecting vortex trains rather than by a self-excited global mode.
\end{abstract}


\maketitle

\section{\label{sec:intro}Introduction}

Two tandem jets in a crossflow pose a two-body problem that a single jet does not. The upstream (front) jet shields the downstream (rear) jet from the crossflow, and the rear jet in turn blocks the flow in the gap and redirects part of it back toward the front jet. In the companion paper~\citep{companion_part1}, which used the same 40\,kHz time-resolved particle image velocimetry (PIV) system over the nine conditions, the front--rear asymmetry was established: the front-jet trajectory lies below the single-jet reference and its near-field segment shortens by up to 40\%, while the shielded rear jet rises above the reference. That study also showed that the two classical candidate causes (the reverse-flow region and the low-pressure attraction) are inconsistent with the observed ordering of the modification. It then condensed these observations into a one-parameter dilution law, $r_{\mathrm{eff}}/r=r/(r+c)$ with $c=3.31$, in which the spacing enters only through the near-field entrainment capacity.

Two questions are addressed here: what carries the modification---the turbulent momentum exchange of the gap flow or a self-excited global mode---and what unsteady dynamics sustains the exchange. The study follows the steady/dynamic division adopted in the two-part study of a single transverse jet by \citet{zhang_flow_2017,yang_zhang_part2_2017}: the mean field, the trajectory extraction and the scaling law are reported in the companion paper~\citep{companion_part1}, while the present measurements are its dynamical counterpart on the same dataset.
The dynamics is organized by a shear layer that separates two streams of opposite streamwise directions. This is the counter-current shear layer (CCSL~\citep{ho_perturbed_1984}), a classical shear-layer configuration. The reverse flow provides a feedback path that makes the layer globally unstable~\citep{huerre_absolute_1985,strykowski_stability_1991}, and its entrainment and mixing are far stronger than in the co-flowing case. This enhancement is documented in the shear layers of the single transverse jet~\citep{getsinger_structural_2014,shoji_origins_2020,nair_near-field_2023}. In the tandem configuration, the rear-jet blockage closes a recirculation bubble in the gap, whose reversed flow sweeps the front-jet leeward shear layer. Enclosed and sustained by the blocked crossflow, which the single jet cannot provide, this bubble turns the two gap-side layers into the CCSL configuration, through which a turbulent momentum exchange couples the two jets.

\begin{figure}
\centerline{\includegraphics[width=\textwidth]{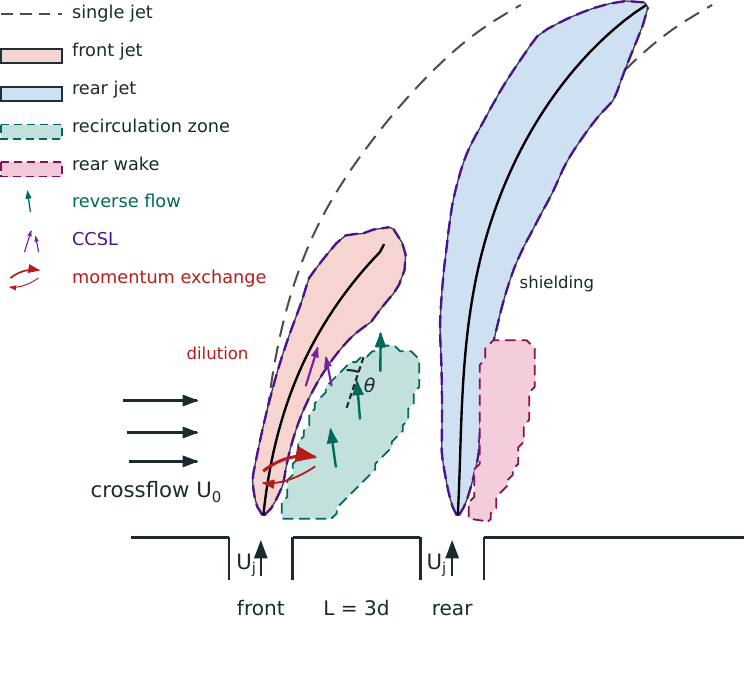}}
\caption{The enclosed-recirculation momentum exchange of the twin-nozzle jet in crossflow: the crossflow (arrows), the front and rear jets, the gap recirculation and the reversed flow it drives, and the single-jet reference (dashed). The rear-jet wake (pink dashed) and the shielding and dilution it produces are also marked, and the sketched nozzle spacing is $L=3d$. The red double arrow marks the turbulent momentum exchange, the purple vectors the counter-current shear layer (CCSL), and $\theta$ the angle of the climbing zone fluid to the front-jet tangent.}
\alttext{Schematic of the twin-nozzle jet in crossflow with the crossflow, the front and rear jets, the gap recirculation zone, the reversed flow it drives and the single-jet reference; the red double arrow marks the turbulent momentum exchange and the purple vectors the counter-current shear layer.}
\label{fig:ccsl-schematic}
\end{figure}

The physical picture is sketched in Fig.~\ref{fig:ccsl-schematic}. The gap recirculation hugs the front-jet leeward layer, and its fluid climbs along the jet at an angle $\theta<90^\circ$.

The zone-top fluid is peeled off and injected into the bending jet, so that high-momentum front-jet fluid is stripped into the zone and replaced by low-momentum gap fluid (entrainment). The front jet is thereby drawn toward the rear jet and its momentum is diluted, while the shielded rear jet rises above the single-jet reference.

Each of these steps is a momentum-mixing process between the jet fluid and the gap fluid. Its strength is naturally measured by the Prandtl mixing length $l_{\rm mix}$ of the gap-side layer, the scale over which the exchanged momentum is mixed. Read in these terms, the picture of Fig.~\ref{fig:ccsl-schematic} makes definite, experimentally testable predictions. The front-jet trajectory should be drawn below the single-jet reference as its momentum is diluted, while the shielded rear jet should rise above it. The velocity distribution should show a deficit on the front-jet centerline and a widened gap-side layer, and the mixing length itself should be enlarged where the exchange acts. Each of these consequences is verified against the measurements below.

The exchange and the dynamics behind these predictions are quantified on the 40\,kHz dataset of Section~\ref{sec:expt}. The Reynolds stress and the Prandtl mixing length, formed from the 400 instantaneous fields of each record without modeling closure, measure the turbulent momentum exchange directly (Section~\ref{sec:ccsl}). The front-jet bending shows a dose--response against the candidate causes: it scales with the absolute exchange, with the pressure force as a common-origin response and the gap pressure anti-correlated (Section~\ref{sec:dose}). Flux-budget diagnostics read the same dilution as a redistribution rather than a mass addition (Section~\ref{sec:flux}). The unsteady shear-layer dynamics is then classified using vortex tracking, dynamic mode decomposition (DMD)~\citep{schmid_dynamic_2010,rowley_spectral_2009} and spectral proper orthogonal decomposition (SPOD)~\citep{sieber_spectral_2016,towne_spectral_2018} (Sections~\ref{sec:liutex} and~\ref{sec:modal}).

This paper makes two contributions. First, it identifies the turbulent momentum exchange of the enclosed gap recirculation as the carrier of the front-jet modification. The absolute exchange is nearly independent of the velocity ratio and exceeds the single-jet leeward zone like-for-like by up to $1.5\times$ (zone peak, across conditions) and $2.8\times$ (area-integrated) at $r\approx7$; the near-field shortening scales with that absolute transport ($\rho=0.68$ at the condition level) rather than with the twin/single enhancement ratio. The flux budget resolves this dilution into a centerline deficit, a widened profile and a momentum loss to the crossflow, rather than a net mass addition. Second, it identifies the dynamical carrier of the exchange as convecting vortex trains. Within the resolvable band and record length, the gap-side layers remain convectively unstable with no self-excited global mode, the two jets shed at comparable pooled frequencies, and the vortex speeds fan out into a material--pattern--phase hierarchy for the twin jets while nearly coinciding for the single jet. The evidence is observational: a consistent account of the measured sequence, not a proof of causation.

\section{\label{sec:expt}Experimental measurement}

\begin{figure}
\centerline{\includegraphics[width=\textwidth]{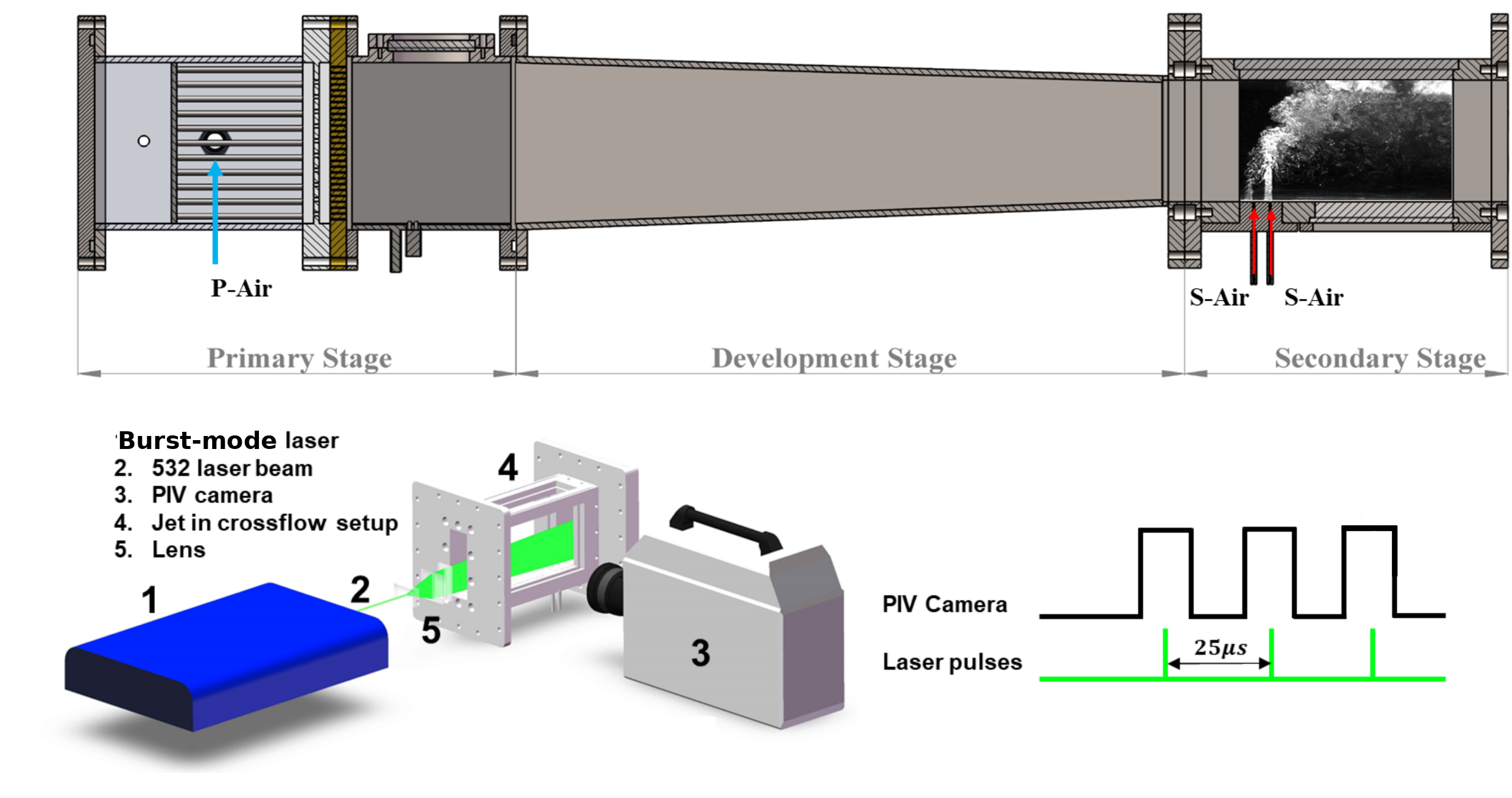}}
\caption{Layout of the twin-nozzle JICF facility (primary, development and secondary stages) with the 40\,kHz burst-mode PIV arrangement and its timing: the camera exposure window and the laser pulse pairs, separated by 25\,$\mu$s.}
\alttext{Schematic of the twin-nozzle test section with its primary, development and secondary stages, the 40 kHz burst-mode PIV laser and camera arrangement, and the 10 ms burst time sequence.}
\label{fig:setup}
\end{figure}

The experiments are conducted in the twin-nozzle jet-in-crossflow (JICF) facility of Fig.~\ref{fig:setup}. It comprises a mainstream supply (primary stage), a contraction (development stage) and the test section (secondary stage: $40$\,mm wide $\times$ $115$\,mm high $\times$ $295$\,mm long, with quartz-glass walls for optical access). The two $d=4$\,mm nozzles are mounted in the test section at spacing $L$. They are the nozzle pair used in the reacting twin-jet experiments of \citet{wang_experimental_2022,wang_lift-off_2023} and discharge to atmosphere.

The experimental conditions comprise the nine combinations of nominal velocity ratio ($r=3$, $5$, $7$) and jet spacing ($L=2d$, $3d$, $4d$). The main-flow velocity varies from about $1.5$ to $3.6$\,m/s, corresponding to a jet Reynolds number of $Re_d\approx2.9\times10^{3}$ and crossflow Reynolds numbers of $U_0d/\nu\approx4.0\times10^{2}$--$1.0\times10^{3}$. The jet and the main flow have the same density (both streams are air at $1.293$\,kg/m$^3$), so the momentum-flux ratio equals $r^2$. In total, 68 twin cases enter the statistics, 5--8 per condition as listed in Table~\ref{table:conditions}, together with the 11-record single-jet baseline (4, 3 and 4 records at $r\approx3$, $5$ and $7$). Each 10\,ms burst of 400 image pairs constitutes one record (case). One record (4d-5r-8) was exported as a time-mean field only and therefore contributes to the mean statistics but not to the stress terms (Table ST1 of the supplementary material).

A flashlamp-pumped Nd:YAG burst-mode laser (Spectral Energies, Quasimodo 1000), operated at 100\,kHz with a 10\,ms burst duration and about 125\,mJ per pulse, provides the illumination. Three cylindrical lenses expand its 532\,nm output into a laser sheet about 45\,mm high and 0.2\,mm thick, which enters the test section from the rear (Fig.~\ref{fig:setup}). A high-speed CMOS camera (Photron SA-Z) records the images, with a Nikon 50\,mm f/1.4\,G normal lens for the far-field view and a Nikon AF-200/4d telephoto lens for the near field of each jet. The flow is seeded with 1\,$\mu$m Al$_2$O$_3$ particles, whose Stokes number ($\approx0.04$) satisfies the tracer-following requirement.

The images are processed with DaVis 8.4 (LaVision) using multi-pass cross-correlation with $24\times24$ pixel interrogation windows and 75\% overlap, sampling the vector fields on a uniform grid with 0.35\,mm spacing. The first resolvable station lies about 0.5\,mm downstream of the exit. The image scale is set by imaging the crossflow alone: with the jets switched off, the measured mean particle displacement divided by the known free-stream velocity and the inter-pulse delay gives the magnification ($0.0575$\,mm per pixel, i.e.\ $0.345$\,mm per vector with the 75\% overlap of the final pass). The residual of this uniform-flow calibration over the common field of view, measured as the departure of the displacement field from a rigid-body fit, is a common-mode error. It shifts the absolute velocities and the velocity ratio $r$, but largely cancels in the front-to-rear and twin-to-single comparisons used below.

Within the DaVis processing, the vector fields are validated by a correlation-peak-ratio threshold, a median test against the $3\times3$ neighborhood filter and a maximum-displacement limit set from the expected local velocity. Rejected vectors and vectors outside the illuminated region are flagged invalid and are never interpolated or filled, so they are excluded from every statistic. The random uncertainty of a validated vector is a few percent of the local velocity, and averaging the 400 fields of a record reduces the random term of the record-mean velocity well below one percent. The mean-field quantities used in the analysis are therefore limited by the systematic calibration term rather than by random noise.

\begin{table*}
\caption{Experimental conditions: the nine combinations of the nominal velocity ratio and jet spacing, numbered 1--9 as referenced in the text, with the number of cases entering the statistics in parentheses. Measured values are $r\approx2.8$, $4.6$ and $6.2$--$6.5$ in the three ratio groups at the documented bulk jet velocity $V_j=10$\,m/s, with crossflow velocities $\approx3.6$, $2.2$ and $1.5$\,m/s and $Re_d\approx2.9\times10^{3}$ (jet, bulk), $U_0d/\nu\approx4.0\times10^{2}$--$1.0\times10^{3}$ (crossflow).}\centering\label{table:conditions}
\alttext{Table of the nine experimental conditions arranged by velocity ratio and jet spacing, with the number of cases entering the statistics in parentheses.}
\begin{tabular}{lccc}
\hline
 & \multicolumn{3}{c}{jet spacing $L$} \\
\cmidrule(lr){2-4}
velocity ratio $r$ & $L=2d$ & $L=3d$ & $L=4d$ \\ \hline
$r=3$ & 1 (8) & 4 (8) & 7 (7) \\ $r=5$ & 2 (8) & 5 (8) & 8 (8) \\ $r=7$ & 3 (8) & 6 (8) & 9 (5) \\ \hline
\end{tabular}
\end{table*}

The near-field shear-layer instability is governed by the centerline exit velocity rather than by the bulk velocity. Because the near-field shedding frequency is set by the peak velocity difference across the shear layer, the near-exit centerline value $u_j\approx14$\,m/s is used for the Strouhal-number and Liutex normalizations; over the 11 single-jet records the measured near-exit centerline velocity is $13.3$--$15.1$\,m/s. In the twin-jet configuration the two jets differ: averaged over the 136 jet exits, the rear-jet near-exit velocity is $15.4\pm0.5$\,m/s, close to the single-jet values, whereas the front-jet value is only $12.8\pm0.7$\,m/s. The front-jet deficit is already present at the first resolvable station and is in part a near-exit signature of the interaction rather than of the supply (Section~\ref{sec:limitations}). The time-resolved analyses use the near-field view, where the two jets interact before merging downstream.

Throughout the paper, the color denotes the velocity-ratio group (green $r\approx3$, orange $r\approx5$, blue $r\approx7$) and the marker shape the jet spacing (diamond $L=2d$, triangle $L=3d$, square $L=4d$); filled and open symbols denote the front and rear jets, respectively, where applicable.

\section{\label{sec:results}Results and discussion}

Because the burst-mode sequence resolves the turbulent fluctuations directly, the measurements provide two quantities beyond the mean fields. The first is the Reynolds stress $\langle u'v'\rangle$, obtained by correlating the velocity fluctuations of the 400 instantaneous fields of each record (Eq.~\eqref{eq:uv}), and the second the Prandtl mixing length $l_{\rm mix}$~\citep{pope_turbulent_2000} (Eq.~\eqref{eq:lmix}, Section~\ref{sec:ccsl}), formed from the measured stress and the mean shear. The stress is directly measured, whereas the mixing length is derived under the Prandtl closure assumption: an increase in the latter is therefore interpreted as a larger momentum-mixing scale rather than as a direct eddy-size measurement.

The presentation follows one chain of reasoning, from the observed trajectory modification to its dynamical carrier, and each step is pushed only as far as the measurement supports it.

\begin{figure}
\centering
\centerline{\includegraphics[width=\textwidth,height=0.78\textheight,keepaspectratio]{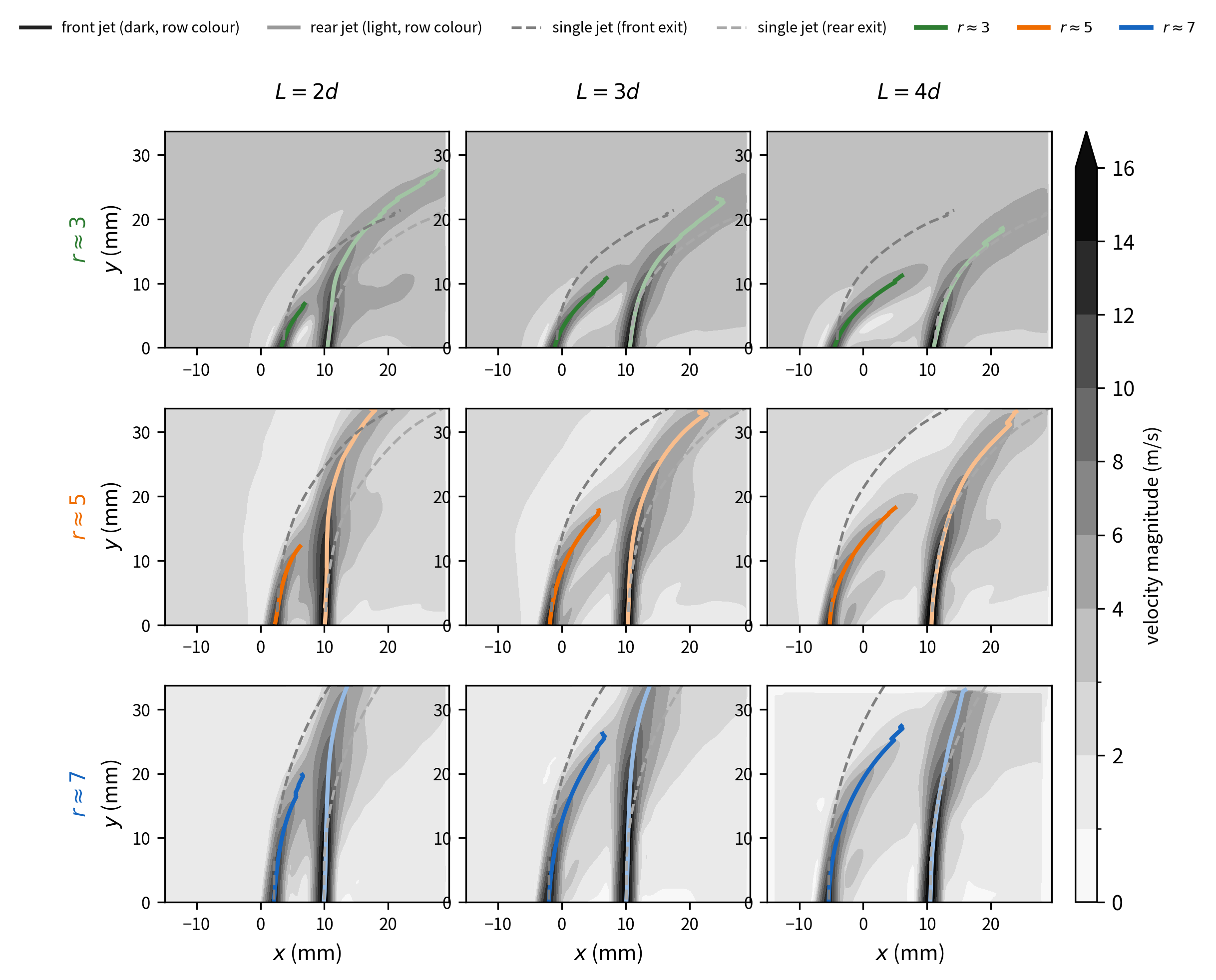}}
\caption{The trajectory modification that motivates this work, reproduced from the companion paper~\citep{companion_part1}: the condition-mean velocity-magnitude clouds and the mean front-jet (dark) and rear-jet (light) trajectories of the nine twin-jet conditions (rows $r\approx3$, 5, 7; columns $L=2d$, $3d$, $4d$), with the matched single-jet trajectories from the same nozzle exits (dashed grey). The front jet is drawn below its single-jet counterpart and the rear jet above it; this deviation is the observable from which the chain of Section~\ref{sec:results} starts.}
\alttext{Nine panels of the condition-mean velocity-magnitude clouds with the mean front-jet and rear-jet trajectories of the twin-jet conditions, against the matched single-jet trajectories from the same nozzle exits.}
\label{fig:traj-deviation}
\end{figure}

\emph{(i) The trajectory is the observable.} The front jet of the twin-jet configuration bends earlier than the matched single jet, its near-field segment is up to 40\% shorter, and its penetration and decay differ accordingly~\citep{companion_part1} (Fig.~\ref{fig:traj-deviation}); the metrics used to quantify this change are recalled in Section~\ref{sec:dose}.

\emph{(ii) The cause is narrowed to the remaining candidate.} Of the two classical candidates, the reverse-flow region does not order with the modification, and the reconstructed mean pressure is inconsistent with a low-pressure attraction (both in the companion paper). What remains is the turbulent momentum exchange across the gap-side shear layers, which is measured directly, rather than inferred, by correlating the fluctuations of the 400 instantaneous fields of each record (Section~\ref{sec:ccsl}).

\emph{(iii) The exchange is then tested as the cause.} It is shown to carry the bending in the dose--response analysis, while the pressure force, tested directly against the bending, runs the wrong way and is therefore excluded as the driver (Section~\ref{sec:dose}); the flux budget reads the resulting dilution as a redistribution of the front-jet momentum rather than a net mass addition (Section~\ref{sec:flux}).

\emph{(iv) The dynamical carrier of that exchange is then identified.} The gap-side layers roll up into convective vortex trains whose shedding frequencies are comparable between the two jets, and whose kinematics is measured independently (Section~\ref{sec:liutex}); the alternative of a self-excited global mode is excluded by the DMD and SPOD analysis (Section~\ref{sec:modal}).

Section~\ref{sec:discussion} then assembles the chain and states its limitations, and Section~\ref{sec:concl} draws the conclusions.

\subsection{\label{sec:ccsl}The exchange: the enclosed-recirculation momentum transfer}

Previous work on the same configuration eliminated both classical candidates~\citep{companion_part1}. The reverse-flow region does not order with the front-jet modification: its raw and momentum-flux-normalized area and strength show no significant correlation with any modification metric (Spearman $\rho=0.08$--$0.50$, $p\ge0.17$ over the nine conditions). The reconstructed mean pressure is likewise inconsistent with a low-pressure attraction: the inter-jet pressure is elevated rather than depressed (rising from $14.8$\,Pa at $L=4d$ to $34.0$\,Pa at $L=2d$), and the lateral pressure difference across the front jet is not enhanced. The analysis therefore starts from the trajectory change itself---the earlier bending and the shortened near-field segment of the front jet---and asks which remaining quantity of the flow can carry it. The candidate measured directly here is the turbulent momentum exchange of the gap flow.

\begin{figure}
\centerline{\includegraphics[width=\textwidth]{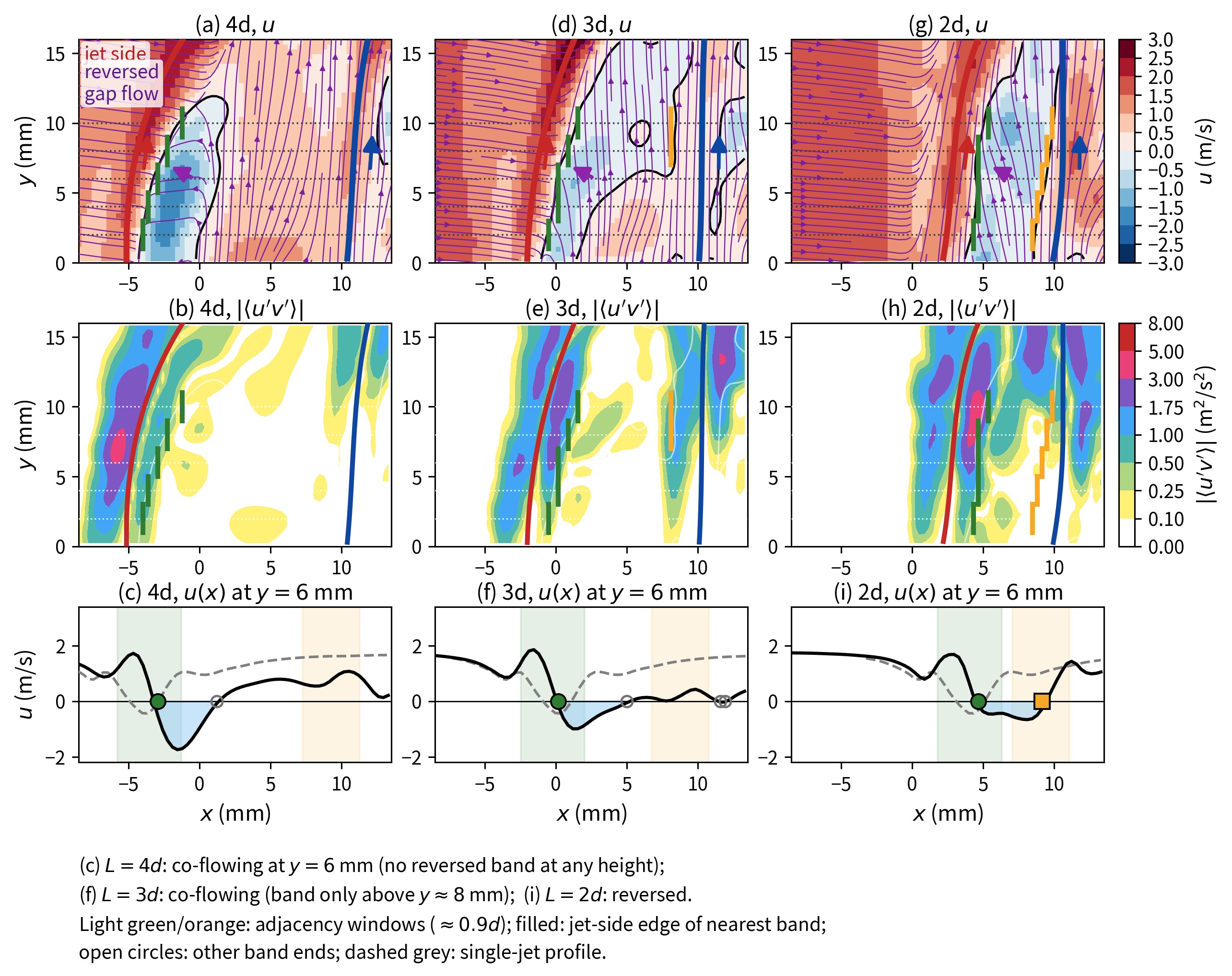}}
\caption{Velocity-field view of the counter-current shear-layer extraction at $r\approx7$ (columns: $L=4d$, $3d$, $2d$): (a), (d), (g) mean $u$ (filled contour bands of $0.5$\,m\,s$^{-1}$ over $\pm3$\,m\,s$^{-1}$, the same absolute scale throughout; purple: mean-flow streamlines) with the $u=0$ contour, the front (red) and rear (blue) trajectories and the shear-layer edges (green/orange bars); (b), (e), (h) $|\langle u'v'\rangle|$ on the fixed absolute scale of Fig.~\ref{fig:sup-stress}, with the same overlays; (c), (f), (i) layer-normal $u(x)$ at $y=6$\,mm (solid: twin; dashed gray: single) with the gap recirculation shaded and the edges (green circle, orange square) where $V_1$ and $V_2$ are read.}
\alttext{Velocity fields and Reynolds-stress fields used to extract the counter-current shear-layer velocities, with the layer-normal profiles of one condition.}
\label{fig:promo-s2}
\end{figure}

In shear-layer terms, the reversed gap flow places the two gap-side layers in the CCSL configuration of the single transverse jet~\citep{huerre_absolute_1985,strykowski_stability_1991,shoji_origins_2020,nair_near-field_2023} (sketched in Fig.~\ref{fig:ccsl-schematic}), characterized by the velocity ratio
\begin{equation}
R_1=\frac{V_1-V_2}{V_1+V_2}
\label{eq:r1}
\end{equation}
Here $V_1$ and $V_2$ are the streamwise velocities on the jet and gap sides of the layer ($R_1>1$ for counter-flow, $R_1<1$ for co-flow), extracted from the condition-mean fields as in Fig.~\ref{fig:promo-s2}.

\begin{figure}
\centerline{\includegraphics[width=\textwidth]{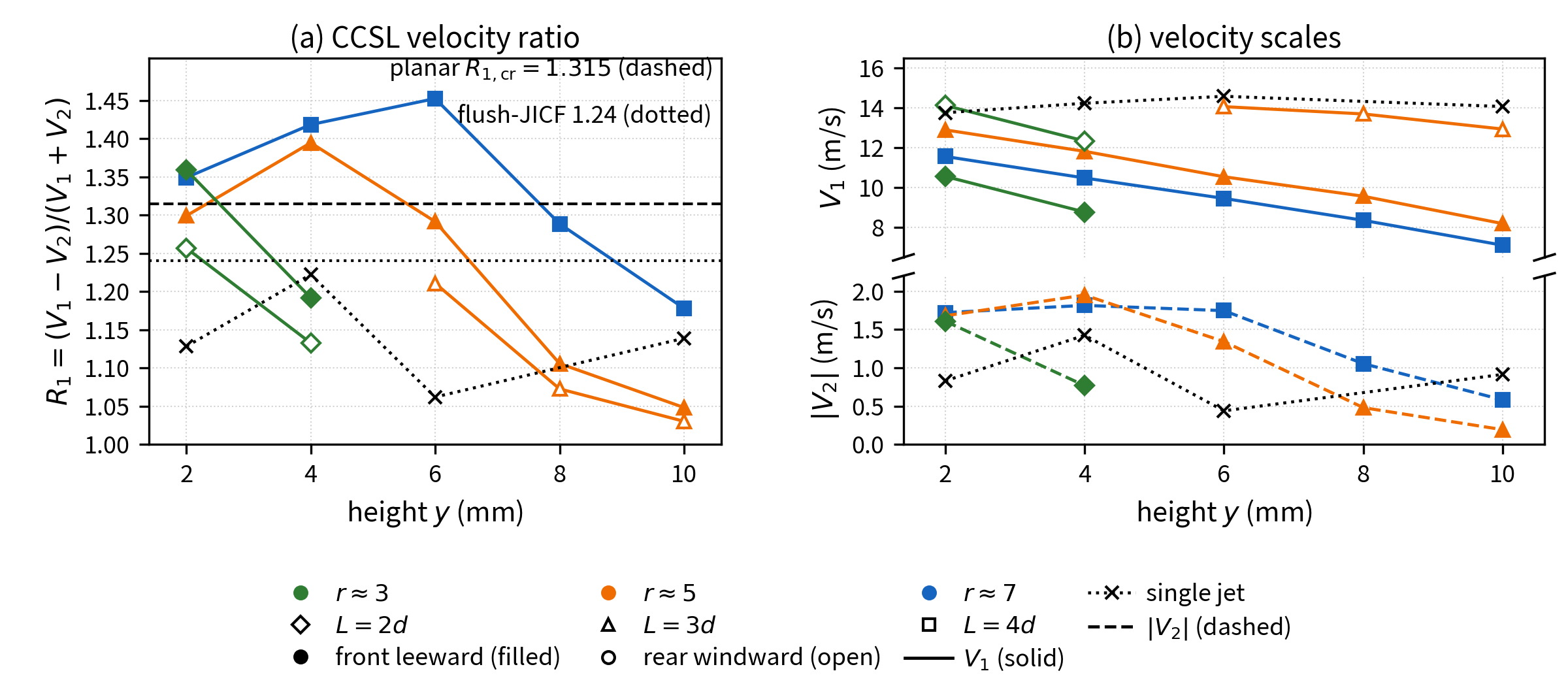}}
\caption{Profiles of the CCSL velocity ratio and its two velocity scales along the height $y$, for one representative condition per velocity-ratio group ($L=4d$, $r\approx7$; $L=3d$, $r\approx5$; $L=2d$, $r\approx3$): (a) $R_1=(V_1-V_2)/(V_1+V_2)$ with the planar counter-flow criterion $R_{1,\mathrm{cr}}=1.315$~\citep{souza_criterion_2021} (dashed) and the flush-JICF value $1.24$ (dotted) marked; (b) the two velocity scales on a broken vertical axis, the jet-side $V_1$ (solid, upper) and the gap-side reversal speed $|V_2|$ (dashed, lower), both in m/s and sharing the height axis. Color: velocity-ratio group ($r\approx3$ green, $5$ orange, $7$ blue); marker shape: spacing (diamond $2d$, triangle $3d$, square $4d$); filled/open: front-jet leeward/rear-jet windward layer; crosses with the dotted line: single-jet leeward reference. The $r\approx3$ layer supplies two stations only ($y=2$ and $4$\,mm), because at this ratio the reversed gap flow is too weak to form an adjacent reversed band above $y=2$\,mm at any spacing; the $L=4d$ front-jet layer has no rear-windward counterpart (co-flowing there). $|V_2|$ is the most negative streamwise velocity of the same gap band for either layer side, so the filled and open values coincide and are drawn once per condition.}
\alttext{Profiles of the counter-current shear-layer velocity ratio R1 and its two velocity scales along the height above the wall for one representative condition per velocity-ratio group, with a broken vertical axis for the two velocity scales.}
\label{fig:promo-s3}
\end{figure}

Over the nine conditions, the front-jet leeward layer is counter-current everywhere; the rear-jet windward layer is counter-current only at $L=2d$ and in the upper gap at $L=3d$ (it is essentially co-flowing at $L=4d$); and the single-jet leeward layer is weakly counter-current as well (representative $R_1$ profiles in Fig.~\ref{fig:promo-s3}).

The counter-current condition is permissive rather than causal. The single jet satisfies it, yet it exchanges momentum an order of magnitude more weakly (see below), so the twin-specific ingredient must be the enclosed, continuously replenished gap recirculation. Part of the front-jet excess in $R_1$ merely reflects the already-weakened front-jet velocity (the front-jet near-exit velocity is $12.8$\,m/s against $15.4$\,m/s for the rear jet, Section~\ref{sec:expt}).

\begin{figure}
\centerline{\includegraphics[width=\textwidth,height=0.70\textheight,keepaspectratio]{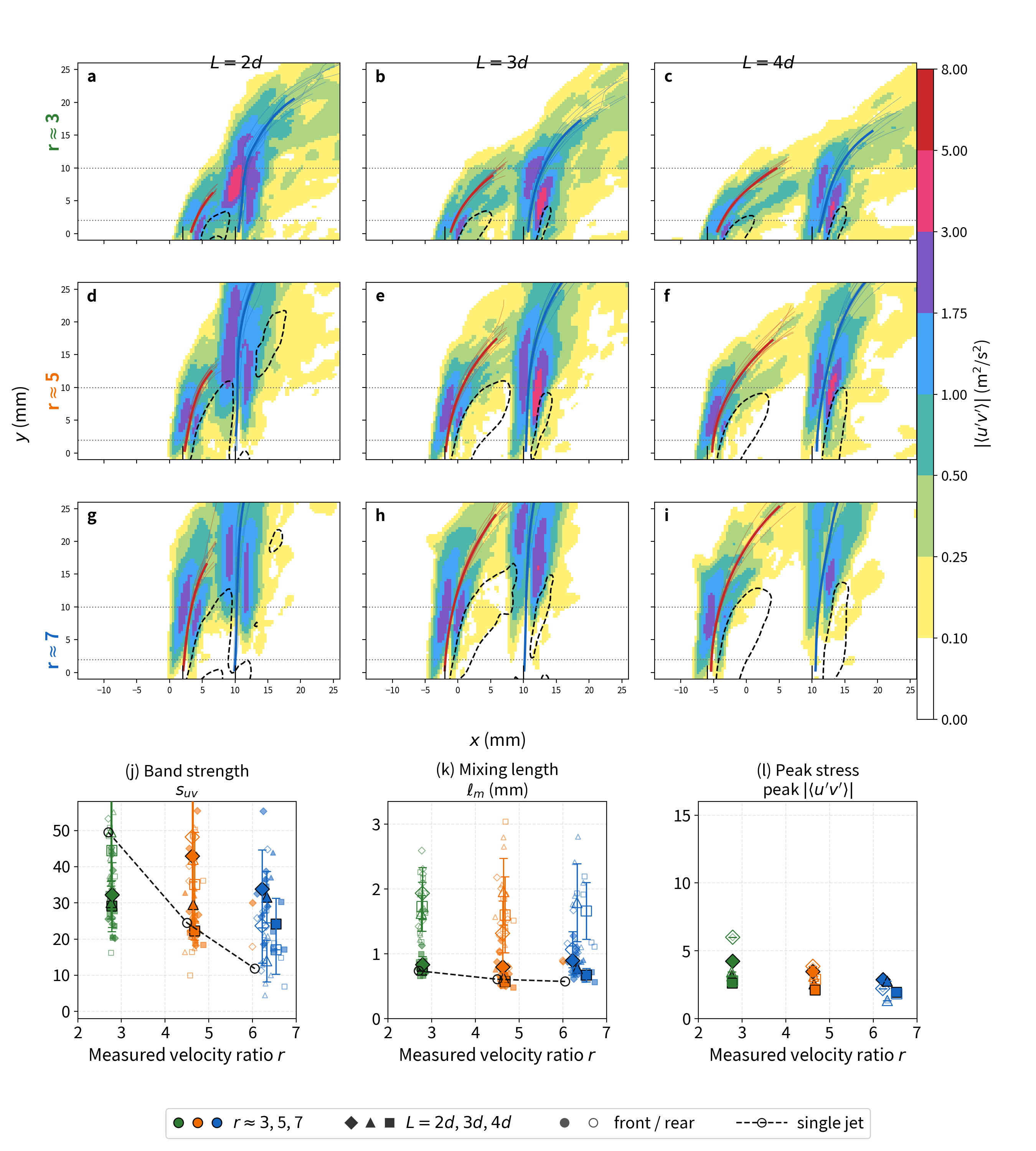}}
\caption{Condition-averaged Reynolds-stress magnitude $|\langle u'v'\rangle|$ in the physical $(x,y)$ coordinates (rows $r\approx3$, 5, 7; columns $L=2d$, $3d$, $4d$), on the fixed absolute scale shared by every $|\langle u'v'\rangle|$ figure of this paper ($0$--$8$\,m$^2$/s$^2$ in eight bands), with the reverse-flow boundary ($u=0$, black dashed), the statistics window (dotted gray) and the condition (thick) and per-case (thin; front red, rear blue) trajectories; the two off-specification records of Fig.~\ref{fig:promo-s5} are excluded. The bottom row gives the gap-side statistics behind these fields against the measured velocity ratio: (j) band strength $S_{uv}$, (k) mixing length $\ell_m$ and (l) peak stress, with the single-jet reference (gray dashed) and the nine-condition symbol convention of Section~\ref{sec:expt}.}
\alttext{Condition-averaged Reynolds-stress magnitude fields in physical coordinates over the nine conditions.}
\label{fig:sup-stress}
\end{figure}

The magnitude $|\langle u'v'\rangle|$ is used throughout; its sign is consistent with momentum transfer from the jet to the gap (the sign convention is documented in the supplementary material). The stress concentrates along the two edges of the recirculation zone of the reversed gap flow, and the full nine-condition fields are shown in Fig.~\ref{fig:sup-stress}.

Geometrically, the front-jet leeward recirculation zone keeps a nearly constant width of $1.5$--$2.5$\,mm at every spacing. As the gap narrows, the zone therefore covers a growing fraction of it. The zone reaches the rear-jet windward side across the full gap height only at $L=2d$ (two-sided coupling); at $L=3d$ it already touches the upper gap. The zone-to-gap ratio thus sets the coupling topology (one-sided, upper-gap, two-sided), and the local zone strength sets the counterflow speed. The zone itself is thin compared with the reversed region upstream of a single flush jet~\citep{shoji_origins_2020}, the rear jet confining the reversal against the front-jet leeward side.

The front-leeward $R_1$ exceeds the planar counter-flow criterion $R_{1,\mathrm{cr}}=1.315$~\citep{souza_criterion_2021} in 88 of the 222 sampled points, at every spacing, so the local criterion flags absolute instability in part of the layer. That criterion assumes a parallel, planar layer and is therefore only a local indication. A zone of local absolute instability is a necessary, not sufficient, condition for a self-excited global mode~\citep{huerre_local_1990,regan_global_2017}; the convective character of the layers is instead examined through the unsteady dynamics of Sections~\ref{sec:liutex} and~\ref{sec:modal}. The enhanced entrainment is consistent with strengthened shear-layer instabilities in single JICFs~\citep{gevorkyan_transverse_2016,gevorkyan_influence_2018,nair_combustion_2022} and with the monotonic entrainment--velocity-ratio dependence~\citep{dimotakis_two-dimensional_1986}.

The exchange is quantified by the transverse Reynolds stress of the 400 instantaneous fields of each record,
\begin{equation}
\langle u'v'\rangle=\frac{1}{N}\sum_{k=1}^{N}\left(u_k-\bar u\right)\left(v_k-\bar v\right), \qquad N=400
\label{eq:uv}
\end{equation}
Its peak within $\pm2$\,mm windows across each layer (Fig. S1a of the supplementary material) defines the window-peak exchange,
\begin{equation}
P_{uv}(y)=\max_{|x-x_e|\le2\,\mathrm{mm}}|\langle u'v'\rangle|(x,y),
\label{eq:puv}
\end{equation}
with $x_e$ the shear-layer edge. Over the nine conditions the front-jet leeward layer transfers streamwise momentum at window peaks $P_{uv}=0.46$--$3.85$\,m$^2$/s$^2$.

The like-for-like comparison with the single-jet leeward zone of the same velocity ratio gives a zone-peak ratio of $0.58$--$1.46$ ($0.80$--$1.18$ at $r\approx7$) and an area-integrated ratio of $0.59$--$2.83$ ($2.04$--$2.83$ at $r\approx7$). The absolute exchange is therefore nearly independent of the velocity ratio, the front-side area integral spanning $22$--$45$\,m$^2$/s$^2$\,mm$^2$, while the single-jet reference weakens with $r$ (from $49.5$ to $11.8$\,m$^2$/s$^2$\,mm$^2$).

A second, contrast-type metric normalizes each window peak by the pooled single-jet leeward zone mean in the near-field band ($0.14$\,m$^2$/s$^2$) rather than by the single-jet peak at the same station. It reaches $27\times$ per height; this value is a peak-to-zone-mean contrast rather than a like-for-like enhancement. Both the like-for-like ratios and the contrast values are quoted below, and the zone statistics behind them are in Table~\ref{tab:natural-uv}. The pooled per-station form of this metric (the more conservative of the two, as it averages over stations and heights) is $11.2\times$ at $r\approx7$, $4.6\times$ at $r\approx5$ and $1.7\times$ at $r\approx3$ (Table~\ref{tab:enhance-verify}).

Four exchange metrics are used throughout the paper: the window-peak $P_{uv}$, the zone peak $P_{uv}^{\rm zone}$ and the area-integrated $S_{uv}$ of Table~\ref{tab:natural-uv}, and the per-height window-peak twin/single ratio with its pooled per-station robustness form. Their different sensitivities matter for the causal reading below; a four-way comparison of the definitions is given in Table ST2 of the supplementary material.

\begin{table}
\centering
\caption{Natural-coordinate gap-side statistics of the turbulent momentum exchange and the reverse-flow zone (the two off-specification records of Fig.~\ref{fig:promo-s5} are excluded from the condition average). $S_{uv}$ is the area-integrated exchange over the gap-side zone (both gap-side layers, $|n|=0.5$--$4$\,mm, $s\le15$\,mm), $A_{rz}$ the reverse-flow area ($u<-0.05$\,m/s), $S_{rz}$ its strength, $P_{uv}^{\rm zone}$ the zone peak, $l_{\rm mix}$ the zone-median mixing length and the last column the zone-peak ratio to the single-jet leeward zone of the same velocity-ratio group; the co-flowing rear-jet layers (the $L=4d$ rows and the $L=3d$, $r\approx3$ row) keep $A_{rz}=S_{rz}=0$ with the other statistics defined over the same zone.}
\alttext{Table of the natural-coordinate gap-side statistics of the momentum exchange and the reverse-flow zone for the nine conditions and the single-jet reference.}
\label{tab:natural-uv}
\footnotesize
\renewcommand{\arraystretch}{0.92}
\begin{tabular}{llcccccc}
\toprule
Condition & Jet & $S_{uv}$ & $A_{rz}$ & $S_{rz}$ & $P_{uv}^{\rm zone}$ & $l_{\rm mix}$ & $P_{uv}^{\rm zone}/P_{uv,\rm s}^{\rm zone}$ \\
 & & (m$^2$/s$^2$\,mm$^2$) & (mm$^2$) & (m/s\,mm$^2$) & (m$^2$/s$^2$) & (mm) & -- \\
\midrule
$L=2d$, $r\approx3$ & fj & 32.219 & 8.639 & 5.200 & 4.228 & 0.833 & 0.9 \\
$L=2d$, $r\approx3$ & rj & 88.885 & 4.739 & 2.630 & 6.000 & 1.935 & 1.3 \\
$L=2d$, $r\approx5$ & fj & 44.574 & 27.300 & 31.790 & 3.475 & 0.789 & 1.5 \\
$L=2d$, $r\approx5$ & rj & 51.992 & 28.479 & 25.938 & 3.856 & 1.368 & 1.6 \\
$L=2d$, $r\approx7$ & fj & 33.383 & 26.102 & 22.262 & 2.871 & 0.898 & 1.2 \\
$L=2d$, $r\approx7$ & rj & 22.940 & 33.358 & 16.970 & 2.218 & 1.046 & 0.9 \\
$L=3d$, $r\approx3$ & fj & 30.125 & 10.986 & 8.035 & 3.157 & 0.752 & 0.7 \\
$L=3d$, $r\approx3$ & rj & 49.621 & 0.000 & 0.000 & 3.441 & 1.624 & 0.8 \\
$L=3d$, $r\approx5$ & fj & 29.600 & 26.321 & 27.322 & 2.578 & 0.633 & 1.1 \\
$L=3d$, $r\approx5$ & rj & 42.032 & 5.880 & 1.370 & 3.129 & 1.960 & 1.3 \\
$L=3d$, $r\approx7$ & fj & 31.602 & 30.429 & 22.268 & 2.767 & 0.775 & 1.1 \\
$L=3d$, $r\approx7$ & rj & 13.985 & 10.181 & 2.004 & 1.336 & 1.790 & 0.5 \\
$L=4d$, $r\approx3$ & fj & 29.147 & 9.705 & 7.616 & 2.625 & 0.801 & 0.6 \\
$L=4d$, $r\approx3$ & rj & 44.481 & 0.000 & 0.000 & 2.803 & 1.731 & 0.6 \\
$L=4d$, $r\approx5$ & fj & 22.334 & 24.213 & 33.310 & 2.108 & 0.574 & 0.9 \\
$L=4d$, $r\approx5$ & rj & 35.083 & 0.000 & 0.000 & 2.906 & 1.603 & 1.2 \\
$L=4d$, $r\approx7$ & fj & 24.147 & 30.765 & 37.649 & 1.963 & 0.672 & 0.8 \\
$L=4d$, $r\approx7$ & rj & 17.120 & 0.363 & 0.044 & 1.834 & 1.665 & 0.8 \\
\midrule
single, $r\approx7$ & single & 11.815 & -- & -- & 2.439 & 0.573 & 1 \\
single, $r\approx5$ & single & 24.442 & -- & -- & 2.386 & 0.608 & 1 \\
single, $r\approx3$ & single & 49.485 & -- & -- & 4.508 & 0.737 & 1 \\
\bottomrule
\end{tabular}
\end{table}

\begin{figure}
\centerline{\includegraphics[width=\textwidth]{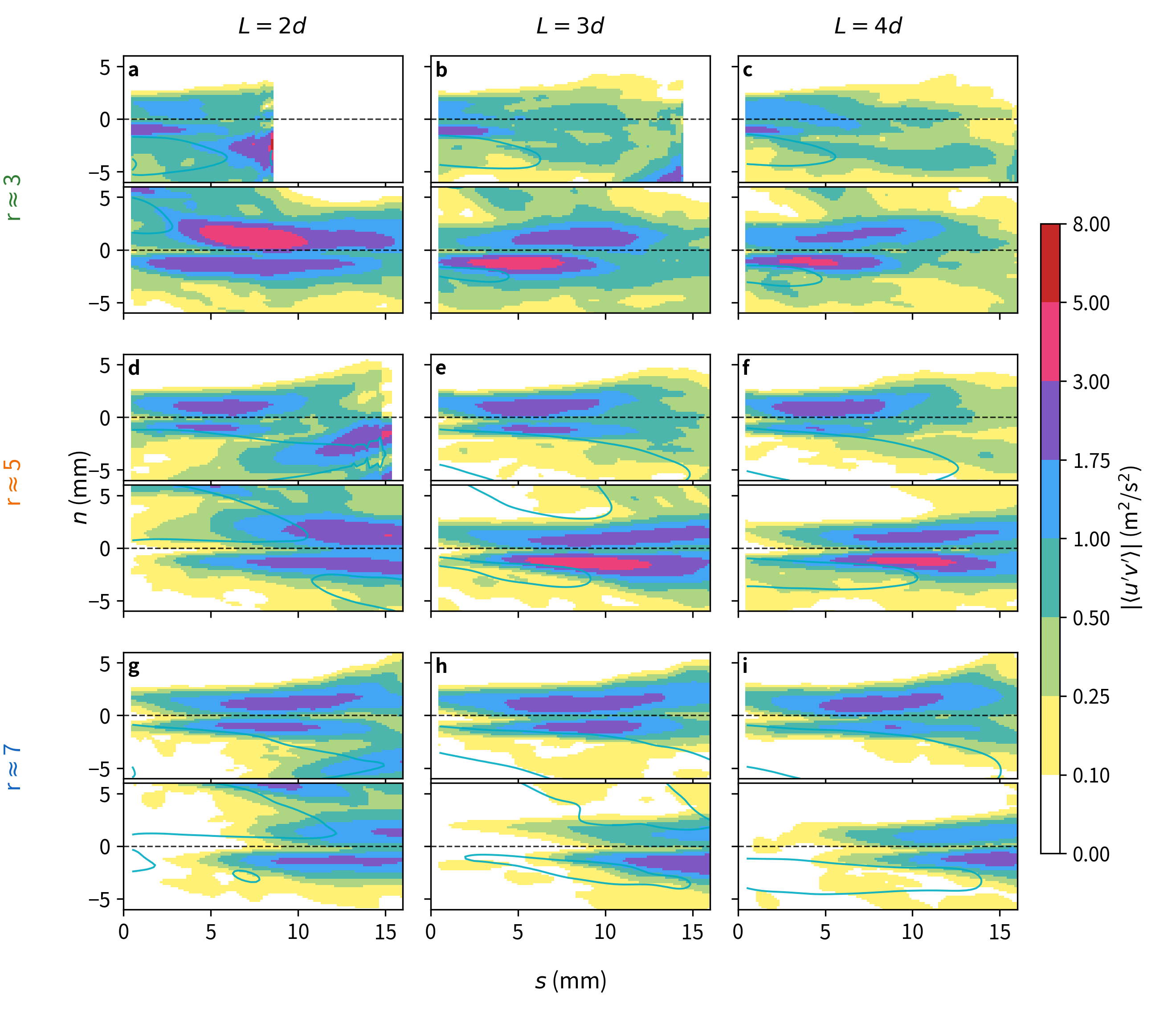}}
\caption{Gap-side turbulent momentum exchange over the nine conditions in the natural coordinates $(s,n)$ of each trajectory (rows $r\approx3$, 5, 7; columns $L=2d$, $3d$, $4d$, the arrangement of Table~\ref{table:conditions}): (a)--(i) condition-mean $|\langle u'v'\rangle|$ fields on the same absolute scale as Fig.~\ref{fig:sup-stress} (front jet in the upper and rear jet in the lower half of every panel), with the $u=0$ contour (cyan) and the trajectory $n=0$ (black dashed); the gap side faces the other jet.}
\alttext{Nine-grid of the condition-mean Reynolds-stress magnitude fields in the natural coordinates of each trajectory, with the zero-velocity contour and the trajectory line.}
\label{fig:exchange-nine}
\end{figure}

The condition-averaged fields (Fig.~\ref{fig:exchange-nine}) concentrate the gap-side exchange along the two shear layers, and the exchange grows as the spacing decreases.

\begin{figure}
\centerline{\includegraphics[width=\textwidth]{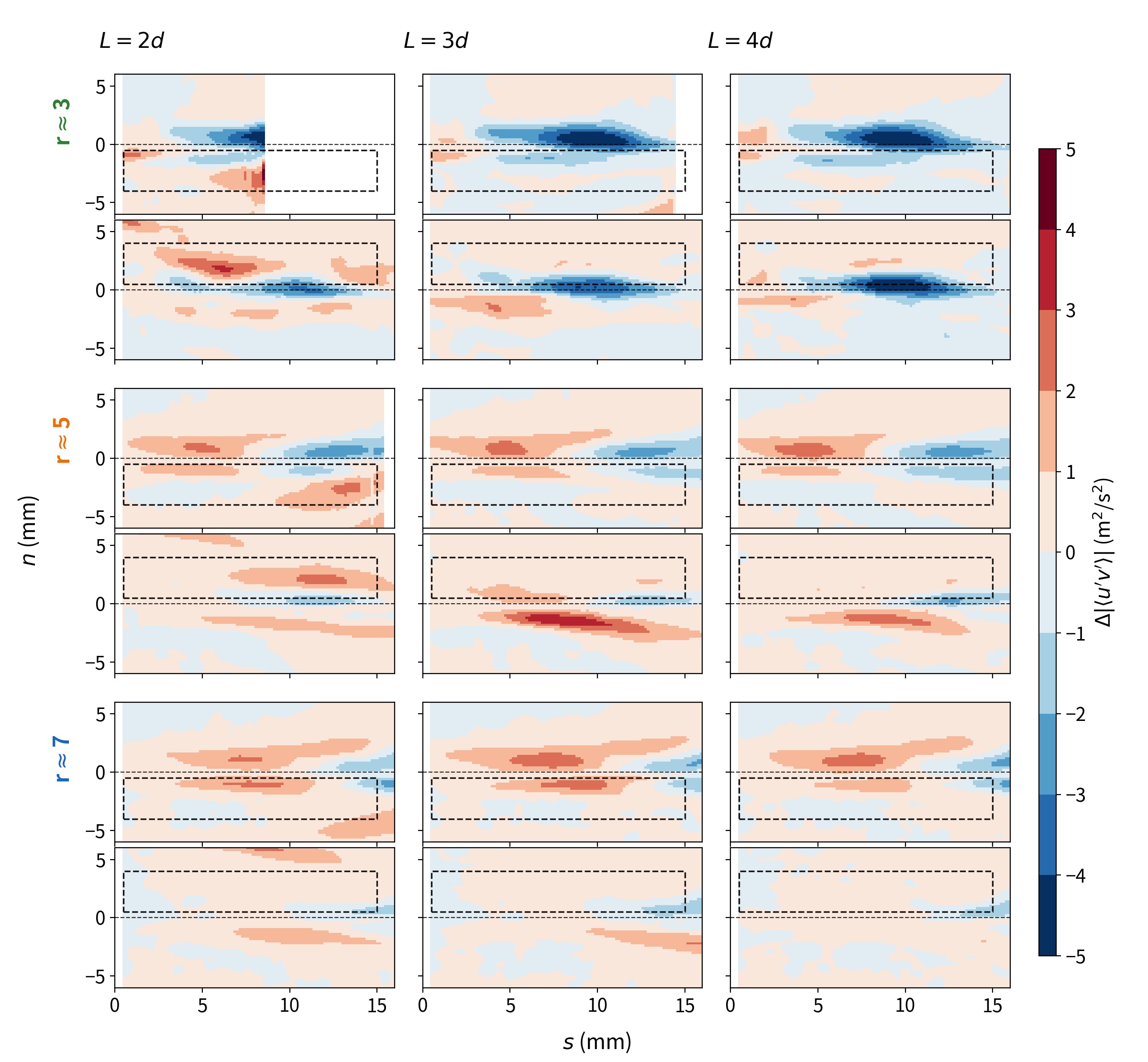}}
\caption{Natural-coordinate difference $|\langle u'v'\rangle|_{\rm twin}-|\langle u'v'\rangle|_{\rm single}$ of the condition-averaged fields against the single jet of the same velocity-ratio group (rows $r\approx3$, 5, 7; columns $L=2d$, $3d$, $4d$; front jet top, rear jet bottom; $s$ arc length, $n$ perpendicular offset; red: twin exceeds single, blue: twin below; fixed $\pm5$\,m$^2$/s$^2$ scale in bands of $1$). Dashed rectangles mark the gap-side zone $|n|=0.5$--$4$\,mm, $s\le15$\,mm over which the exchange integrals of Table~\ref{tab:natural-uv} are evaluated; the two off-specification records, with nozzles at $x\approx9.4$--$9.6$\,mm instead of 10\,mm, are excluded.}
\alttext{Natural-coordinate difference fields of the Reynolds stress between the twin jets and the single-jet reference over the nine conditions.}
\label{fig:promo-s5}
\end{figure}

Figure~\ref{fig:promo-s5} compares the twin fields with the single-jet reference in the natural coordinates of each trajectory and marks the gap-side zone over which the area-integrated statistics of Table~\ref{tab:natural-uv} are evaluated.

The difference fields show the enhancement filling the gap-side zone of both jets at $r\approx5$ and 7, and growing as the spacing decreases. At $r\approx3$ the twin values fall to or slightly below the single-jet reference, whose wake is itself strongly reversed at that ratio (Table~\ref{tab:natural-uv}).

The band-resolved decomposition of the gap-side kinetic energy is given in Fig. S2 of the supplementary material.

\begin{figure}
\centerline{\includegraphics[width=\textwidth]{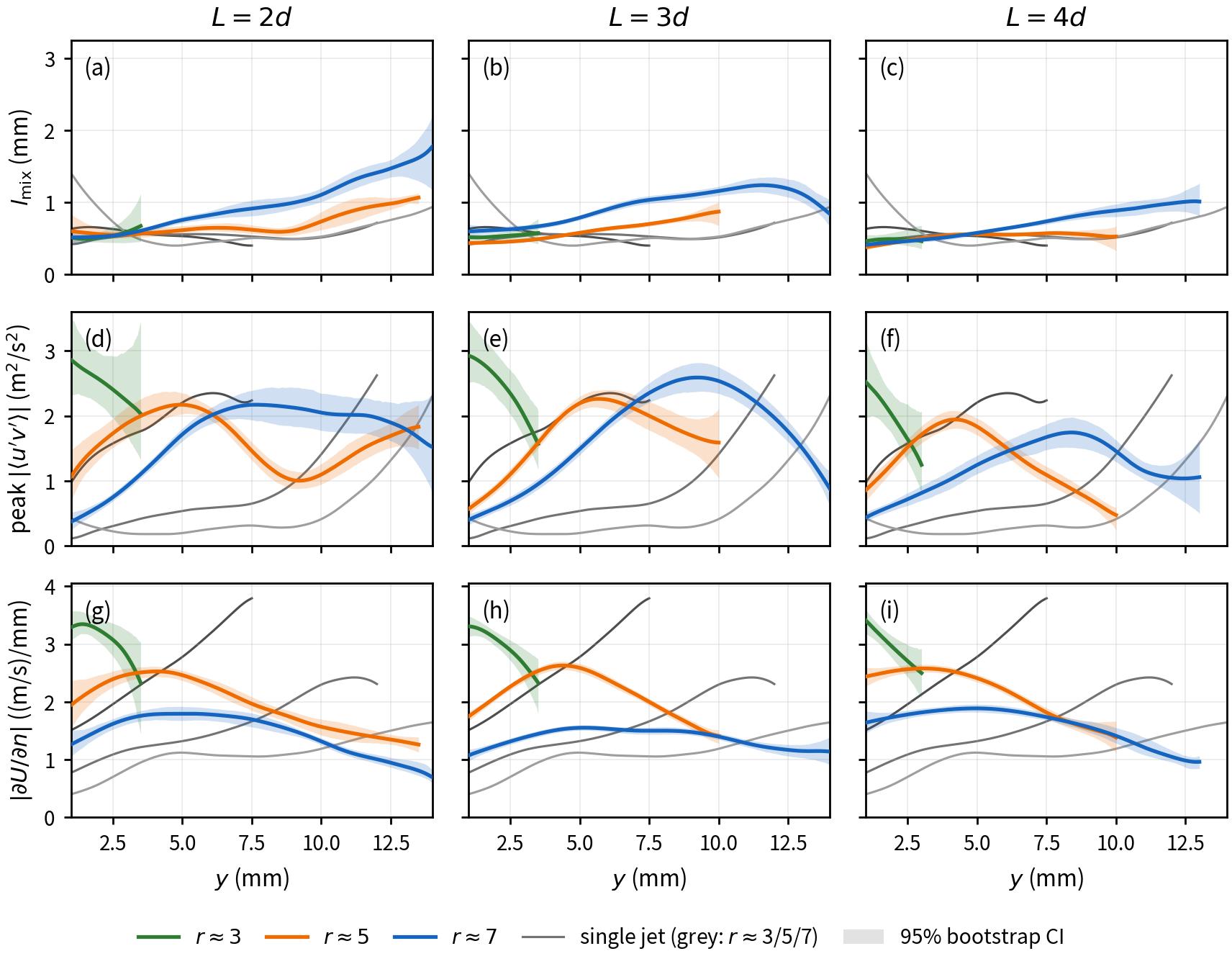}}
\caption{Turbulent momentum exchange of the gap-side shear layers, versus the height $y$: rows (a)--(c) Prandtl mixing length $l_{\rm mix}=\sqrt{|\langle u'v'\rangle|}/|\partial U/\partial n|$~\citep{pope_turbulent_2000}, (d)--(f) peak exchange $|\langle u'v'\rangle|$ and (g)--(i) mean shear $|\partial U/\partial n|$ across the layer; columns: spacing $L=2d$, $3d$, $4d$. Each panel shows the front-jet leeward layer of the three velocity-ratio groups (color: $r\approx3$, $5$, $7$) and, as thin gray curves, the single-jet leeward reference of the same ratio group (dark to light with increasing $r$; identical in every column, the single-jet records having no spacing dimension); solid lines: local-linear (LOWESS-type) fits through the per-station means, shaded: $95\%$ bootstrap confidence interval of the fitted mean (1000 resamples of the raw station data). Fits are not extrapolated beyond the measured $y$ range, which is shorter for the $r\approx3$ group ($y\le3.5$\,mm) and for several $L=3d$ and $4d$ groups ($y\le10$\,mm). The rear-jet windward layers are not drawn, their station coverage being too uneven across the grid (only $L=2d$ at $r\approx3$, $L=2d$ and $3d$ at $r\approx5$ and $7$, none at $L=4d$); the rear-jet statistics are given in Table~\ref{tab:natural-uv}.}
\alttext{Nine-panel grid of the Prandtl mixing length, peak exchange and mean shear of the gap-side layers versus height, arranged by measured quantity (rows) and spacing (columns), with local-linear fits, bootstrap confidence bands and single-jet reference curves for the three velocity-ratio groups.}
\label{fig:promo-s7}
\end{figure}

\begin{figure}
\centering
\centerline{\includegraphics[width=0.94\textwidth,height=0.72\textheight,keepaspectratio]{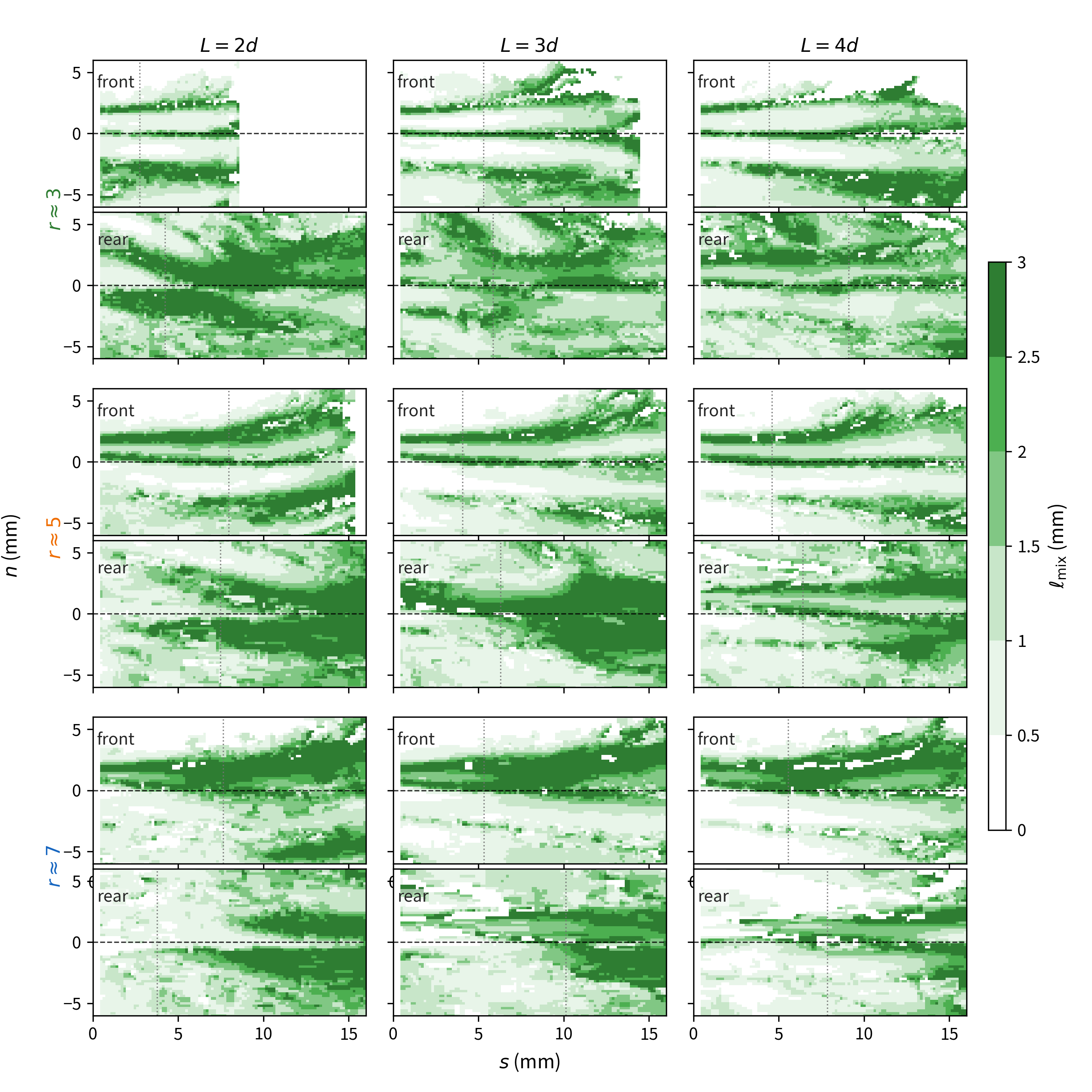}}
\caption{Condition-mean fields of the Prandtl mixing length $l_{\rm mix}=\sqrt{|\langle u'v'\rangle|}/|\partial U/\partial n|$ of Eq.~\eqref{eq:lmix} in the natural coordinates $(s,n)$ of each trajectory (rows $r\approx3$, 5, 7; columns $L=2d$, $3d$, $4d$; upper and lower panels of every cell: front and rear jet). The fields are the means over the 5--8 twin cases of each condition, on the fixed scale $0$--$3$\,mm in bands of $0.5$\,mm; the dashed line marks $n=0$ and the dotted line the near-field transition $s_{tn}$. The mixing length reaches $2$--$3$\,mm in a narrow band along the two shear layers and drops towards the jet axis and the outer edge, so the enlarged mixing scale read from Fig.~\ref{fig:promo-s7} occupies the layer itself rather than the surrounding field.}
\alttext{Nine panels of the condition-mean Prandtl mixing-length fields in the natural coordinates of the trajectories, showing the enlarged mixing scale along the two shear layers.}
\label{fig:lmix-fields}
\end{figure}

The enhancement is carried by the mixing scale rather than by a larger velocity difference across the layer (Fig.~\ref{fig:lmix-fields} shows the condition-mean fields). The zone-median mixing length $l_{\rm mix}$ on the front-jet gap side exceeds the single-jet leeward value in five of the six conditions at $r\approx5$ and 7 (up to $1.6\times$ in the zone-median convention of Table~\ref{tab:natural-uv}, the exception being the $L=4d$, $r\approx5$ condition at $0.94\times$); at $r\approx7$ the per-station values are more than doubled, $0.86\pm0.29$ against $0.39\pm0.09$\,mm (Fig.~\ref{fig:promo-s7}).

The difference field of the mixing length (Fig. S4 of the supplementary material) occupies the same gap-side zone at $r\approx5$ and 7. At $r\approx3$ the difference becomes small and mixed in sign, because the weakened shear there keeps the zone-median $l_{\rm mix}$ near the single-jet value (Table~\ref{tab:natural-uv}) without enhancing the exchange.

The enlarged mixing scale is not accompanied by a stronger mean shear. The implied scale $\sqrt{P_{uv}^{\rm zone}}/l_{\rm mix}$ of the front layer is comparable to or lower than that of the single-jet leeward zone (ratio $0.70$--$1.00$ over the nine conditions), so the exchange increase is carried by the mixing scale and by the persistence of the layer rather than by a steeper shear. Since $l_{\rm mix}$ is derived from the stress and the mean shear, this decomposition is kinematic bookkeeping. The independent thickness $\delta_\omega$ ($2.00$ versus $1.69$\,mm) supports the enlargement, but only by $1.2\times$, and the peak vorticity is lower in the twin layer.

In the Prandtl mixing-length picture~\citep{pope_turbulent_2000} ($l_{\rm mix}$ follows the standard notation and should not be confused with the momentum length $\ell_M$ of the companion paper),
\begin{equation}
l_{\rm mix}=\sqrt{\frac{|\langle u'v'\rangle|}{(\partial U/\partial n)^2}},
\label{eq:lmix}
\end{equation}
characterizes the scale of the momentum mixing across the layer, with $n$ denoting the layer-normal direction (approximated by the horizontal $x$ in the quasi-vertical near field).

Figure~\ref{fig:exchange-nine} and Table~\ref{tab:natural-uv} quantify how the exchange depends on the velocity ratio and the spacing; the four statistics are shown in Fig. S20 of the supplementary material.

The gap-side area-integrated exchange $S_{uv}=\iint|\langle u'v'\rangle|\,ds\,dn$ of the front jet grows as the spacing decreases; it exceeds the single-jet reference at $r\approx7$ for every spacing and at $r\approx5$ for $L=2d$ and $3d$. At $r\approx3$ the single-jet wake itself exchanges momentum most strongly, consistent with the near-disappearance of the natural-coordinate enhancement (Fig.~\ref{fig:promo-s5}).

The single jet also carries a leeward recirculation, yet its trajectory follows the standard power law. Its recirculation is self-induced: the wake is left behind by the already-bent jet, and it is therefore a consequence of the bending set by the crossflow momentum exchange rather than its cause. The front-jet gap recirculation is instead externally forced by the rear-jet blockage and enclosed between the two jets, and its layers exchange momentum an order of magnitude more strongly (above). The front jet therefore bends as a single jet would, with an extra recirculation-driven deflection, the near-field transition up to 40\% earlier than for the single jet.

Three coupled effects carry the enhancement. (i) The reversed configuration itself: its shear-layer entrainment ratio grows as the gap-side stream turns from co-flowing to reversed~\citep{dimotakis_two-dimensional_1986}. (ii) The weakened front jet amplifies the entrainment further. (iii) The rear jet encloses the bubble and continuously replenishes its top, so the zone fluid climbs along the jet at essentially every sampled point ($\theta<90^\circ$ at 99--100\% of the points versus 84--100\% for the single jet, Fig. S1b) and the entrainment never stalls.

Frame-resolved, the persistence of the reversal is measured by the reverse-flow probability $P$ sampled in the same entrainment zone. The zone mean grows with the velocity ratio in the twin-jet configuration while the single-jet zone mean remains far below the twin value. The most persistent zone points (the 90th percentile of the frame-resolved reverse-flow probability over the zone, $P_{90}$) reach $0.72$ at $r\approx7$ versus $0.07$ for the single jet. At $r\approx3$ the single-jet wake is the more persistent of the two, matching the near-disappearance of the exchange enhancement there.

The entrainment also has a purely geometric condition, which the measurements satisfy: the reverse-flow zone fluid must climb along the front jet instead of returning downward. That is, the angle
\begin{equation}
\cos\theta=\frac{\mathbf v_b\cdot\hat{\mathbf t}}{|\mathbf v_b|}, \qquad u<0,
\label{eq:theta}
\end{equation}
between the zone velocity $\mathbf v_b=(u,v)$ and the front-jet tangent $\hat{\mathbf t}$ at the nearest trajectory point must stay below $90^\circ$.

$\theta<90^\circ$ holds at essentially every sampled point of the twin gap zone (99--100\%), the condition-median angles decreasing with both the velocity ratio and the spacing as the closest spacing turns the reversed flow most fully into the jet direction. The single-jet leeward zone satisfies the same condition only partially (84--100\% of its sampled points). The acute angle is thus a prerequisite for the entrainment rather than the source of the enhancement (Fig. S21 of the supplementary material). The nine-condition statistics of the entrainment angle are given in Fig. S22 of the supplementary material.

Of the three leading terms of the vertical momentum balance (convection, turbulent transport, pressure force; Fig. S1c,d), the mean convection term ($5.0$--$8.1$\,m/s$^2$) is the largest. It is interpreted as the response of the mean flow to the bending of the trajectory rather than as a driving force.

Table~\ref{tab:enhance-verify} verifies the pooled per-station contrast at the velocity-ratio-group level. The per-station peak of the front gap-side layer exceeds the pooled single-jet value in every velocity-ratio group (up to $11.2\times$, marginally at $r\approx3$), with the layer thicker and its peak vorticity not enhanced.

\begin{table}
\centering
\caption{Verification of the exchange enhancement of the front gap-side layer against the single-jet leeward layer (per-station-mean peak $|\langle u'v'\rangle|$ in $\pm2$\,mm windows at $y=2$--$10$\,mm; Mann--Whitney): the ratio exceeds unity in every velocity-ratio group and grows with the velocity ratio. This pooled per-station ratio is a robustness metric of the primary per-height window-peak ratio of Section~\ref{sec:ccsl} (up to $27\times$ over the pooled single-jet leeward layer across the nine conditions): the pooling over stations and heights lowers the ratio relative to the per-height comparison. $\delta_\omega=(V_1-V_2)/|\omega_z|_{\max}$ is the vorticity thickness and $\omega_z^*=\omega_z\,d/u_j$ the normalized peak layer vorticity (mean $\pm$ std over the cases; the two lower rows are pooled over the nine conditions).}\label{tab:enhance-verify}
\alttext{Table verifying that the twin/single peak exchange ratio of the front gap-side layer exceeds unity in every velocity-ratio group, with the vorticity thickness and normalized peak vorticity of the two layers.}
\begin{tabular}{llll}
\toprule
velocity-ratio group & twin/single peak ratio & $p$ & \\
\midrule
$r\approx7$ & $11.2\times$ & $3\times10^{-7}$ & \\
$r\approx5$ & $4.6\times$ & $4\times10^{-6}$ & \\
$r\approx3$ & $1.7\times$ & $0.05$ & \\
\midrule
quantity & twin & single & $p$ \\
$\delta_\omega$ (mm) & $2.00\pm0.30$ & $1.69\pm0.26$ & $3\times10^{-9}$ \\
$\omega_z^*$ & $1.73\pm0.44$ & $2.65\pm0.38$ & $10^{-20}$ (two-sided) \\
\bottomrule
\end{tabular}
\end{table}

The measured exchange is the force-level counterpart of the entrainment invoked in the single-jet shear-layer studies---the entrainment ratio of \citet{dimotakis_two-dimensional_1986}, the momentum and vorticity thicknesses of \citet{nair_combustion_2022} and \citet{nair_near-field_2023}---quantities that those studies employ to scale the layer without evaluating the exchange itself.

Physically, the vertical momentum is drawn into the gap recirculation zone not by the mean pressure but by the roll-up of the gap-side layer: the shear-layer vortices growing on the front-jet leeward layer wrap the reversed gap fluid into the jet. The transverse Reynolds stress $\langle u'v'\rangle$ therefore decays from the jet side toward the zone interior. Its gradient, together with that of $\langle v'v'\rangle$, then acts as a net upward force on the zone-top fluid, which is continuously peeled off and injected into the bending jet. The mass deficit is replenished by the reversed flow returning from downstream and by ambient fluid entrained from above, closing the recirculation loop.

The chain through which this exchange bends the jet earlier is therefore a momentum dilution rather than a lateral push. In terms of the momentum-length scaling of the companion paper ($\ell_M\sim rd$), it acts as an effective reduction of the velocity ratio, $r_{\mathrm{eff}}<r$---the same effective-velocity-ratio idea used by \citet{new_trajectory_2015} to collapse merged tandem-jet trajectories. The jet then reaches its momentum length $\ell_M\sim r_{\mathrm{eff}}d$ sooner, the crossflow momentum dominates earlier, and the jet bends before the single jet does.

The measurements assembled here---the rear-jet blockage, the enclosed recirculation it sustains, the exchange that dilutes the front jet, and the earlier bending this produces---are condensed into the one-parameter velocity-dilution scaling of the companion paper~\citep{companion_part1}; the chain is tested step by step in the following subsections.

\subsection{\label{sec:dose}The cause of the bending: the pressure force excluded}

\begin{figure}
\centering
\centerline{\includegraphics[width=\textwidth]{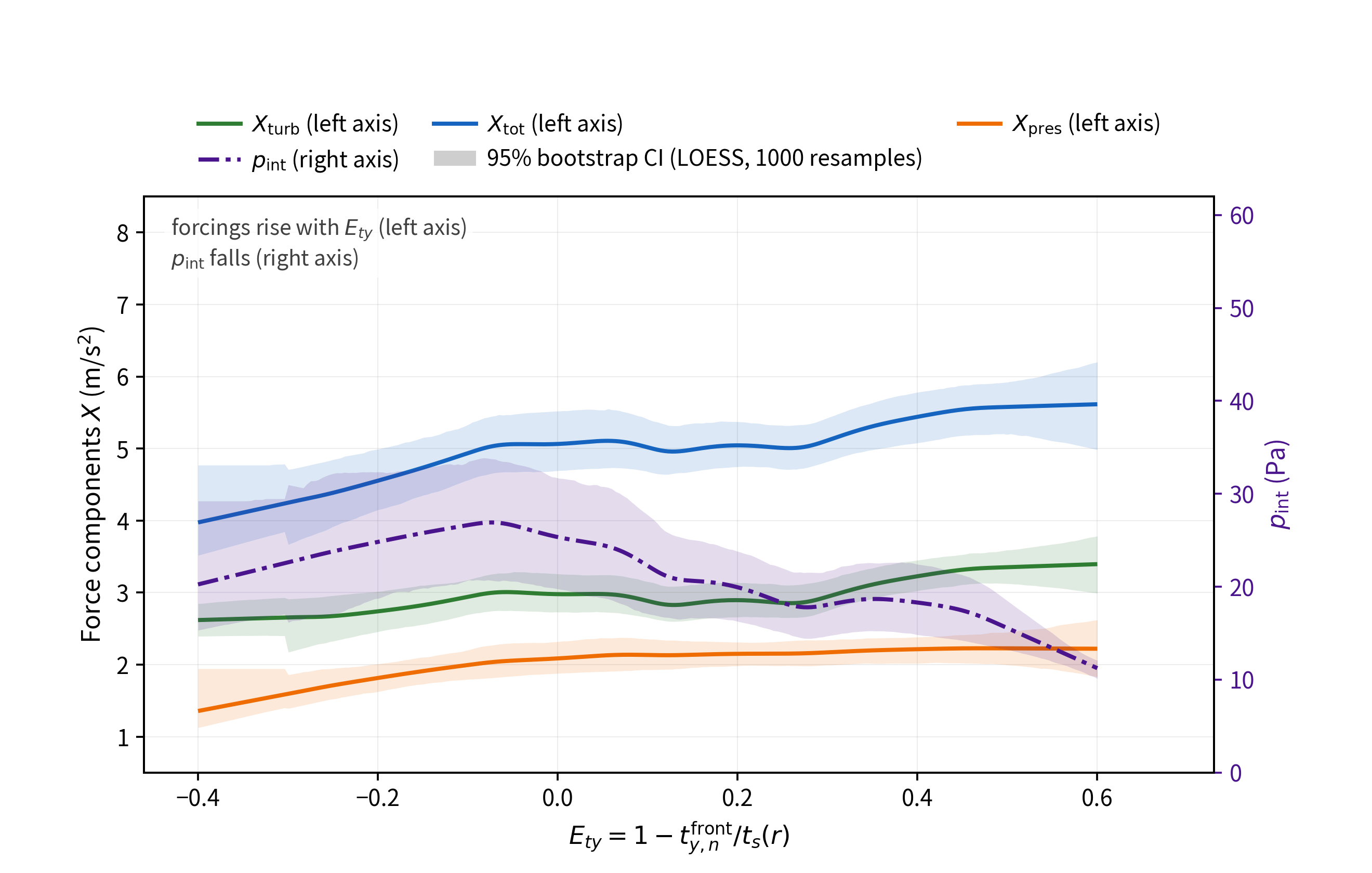}}
\caption{Dose--response of the front-jet bending against its candidate causes, over the 56 cases with a detectable front-jet near-field transition. A single panel with two ordinates plots the three resolved forcing terms---turbulent transport $X_{\rm turb}$, total forcing $X_{\rm tot}=X_{\rm turb}+X_{\rm pres}$ and pressure force $X_{\rm pres}$ (left axis, m$^2$/s$^2$; $s=2$--$12$\,mm zone means)---and the closest-approach gap pressure $p_{\rm int}$ (dash-dotted, right axis, Pa) against the near-field shortening $E_{ty}$ of Eq.~\eqref{eq:ety}. The trends are local-linear (LOWESS-type) fits to the case means on a $0.20$ grid of $E_{ty}$ (tricube kernel, bandwidth $0.30$), with shaded $95\%$ bootstrap confidence bands ($1000$ resamples of the 56 cases); no binned symbols are drawn. All three forcing magnitudes grow with the shortening, while the pressure level runs the other way.}
\alttext{Dose-response figure with two ordinates: the near-field turbulent transport, total forcing and pressure force on the left axis and the gap pressure on the right axis, each shown as a local-linear fit with a 95\% bootstrap confidence band against the near-field shortening.}
\label{fig:dose-response}
\end{figure}

The exchange characterized above is only a candidate carrier until the alternative causes are ruled out; each candidate is therefore examined against the trajectory modification at the condition level over the nine conditions and per case over the 56 transition-reliable cases (Table~\ref{tab:bend-exchange}). The pressure force is examined first. A low-pressure attraction---the classical reading of twin-jet bending---requires a gap pressure depressed below ambient and lowest wherever the shortening $E_{ty}$ is largest. The measured closest-approach gap pressure does fall, from about $28$\,Pa at $E_{ty}\approx-0.2$ to about $12$\,Pa at $E_{ty}\approx0.6$ (Fig.~\ref{fig:dose-response}), so it is lowest exactly where the shortening is largest; it never becomes sub-atmospheric, however, and the gap is pressurized at every condition (Section~\ref{sec:ccsl}). With no suction to drive the bending, and a condition-level correlation that is negative but not significant ($\rho=-0.35$, $p=0.36$, $n=9$), the pressure force is excluded as the cause of the bending.

The trajectory analysis follows the extraction procedure of the companion paper~\citep{companion_part1}: an automated row-tracking algorithm yields each trajectory from the condition-mean velocity-magnitude field, the near- and far-field transitions are detected along it, and the two segments are fitted separately---the near field as $x-x_0=A_n y^{B_n}$ and the far field as $y/(rd)=A_f\,[x/(rd)]^{B_f}$. The near-field transition height $t_{y,n}$ is one of the fitted parameters, and the single-jet records of the same velocity ratio supply the reference $t_s(r)$ ($4.77$, $7.94$ and $9.17$\,mm at $r\approx3$, $5$ and $7$). The shortening of the front-jet near-field segment is measured by

\begin{equation}
E_{ty}\equiv1-\frac{t_{y,n}^{\rm front}}{t_s(r)},
\label{eq:ety}
\end{equation}

with negative values meaning that no shortening is detected. The penetration loss $A_{\rm loss}\equiv1-A_f^{\rm front}/A_s(r)$, the decay-rate excess $E_{\rm lam}\equiv\lambda/\lambda_s-1$ and the penetration-depth loss $H_{\rm loss}\equiv1-H/H_s(r)$ are defined in the same relative form. The single-jet references are $A_s(r)=0.1433r+1.2384$, $H_s=21.89$, $37.79$ and $38.12$\,mm and $\lambda_s=0.0912$, $0.0726$ and $0.0572$\,mm$^{-1}$ at $r\approx3$, $5$ and $7$~\citep{companion_part1}.

\begin{table}
\centering
\caption{Condition-mean pairing of the front-jet bending with the gap-side exchange behind the correlations of Section~\ref{sec:dose}: the near-field transition height $t_{y,n}^{\rm front}$, its fractional shortening $E_{ty}$ (single-jet reference $t_s(r)=4.77$, $7.94$, $9.17$\,mm; negative = no detected shortening), the near-field turbulent transport $X_{\rm turb}$ and pressure force $X_{\rm pres}$ ($s=2$--$12$\,mm zone means) and the area-integrated exchange $S_{uv}$ of Table~\ref{tab:natural-uv}.}
\alttext{Table pairing the front-jet near-field shortening with the gap-side turbulent transport, pressure force and area-integrated exchange for the nine conditions.}
\label{tab:bend-exchange}
\footnotesize
\renewcommand{\arraystretch}{0.92}
\begin{tabular}{lcccccc}
\toprule
Condition & $r$ & $t_{y,n}^{\rm front}$ (mm) & $E_{ty}$ & $X_{\rm turb}$ (m/s$^2$) & $X_{\rm pres}$ (m/s$^2$) & $S_{uv}$ (m$^2$/s$^2$\,mm$^2$) \\
\midrule
$L=2d$, $r\approx3$ & 2.79 & 2.850 & 0.402 & 3.817 & 2.718 & 32.219 \\
$L=2d$, $r\approx5$ & 4.62 & 7.980 & $-0.005$ & 3.191 & 1.840 & 44.574 \\
$L=2d$, $r\approx7$ & 6.22 & 7.571 & 0.175 & 2.768 & 1.693 & 33.383 \\
$L=3d$, $r\approx3$ & 2.76 & 4.864 & $-0.021$ & 2.286 & 1.898 & 30.125 \\
$L=3d$, $r\approx5$ & 4.64 & 5.279 & 0.335 & 3.669 & 2.355 & 29.600 \\
$L=3d$, $r\approx7$ & 6.32 & 6.241 & 0.320 & 3.333 & 2.132 & 31.602 \\
$L=4d$, $r\approx3$ & 2.78 & 4.372 & 0.083 & 2.513 & 1.683 & 29.147 \\
$L=4d$, $r\approx5$ & 4.67 & 4.730 & 0.405 & 3.317 & 2.551 & 22.334 \\
$L=4d$, $r\approx7$ & 6.53 & 5.650 & 0.384 & 2.858 & 2.914 & 24.147 \\
\bottomrule
\end{tabular}
\end{table}

The bending effect, measured by the shortening $\Delta t=t_s-t_{y,n}^{\rm front}$ and its fractional form $E_{ty}$ of Eq.~\eqref{eq:ety}, scales with the near-field exchange: at the condition level the Spearman coefficients are $\rho=0.68$ ($p=0.042$) for $E_{ty}$ against the turbulent transport and $0.87$ ($p=0.003$) against the total forcing ($\rho=0.42$ and $0.48$ per case). The pressure force alone correlates per case at $\rho=0.39$ ($p=0.0027$, Table~\ref{tab:degree-matrix}), whereas the closest-approach gap pressure is the counter-example: its condition-level rank correlation is negative but not significant ($\rho=-0.35$, $p=0.36$, $n=9$; per case $-0.45$). The local-linear dose--response (Fig.~\ref{fig:dose-response}) confirms the pattern---the three forcing magnitudes rise with the shortening while the pressure level falls---and the effect sizes are moderate, $E_{ty}$ accounting for about 14\% of the variance case by case, so these are dose--response relations in the mean rather than tight relations; the per-case scatter behind these fitted trends is given in Fig.~S3 of the supplementary material.

\begin{table}
\centering
\small
\caption{Spearman correlations of the modification metrics of the companion paper---the percentage-form trajectory and distribution parameters, positive values denoting a stronger modification (the turning-angle row uses the deficit $\Delta\theta^{\rm single}-\Delta\theta^{\rm front}$)---with the four forcing and pressure quantities, per case and at the condition level (nine conditions; eight for the far-field metrics, condition 1 having no far-field fit; $B_n$ is condition-level only). Significance: $^*p<0.05$, $^{**}p<0.01$, $^{***}p<0.001$, without multiplicity correction; the per-case sample is the 56 cases with a detectable front-jet near-field transition (the $p_{\rm int}$ column uses the 56 cases and nine conditions for which the closest-approach pressure is available; its condition-level entry is the nine-condition Spearman). Only the near-field transition family reproduces the dose--response pattern (positive against the force-level quantities, negative against the pressure level); the far-field metrics track the velocity-ratio and merging axes, and the decay excess is mixed.}\label{tab:degree-matrix}
\alttext{Table of the Spearman correlations of the modification metrics against the turbulent transport, total forcing, pressure force and gap pressure, per case and at the condition level with significance stars.}
\footnotesize
\resizebox{\textwidth}{!}{%
\begin{tabular}{llcccc}
\toprule
Degree metric & $n$ (case/cond) & $X_{\rm turb}$ & $X_{\rm turb}+X_{\rm pres}$ & $X_{\rm pres}$ & $p_{\rm int}$ \\
\midrule
$E_{ty}$ (near-field shortening) & 56/9 & $+0.42^{**}/+0.68^{*}$ & $+0.48^{***}/+0.87^{**}$ & $+0.39^{**}/+0.78^{*}$ & $-0.45^{***}/-0.35$ \\
$\Delta t$ (absolute shortening) & 56/9 & $+0.39^{**}/+0.53$ & $+0.47^{***}/+0.68^{*}$ & $+0.41^{**}/+0.70^{*}$ & $-0.52^{***}/-0.67^{*}$ \\
$B_n$ (near-field exponent) & --/9 & --$/+0.25$ & --$/+0.43$ & --$/+0.77^{*}$ & --$/-0.70^{*}$ \\
$\Delta\theta^{\rm single}-\Delta\theta^{\rm front}$ & 67/9 & $-0.05/-0.07$ & $-0.11/-0.15$ & $-0.19/-0.30$ & $+0.57^{***}/+0.68^{*}$ \\
$A_{\rm loss}$ (penetration loss) & 59/8 & $-0.43^{***}/-0.52$ & $-0.37^{**}/-0.33$ & $-0.16/-0.05$ & $+0.46^{***}/+0.24$ \\
$H_{\rm loss}$ (depth loss) & 67/9 & $+0.08/+0.05$ & $-0.03/-0.07$ & $-0.18/-0.28$ & $+0.75^{***}/+0.78^{*}$ \\
$\varphi/\varphi_s-1$ (deflection excess) & 67/9 & $-0.27^{*}/-0.23$ & $-0.10/-0.12$ & $+0.16/+0.23$ & $-0.46^{***}/-0.53$ \\
$B_f/B_s$ (far-field exponent ratio) & 59/8 & $-0.35^{**}/-0.31$ & $-0.34^{**}/-0.17$ & $-0.20/+0.02$ & $-0.07/-0.07$ \\
$E_{\rm lam}$ (decay-rate excess) & 68/9 & $+0.30^{*}/+0.62$ & $+0.22/+0.45$ & $-0.01/+0.08$ & $+0.38^{**}/+0.37$ \\
\bottomrule
\end{tabular}}
\end{table}

The trends are robust: they survive a Bonferroni correction for the four quantities, are leave-one-out stable, and persist when the measured velocity ratio $r$ is partialled out ($+0.35$, $+0.33$ and $-0.33$; $p=0.007$--$0.015$), so they are not an artifact of comparing the three velocity-ratio groups. Within the individual groups the gap pressure keeps its negative sign where the sample is largest ($r\approx5$: $\rho=-0.72$; $r\approx7$: $\rho=-0.55$), whereas the turbulent transport is only marginally significant within a group, so part of its correlation comes from the between-group contrast; at the per-case level the transport--shortening link survives partialling out the spacing and the velocity ratio, while the pressure-force link drops below significance once the transport is controlled for.

The pressure force therefore co-varies with the transport without reviving the pressure mechanism: the two resolved forcing terms are correlated ($\rho=+0.42$), and the pressure is reconstructed with the turbulent source included (the companion paper), so it is not an independent regressor; both are outcomes of the same exchange dynamics, which is why the total forcing, their sum, ranks highest of the three. What scales with the shortening is the absolute transport, not the twin/single enhancement ratio ($\rho=0.68$ against $+0.13$ at the condition level). The condition-level coefficients of the two force terms ($+0.68$ and $+0.78$, $n=9$) are not statistically distinguishable on this sample, so the assignment rests on this common-origin argument rather than on the coefficient ranking. The partial-correlation and mediation analysis (Fig.~S16 of the supplementary material) confirms the transport--shortening link and locates the turning-angle change on the enhancement ratio ($\rho=+0.77$), as expected along the velocity-ratio axis.

A leading-term comparison of the vertical momentum balance restates the same assignment at the force level. The steady incompressible $y$-momentum equation reads

\begin{equation}
U\frac{\partial V}{\partial x}+V\frac{\partial V}{\partial y}
=-\frac{1}{\rho}\frac{\partial P}{\partial y}
-\frac{\partial\langle u'v'\rangle}{\partial x}
-\frac{\partial\langle v'v'\rangle}{\partial y}
-\frac{\partial\langle v'w'\rangle}{\partial z},
\label{eq:ybudget}
\end{equation}

with the viscous term neglected. Sampled in the gap-side zone ($0\le n\le4$\,mm) of each jet (Fig.~S17 of the supplementary material), the resolved terms locate the exchange: the front-jet gap-side turbulent transport ($2.3$--$3.8$\,m/s$^2$) exceeds that of the rear jet ($0.7$--$1.6$\,m/s$^2$) and of the single jet ($0.47$\,m/s$^2$) at every spacing, and it exceeds the pressure term in eight of the nine conditions ($|X_{\rm turb}|/|X_{\rm pres}|=1.2$--$1.7$), becoming comparable only at $L=4d$, condition 9 being the only pressure-dominated near-field mean, consistent with the strong near-wall pressure field of the wider gap recirculation (the companion paper). The resolved turbulent term is therefore the larger of the two resolved terms at the narrower gaps, the spanwise term being unmeasured, and the pressure force becomes comparable only at the widest gap, where the bending nonetheless scales with the exchange. The comparison is one of magnitudes rather than a closed budget: the spanwise term is unavailable in the center plane, and the pressure force is not an independent regressor (the companion paper).

Across the parameter set the near-field proxy of the effective ratio, $1-E_{ty}\approx0.60$--$1.02$, gives a transition height up to 40\% shorter than the single-jet value in seven of the nine conditions. This proxy is operational rather than exact: the whole-trajectory effective ratio is obtained by curve matching against the single-jet library, and the family follows the one-parameter dilution law $r_{\mathrm{eff}}/r=r/(r+c)$ with $c=3.31$~\citep{companion_part1}.

To separate the two contributions, recall the exit-velocity baseline $r_{\rm cf}/r\approx0.83$ of the companion paper: the part of the proxy above this baseline ($0.83$--$1.02$) is compatible with the exit deficit alone, and the reduction below it (down to $0.60$) is the interaction contribution. In the strongest-shortening conditions the measured shortening ($33$--$40\%$) exceeds the $17\%$ exit-deficit bound by a factor of $1.9$--$2.4$, and $52\%$ of the 56 transition-reliable cases fall below the baseline (12 of the 68 twin cases have no detectable near-field transition and are left out of the per-case statistics). The decomposition splits the near-field proxy $S\equiv1-E_{ty}$ into the exit-deficit bound $S_{\rm exit}=1-0.83=0.17$ and the interaction part $S_{\rm int}=S-S_{\rm exit}$, which carries $51$--$71\%$ of $S$ in every condition, in agreement with the counterfactual analysis of the companion paper. The low-momentum prediction shortens $t_{y,n}$ by at most $1.5$\,mm ($16\%$ of $t_s(r)$ at $r\approx7$) against the up-to-$40\%$ shortening measured in the same dataset, and the baseline is conservative: the near-exit velocity is measured at the first resolvable station ($\approx0.5$\,mm), where the front jet already decelerates under the gap-side interaction, so part of the $12.8$ versus $15.4$\,m/s difference is the interaction's near-exit signature rather than a supply effect. The true supply-side deficit is therefore smaller than $17\%$, and the interaction fractions quoted above are lower bounds. The pressure-side controls thus leave the pressure force without a driving role; the next subsection asks whether the exchange instead dilutes the front-jet momentum as a redistribution rather than as a mass addition (Section~\ref{sec:flux}).

\subsection{\label{sec:flux}The dilution as a redistribution}

The curve-matching estimator of the companion paper measures a trajectory shape. This subsection asks what the momentum and mass flux budgets of the front jet say about the same dilution. If the exchange is the primary cause, the dilution should appear as a redistribution of front-jet momentum---a centerline deficit and a momentum-flux loss---rather than as an addition of mass, and the flux budget tests exactly this expectation. Natural-coordinate profiles $(s,n)$ are built from the instantaneous fields (400 frames per record; the mean-only record 4d-5r-8 contributes no stress terms). At each station the excess velocity $\Delta u=u_s-U_\infty t_x(s)$ is integrated over the contiguous jet region, defined by $\Delta u>\max(0.08\,\Delta u_{\rm peak},\,0.2\ {\rm m/s})$ and truncated at the gap-side valley, at the windward blockage minimum and at the merging station. The analysis window is the first contiguous pre-merging segment ($\tilde u=s/(rd)\le0.35$ unless merging intervenes earlier).

The truncation at the gap-side valley biases the excess-mass growth downward, because the widened gap-side flank of the diluted profile falls outside the contiguous jet region. The redistribution reading below nevertheless rests primarily on the centerline deficit and the momentum-flux loss, neither of which depends on the truncation choice; a relaxed truncation can only strengthen the mass-growth side without touching the deficit side (Section~\ref{sec:limitations}).

The bookkeeping is first validated on the single-jet records: the near-exit volume flux $Q(s_0)=\int\Delta u\,dn$ averages $41.3$ against the nominal $V_j d=40$ (mm\,m/s per unit span), and the leave-one-out momentum-to-mass ratio $u_{\rm eff}=J_{\rm adv}/Q$, with $J_{\rm adv}=\int\Delta u^2\,dn$, recovers the single-jet family at $u_{\rm eff}/V_j=1.018$ (range $0.71$--$1.19$).

\begin{figure}
\centerline{\includegraphics[width=\textwidth]{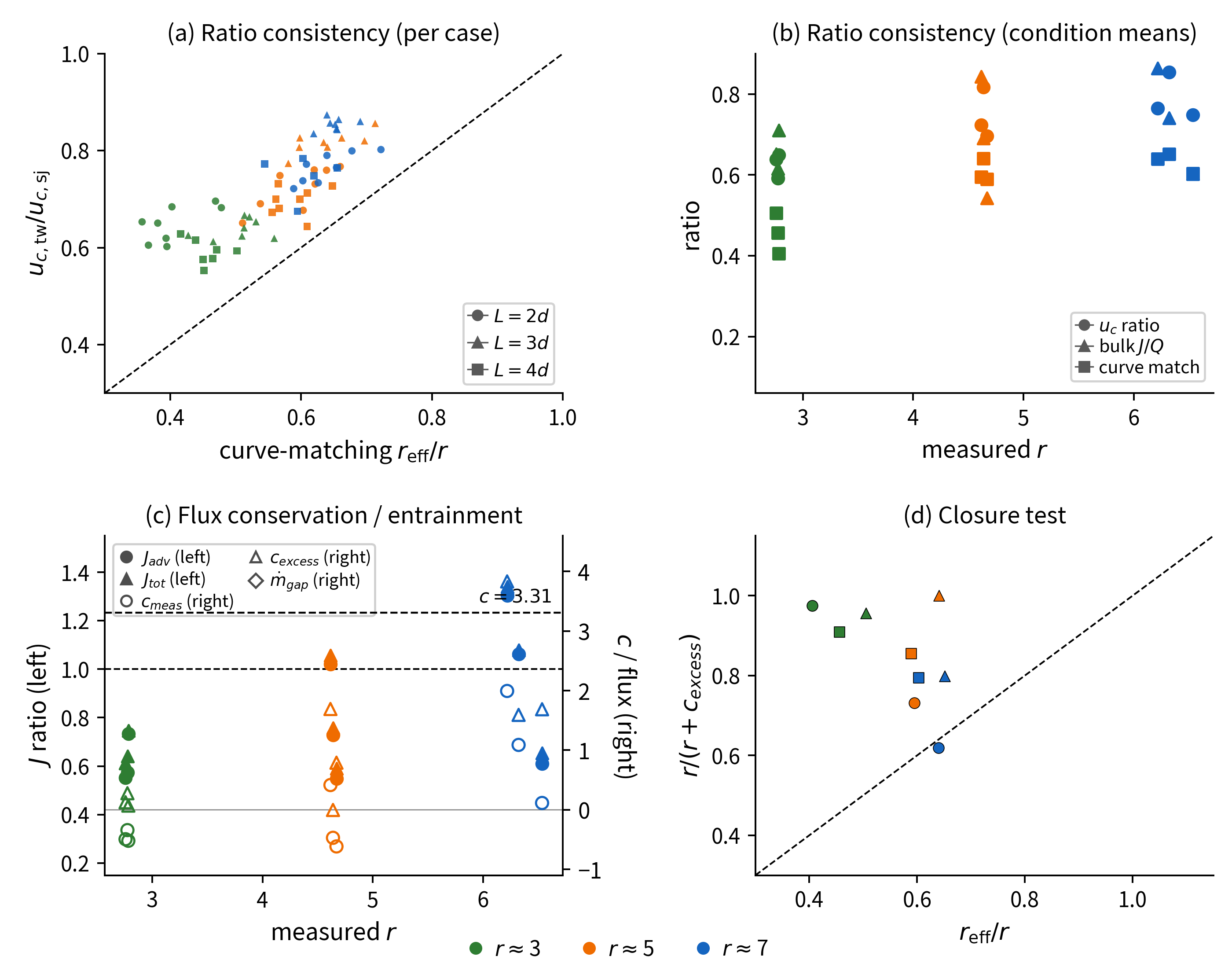}}
\caption{Flux-budget diagnostics. (a) Ratio consistency per case: the centreline excess-velocity ratio against the curve-matching ratio of the companion paper (dashed: 1:1). (b) The same three ratios as condition means versus $r$ (colour: velocity-ratio group). (c) Flux conservation and entrainment: the momentum-flux conservation ratio with (triangles) and without (circles) the normal stress on the left axis (dashed: unity), and the three direct entrainment readings on the right axis (open markers; dashed: the fitted $c=3.31$). (d) Closure test $r/(r+c_{\rm excess})$ versus the curve-matching ratio (dashed: 1:1). The example flux evolutions $Q/Q_0$, $J/J_0$ are deferred to Fig.~S18 of the supplementary material.}
\alttext{Six-panel flux-budget figure: the centerline excess-velocity ratio against the curve-matching effective ratio, the condition means, the momentum-flux conservation ratio, the three entrainment readings against the fitted capacity constant, the closure test and example flux evolutions.}
\label{fig:fluxbudget}
\end{figure}

Three results follow (Fig.~\ref{fig:fluxbudget}).

First, the centerline excess ratio $u_{c,{\rm tw}}/u_{c,{\rm sj}}$ at matched $\tilde u$ confirms the dilution ranking independently of the trajectory fit (Pearson $+0.82$ per case, $+0.87$ per condition against the curve-matching ratios of the companion paper). It sits at a systematically shallower level ($0.72$ against $0.565$), because the trajectory integrates the bending history beyond the local velocity ratio. Second, the bulk readout $u_{\rm eff}=J_{\rm adv}/Q$ does not reproduce the per-case ratios (Pearson $+0.33$); the along-trajectory momentum flux is conserved in only two of the nine conditions ($J(\tilde u^*)/J(\tilde u_0)=1.02$ and $1.06$ at $L=2d$, $r\approx5$ and $L=3d$, $r\approx7$). The $L=2d$, $r\approx7$ value of $1.30$ carries a known windward-truncation artifact and is not read as conservation; the ratio falls to $0.55$--$0.73$ where the bending is strong, with the $r\approx3$ and the wide-gap $L=4d$ conditions included. The crossflow therefore removes along-trajectory momentum as drag: the momentum-conservation premise of the dilution bookkeeping holds only where the whole-trajectory dilution of the companion paper is weakest, not simply where the near-field shortening is smallest.

Third, the excess mass flux does not grow at the rate implied by the fitted capacity (Fig.~\ref{fig:fluxbudget}d). The total growth gives $c_{\rm meas}=(Q(\tilde u^*)-Q(\tilde u_0))/(U_\infty d)=0.13$ per condition on average, and the excess over the matched single-jet growth averages $c_{\rm excess}=1.12$ (range $-0.00$--$3.83$, the largest at $L=2d$, $r\approx7$); the mean-flow influx across the gap-side edge is small and of mixed sign ($-0.57$--$0.14$). On average these readings fall well below the fitted $c=3.31$ of the companion paper, although the largest single-condition excess exceeds it; the closure $r/(r+c_{\rm excess})$ fails against the measured ratios (Pearson $-0.55$).

Within the resolvable window the dilution is therefore a redistribution (centerline deficit, profile widening, momentum loss to the crossflow) rather than a net mass addition. The mass-addition bookkeeping that motivates the hyperbolic form in the companion paper remains valid only as an integrated description of the post-merging evolution, which the present fields cannot resolve. The companion paper accordingly reports $r_{\rm eff}$ and $c$ as effective parameters of the trajectory family, certified by the agreement of its independent readouts, and states the uncertainty of $c$ as the total interval $[3.03,\ 4.81]$.

\subsection{\label{sec:liutex}The carrier: convective vortex trains}
The time-resolved measurements test two expectations raised by the exchange mechanism of Section~\ref{sec:ccsl}: that the gap-side shear layers, being convectively unstable, roll up into trains that are convected downstream without a self-excited global mode, and that the two jets shed at comparable frequencies. This connection is not incidental. The exchange enhancement of Section~\ref{sec:ccsl} is carried by an enlarged mixing scale, and the mixing across a shear layer is executed by its large-scale vortices, so the vortex dynamics below is the direct test of that carrier. The CCSL configuration is moreover the canonical route toward absolute and global instability~\citep{ho_perturbed_1984,huerre_absolute_1985}, and a self-excited global mode is the one alternative that could replace the directly measured turbulent exchange as the carrier of the bending; the modal analysis of Section~\ref{sec:modal} finds no evidence for it within the resolvable band and record length. Finally, the bending of Section~\ref{sec:dose} scales with the absolute transport of the exchange, whose microscopic realization is the momentum advected by the convecting vortices. The material--pattern--phase hierarchy below distinguishes the structures that actually carry that momentum. The kinematics of the train---convection speed $U_c$, vortex spacing $\lambda$ and shedding frequency $f_0$---are measured independently of one another, so that their scale relation $U_c=\lambda f_0$ is checked rather than assumed.

\begin{figure}
\centerline{\includegraphics[width=\textwidth,height=0.70\textheight,keepaspectratio]{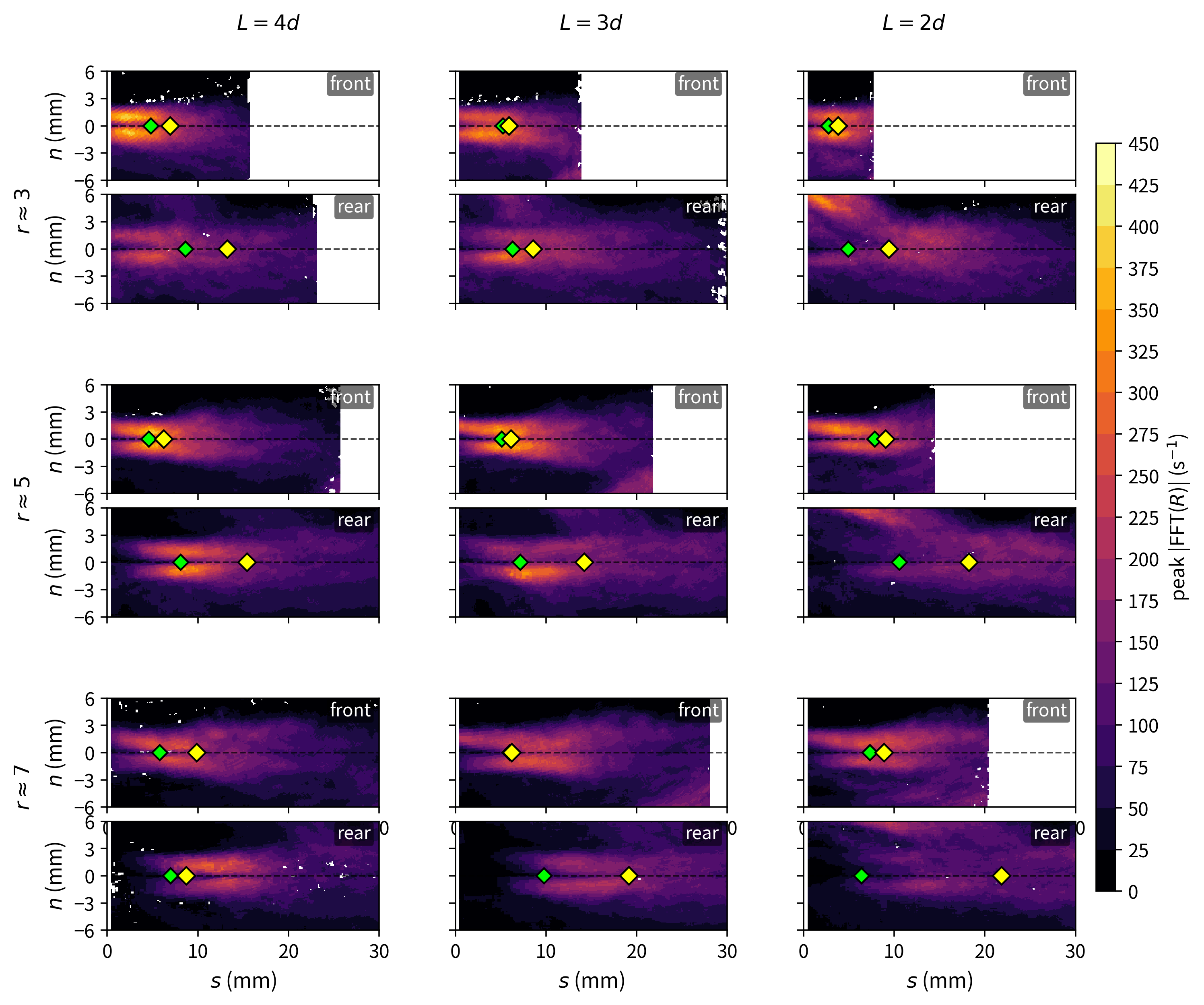}}
\caption{Condition-mean Liutex amplitude fields of the nine conditions in the natural coordinates $(s,n)$ of their trajectories (rows $r\approx3$, 5, 7, with the front and rear jet of each cell; columns $L=4d$, $3d$, $2d$), drawn as the peak of the FFT of the time-resolved Liutex field along the trajectory on the fixed absolute scale $0$--$450$\,s$^{-1}$ in bands of $25$\,s$^{-1}$; $s$ arc length, $n$ perpendicular offset, $n=0$ dashed black, the near-field (green diamond) and far-field (yellow diamond) transitions marked, and white regions where no valid data exist. The front-jet vortex motion is confined to a short near-field segment and fades downstream, whereas the rear-jet motion persists along the full trajectory.}
\alttext{Condition-mean Liutex amplitude fields of the nine conditions in the natural coordinates of the trajectories, with the near-field and far-field transitions.}
\label{fig:amp-fields}
\end{figure}

The shear-layer vortices are identified through the Liutex magnitude $R(t)$ (the rigid-rotation part of the velocity gradient~\citep{liu_third_2019}) in the single-transverse-jet normalization $R^*=R\,d/u_j$; the condition-mean amplitude profile along the trajectory is shown in Fig.~S7 of the supplementary material and the corresponding fields in Fig.~\ref{fig:amp-fields}.

\begin{figure}
\centerline{\includegraphics[width=\textwidth]{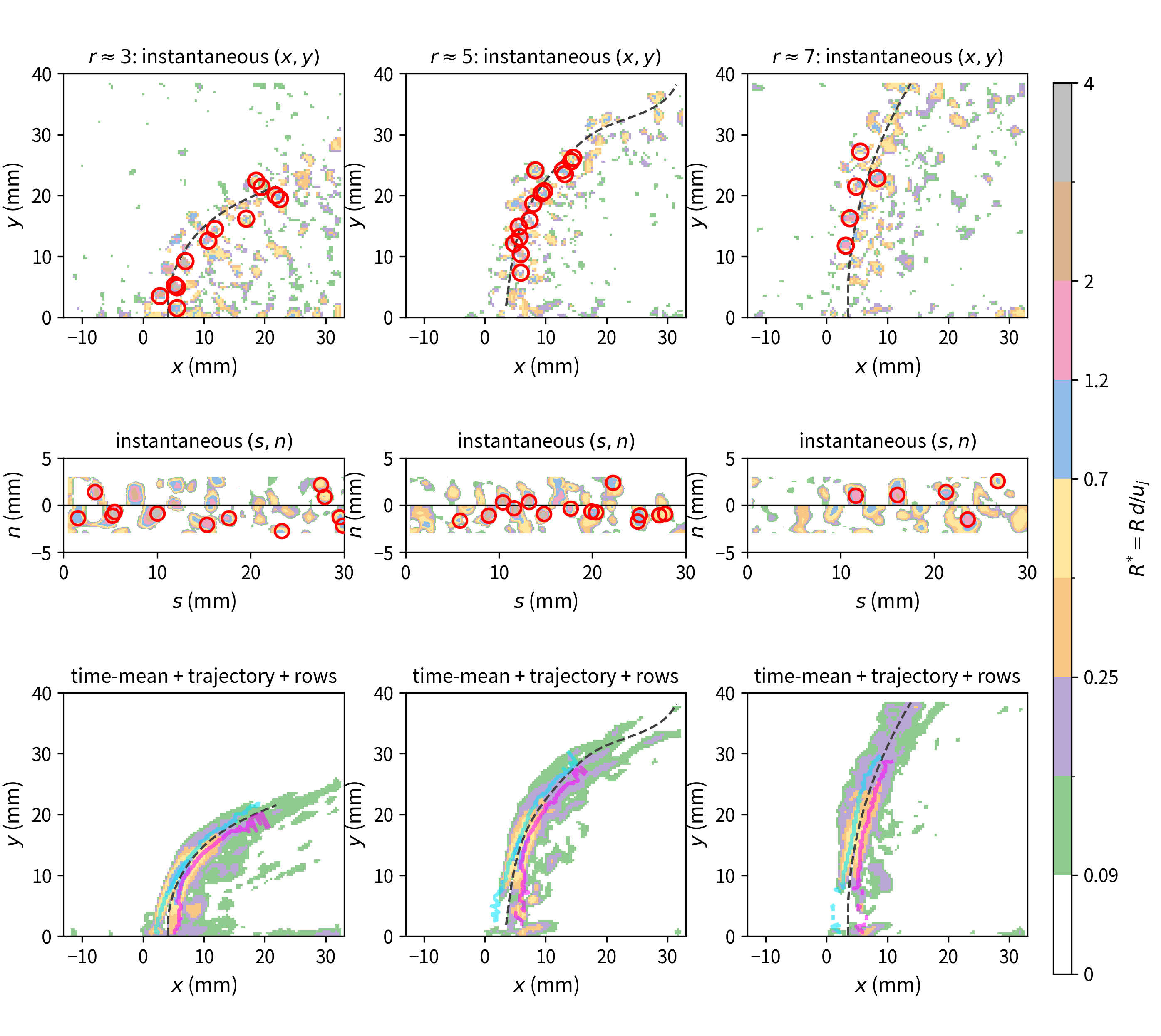}}
\caption{Vortex-motion tracking of the single jet (columns $r\approx3$, 5, 7; the single-jet counterpart of the nine-condition tracking summary): instantaneous Liutex field in the physical $(x,y)$ coordinates with the detected vortex cores and in the straightened $(s,n)$ coordinates, and the time-mean field with the extracted trajectory ($n=0$) and the two mean vortex-row trajectories (cyan windward, magenta leeward); space--time diagrams in Figs.~\ref{fig:sup-spacetime-single} and~\ref{fig:sup-spacetime-twin}. The Liutex fields are drawn on the fixed absolute scale shared by every $R^{*}$ figure of this paper (white below $R^{*}=0.09$, then eight non-uniform bands up to $R^{*}=4$); white regions thus mark $R^{*}<0.09$.}
\alttext{Vortex-motion tracking of the single jet for the three velocity ratios.}
\label{fig:sup-single-track}
\end{figure}

Four independent estimators measure the vortex-motion speed: Liutex-PIV (a windowed-deformation fit to the measured Liutex field), core tracking (the $(s,n,t)$ ridges of the tracked cores), the cross-spectral phase along the layer, and the optical flow of the particle images. The four estimates nearly coincide for the single jet (within the $\approx20\%$ spread of Tables ST4 and ST6; the single-jet tracking is shown in Fig.~\ref{fig:sup-single-track}) and fan out into a material--pattern--phase hierarchy (about $5$ to $9$\,m/s) for the twin jets.

Core tracking (material speed), optical flow (pattern speed) and the cross-spectral phase speed rank in that order; the Liutex-PIV estimate sits at or below the tracked core speed in the near field and above it in the rear-jet far field (Tables ST4 and ST6). The method-by-method detail is given in the supplementary material (Figs. S5 and S10; Tables ST4 and ST6).

\begin{figure}
\centerline{\includegraphics[width=0.95\textwidth,height=0.62\textheight,keepaspectratio]{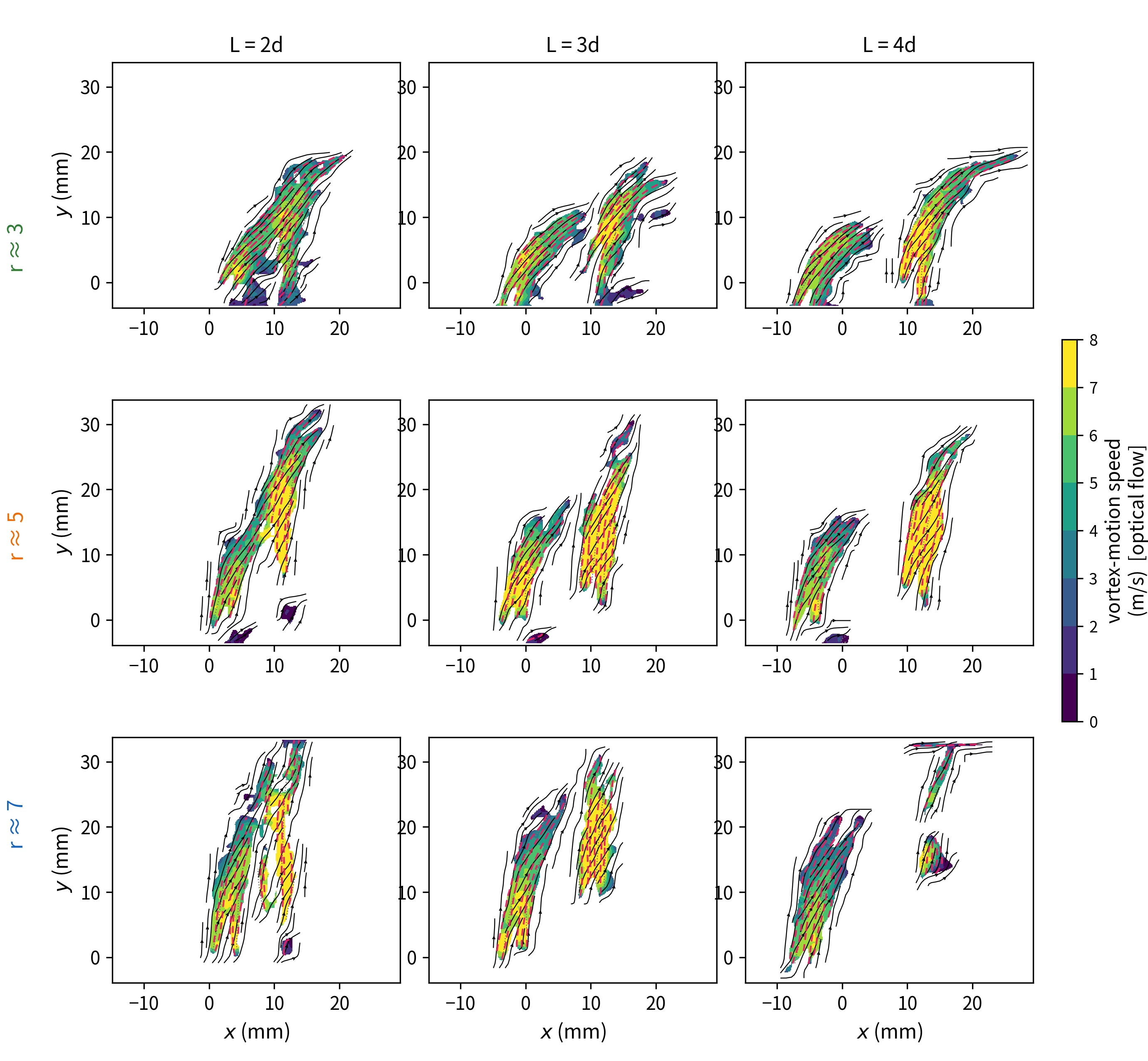}}
\caption{Velocity of the Liutex field in the physical $(x,y)$ coordinates for the nine conditions (rows $r\approx3$, 5, 7; columns $L=2d$, $3d$, $4d$), comparing three estimators restricted to the strong-Liutex region (90th percentile of the time-mean $R$): background colour, the dense optical-flow speed on the fixed absolute scale $0$--$8$\,m\,s$^{-1}$ in bands of $1$\,m\,s$^{-1}$ (colour bar); black lines, the OpenPIV cross-correlation streamlines of the Liutex pattern; dotted gray, the streamlines of the mean velocity field. The three families of streamlines run parallel to the shear layers; the OpenPIV estimate of the same pattern is $1.1$--$1.4$ times larger than the optical-flow speed and correlates less well with it.}
\alttext{Comparison of the optical flow, OpenPIV cross-correlation and mean-flow streamlines of the Liutex field in physical coordinates for the nine conditions.}
\label{fig:liutex-xy}
\end{figure}

A second, finer velocity field for the vortex-motion estimators of this section comes from the optical flow, computed on the raw particle images with the Farneb\'ack method (pyramid level 3, $15$\,px interrogation with Gaussian weighting; Fig.~\ref{fig:liutex-xy}).

\begin{figure}
\centerline{\includegraphics[width=\textwidth]{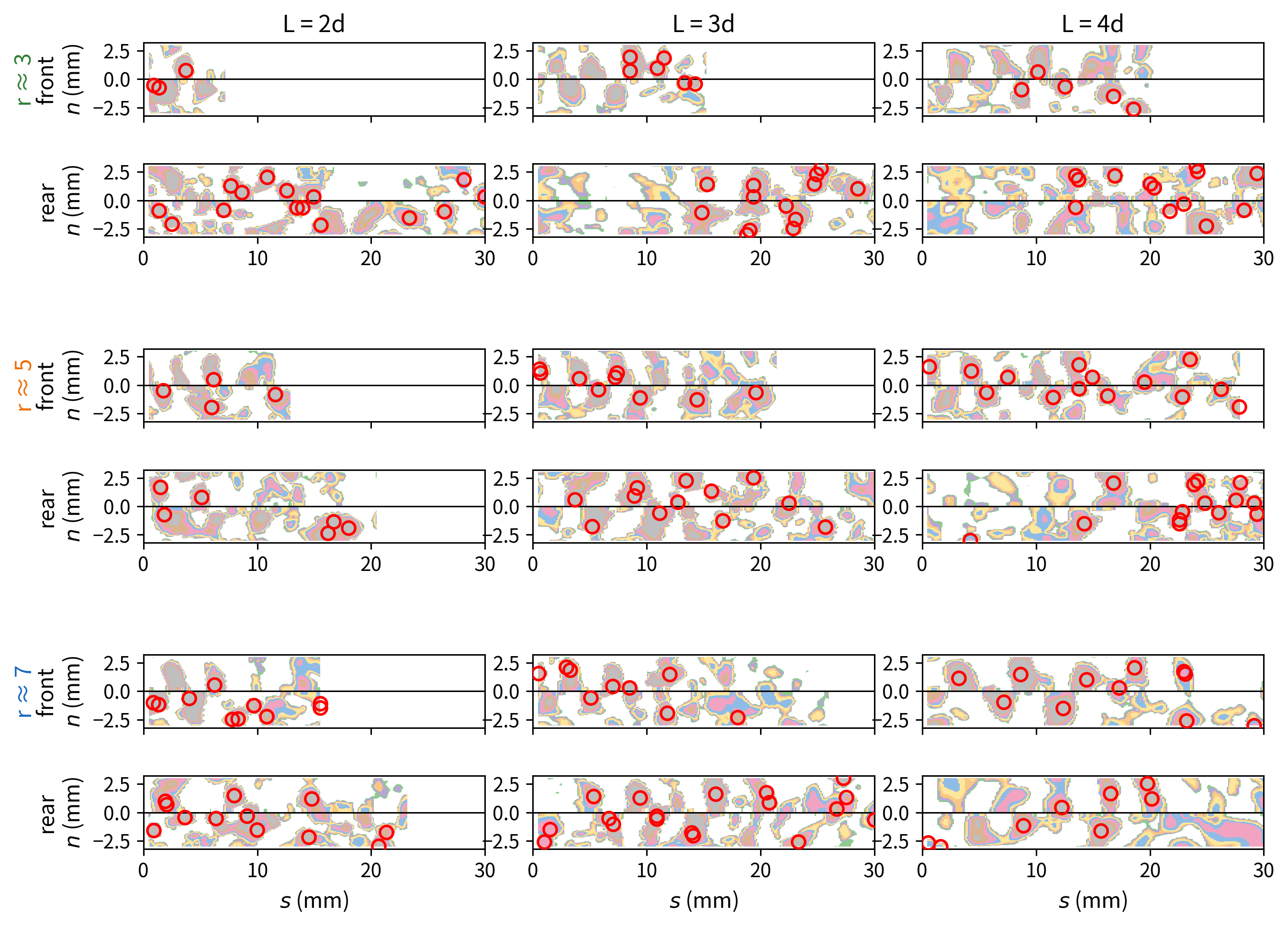}}
\caption{Instantaneous Liutex field of the nine conditions straightened onto the natural coordinates $(s,n)$ of each trajectory (rows $r\approx3$, 5, 7; columns $L=2d$, $3d$, $4d$, as in Table~\ref{table:conditions}; the upper and lower panels of every cell show the front and rear jets): the vortex cores detected in the two shear layers (red open circles) are marked together with the trajectory centre $n=0$ (black line), and the colour scale is the fixed absolute one of Fig.~\ref{fig:sup-spacetime-single} (white below $R^{*}=0.09$, then $R^{*}$ up to 4). Straightening removes the jet bending and exposes the quasi-periodic core train along the layer.}
\alttext{Nine-grid of instantaneous Liutex fields in the natural coordinates of each trajectory, with the detected vortex cores in the front-jet and rear-jet shear layers.}
\label{fig:twin-track}
\end{figure}

The optical-flow field is also an independent cross-check of the three other estimators. Figure~\ref{fig:twin-track} first shows the instantaneous Liutex fields of all nine conditions with the vortex cores detected in the two shear layers.

\begin{figure}
\centerline{\includegraphics[width=0.95\textwidth]{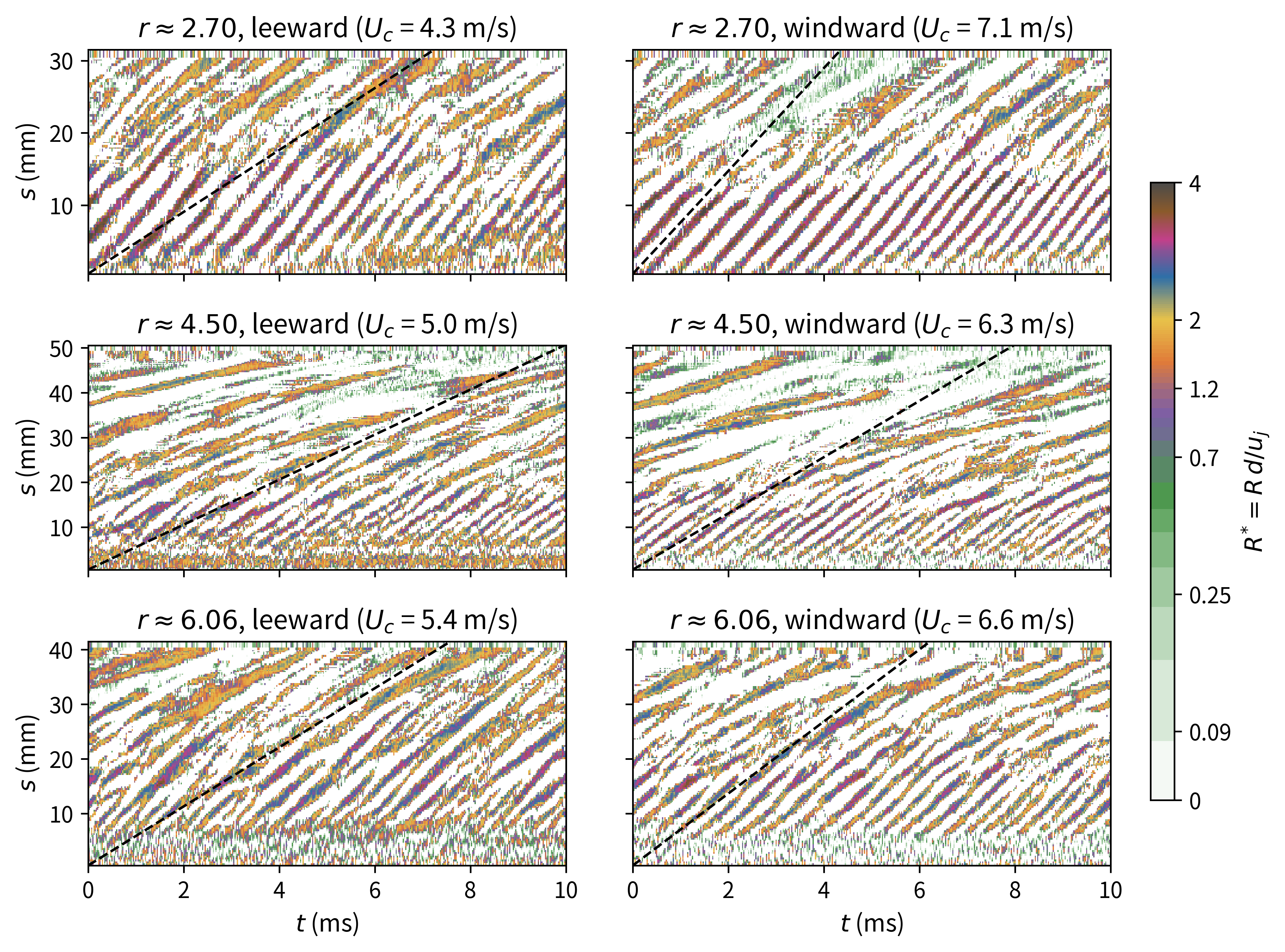}}
\caption{Space--time diagrams of the shear-layer vorticity $R^{*}=|R|\,d/u_j$ of the \emph{single-jet baseline}, on the fixed absolute Liutex scale shared by every $R^{*}$ figure of this paper (white below $R^{*}=0.09$, then eight non-uniform bands up to $R^{*}=4$), so that every panel of Fig.~\ref{fig:sup-spacetime-twin} is read the same way: one record per condition (rows $r\approx2.70$, 4.50, 6.06; columns leeward, windward) with the dashed line of slope $U_c$.}
\alttext{Six single-jet space-time diagrams of the shear-layer vorticity, three velocity ratios by two layer sides, with the convection-speed line.}
\label{fig:sup-spacetime-single}
\end{figure}

The corresponding time-mean fields are given in Fig. S8 of the supplementary material, where the two shear-layer branches appear as distinct vortex rows whose separation sets the layer-normal band over which the convection speed is measured and whose downstream fading bounds the window in which the tracking is reliable. The same fields straightened onto the natural coordinates of each trajectory are shown in Fig.~\ref{fig:amp-fields}, and the underlying space--time diagram of the single jet in Fig.~\ref{fig:sup-spacetime-single}.

\begin{figure}
\centerline{\includegraphics[width=\textwidth]{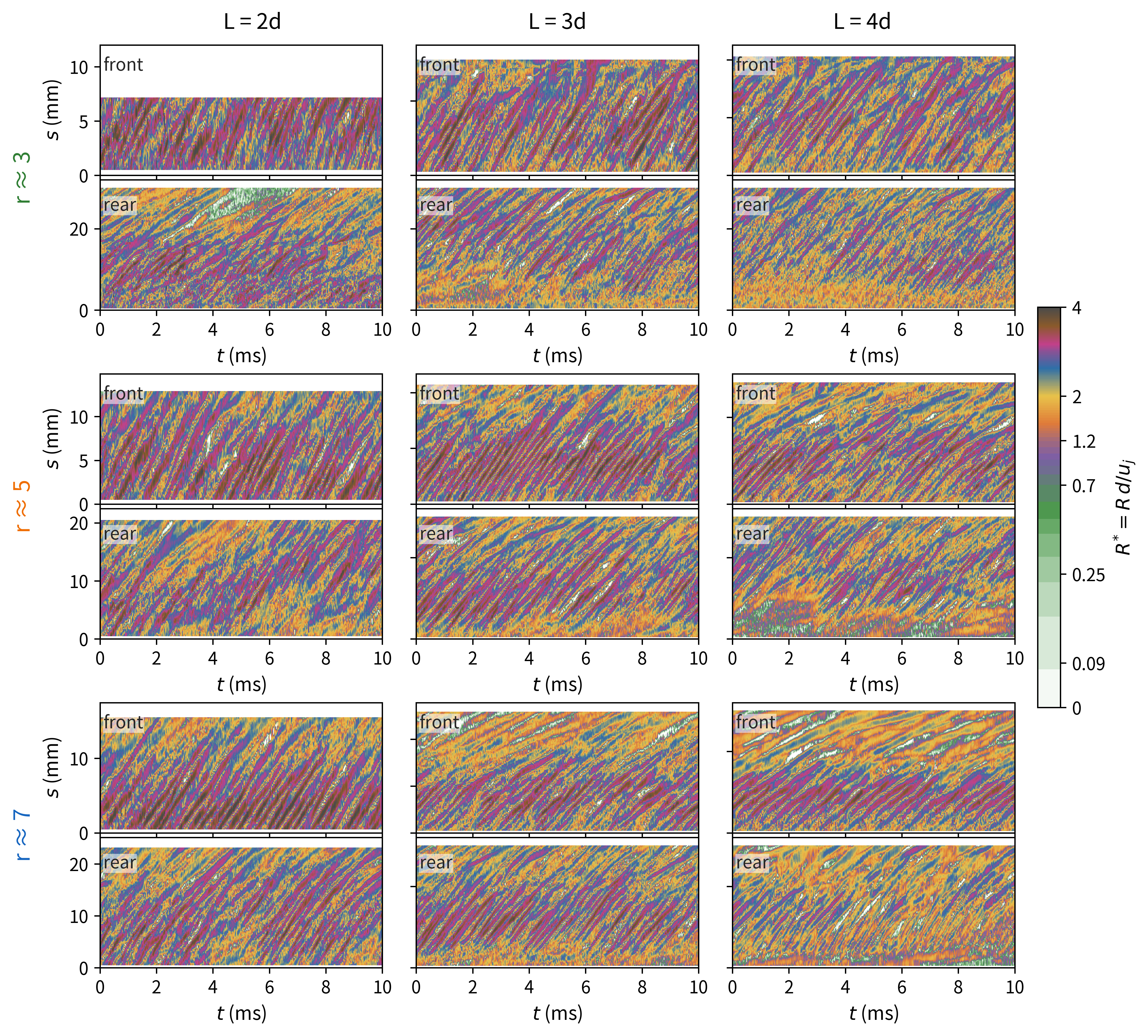}}
\caption{Space--time diagrams of the shear-layer vorticity $R^{*}=|R|\,d/u_j$ of the \emph{twin jets} on the same discrete scale as Fig.~\ref{fig:sup-spacetime-single}: one representative record per condition (rows $r\approx3$, 5, 7; columns $L=2d$, $3d$, $4d$), the upper and lower panels of each cell showing the front and rear jets, respectively.}
\alttext{Nine-condition space-time diagrams of the twin-jet shear-layer vorticity, with the front-jet and rear-jet panels of each cell stacked tightly.}
\label{fig:sup-spacetime-twin}
\end{figure}

The twin-jet diagrams are given in Fig.~\ref{fig:sup-spacetime-twin}.

\begin{figure}
\centerline{\includegraphics[width=\textwidth,height=0.72\textheight,keepaspectratio]{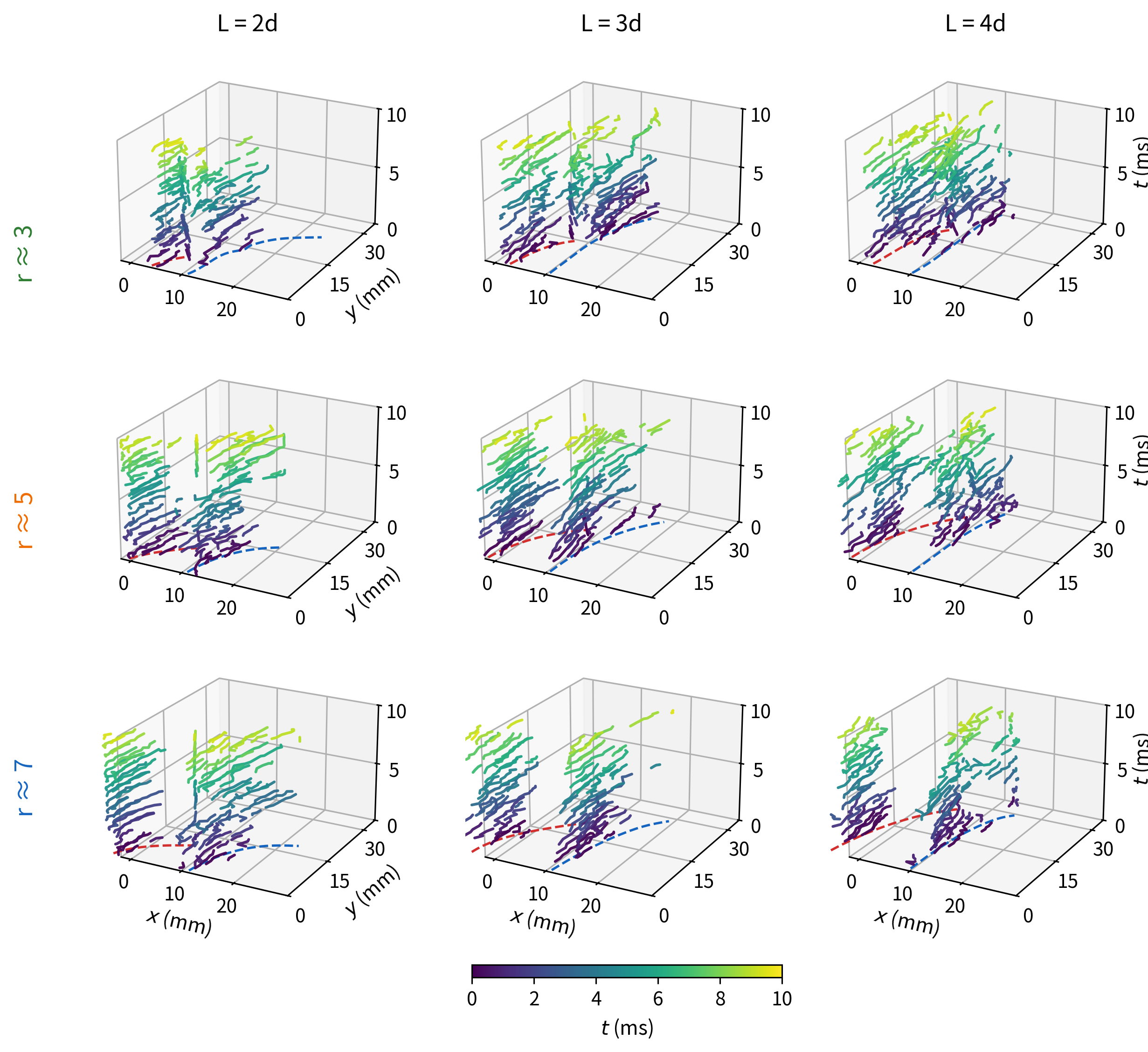}}
\caption{Space--time vortex tubes of the tracked cores of both jets in the physical $x$--$y$--$t$ space, for the nine conditions (rows $r\approx3$, 5, 7; columns $L=2d$, $3d$, $4d$, as in Table~\ref{table:conditions}), coloured by time over the 10\,ms record; the dashed lines on the floor are the mean trajectories (front jet red, rear jet blue). The two jets occupy different $x$ bands, so the two bundles of tubes separate without any coordinate transformation: the front-jet tubes are dense in the near field and fade beyond $x\approx15$\,mm, while the rear-jet tubes persist through the whole record and reach further downstream.}
\alttext{Nine-grid of the space--time vortex tubes of the tracked cores of both jets, coloured by time, with the mean trajectories on the floor plane.}
\label{fig:tubes}
\end{figure}

The space--time vortex tubes of the tracked cores are shown in Fig.~\ref{fig:tubes}.

\begin{figure}
\centerline{\includegraphics[width=\textwidth,height=0.67\textheight,keepaspectratio]{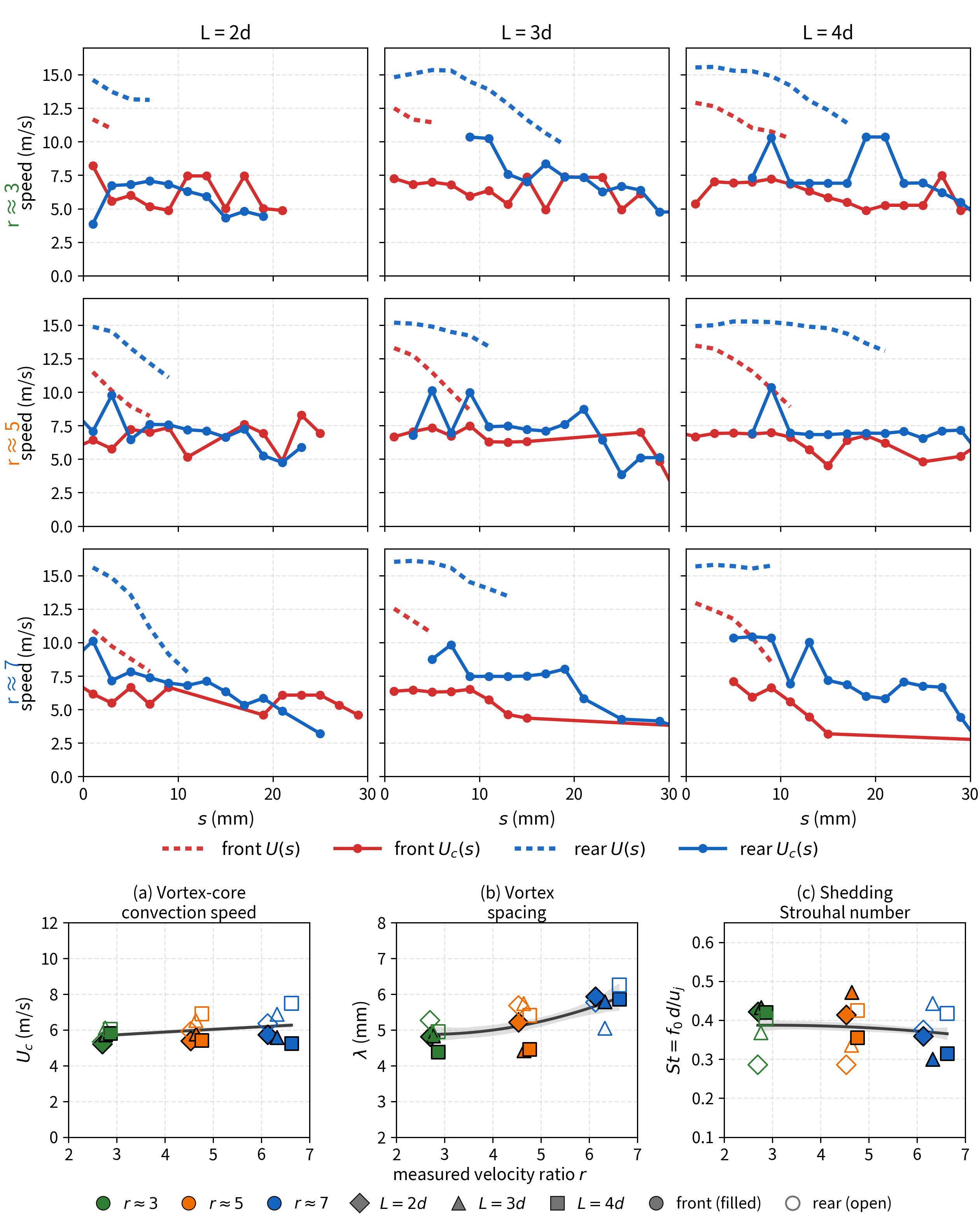}}
\caption{Local centerline speed $U(s)$ (dotted) and vortex-core convection speed $U_c(s)$ (solid with markers), obtained from the tracked cores as $\Delta s/\Delta t$, along the trajectory for the nine conditions (rows $r\approx3$, 5, 7; columns $L=2d$, $3d$, $4d$; front jet red, rear jet blue); bottom row: (a) the vortex-core convection speed $U_c$, (b) the inter-vortex spacing $\lambda$ and (c) the shedding Strouhal number $St=f_0d/u_j$ against the measured velocity ratio (condition means; symbol convention of Section~\ref{sec:expt}), the lines are quadratic least-squares fits to the 18 condition--jet means and the shaded bands are $95\%$ bootstrap confidence bands from 1000 resamples of the cases within each condition cell. A representative kymograph behind these profiles is given in Fig. S6 of the supplementary material.}
\alttext{Nine-grid of the local centerline speed and the vortex-core convection speed along each trajectory, with the bottom panel showing the convection speed against the measured velocity ratio.}
\label{fig:uc-profiles}
\end{figure}

The vortex-motion estimates rest on this tracking chain. Figure~\ref{fig:uc-profiles} gives the recovered speed condition by condition and resolves it in position, station by station; the cross-case spread is shown in Fig. S9 and a representative kymograph with the convection fits in Fig. S6 of the supplementary material.

\begin{figure}
\centerline{\includegraphics[width=0.95\textwidth,height=0.58\textheight,keepaspectratio]{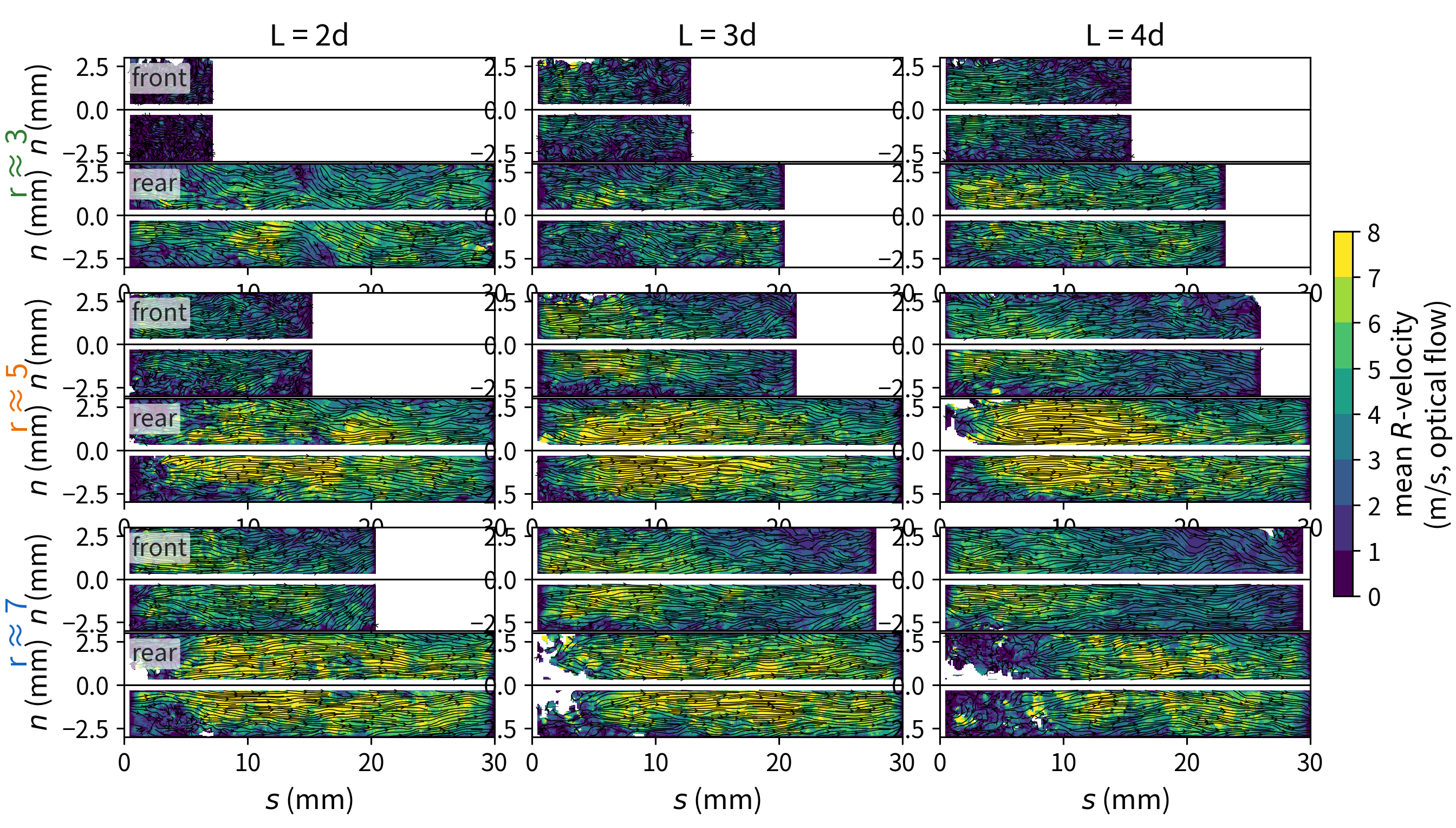}}
\caption{Velocity of the Liutex field itself---the speed at which the vortex structures of the two jets are convected---in the natural coordinates $(s,n)$ of the trajectories, from dense optical flow (frame pairs four frames apart, $100\,\mu$s): rows, the three velocity-ratio groups with the front/rear jet pairs; columns, the three spacings (Table~\ref{table:conditions}); colour: fixed absolute scale of $0$--$8$\,m\,s$^{-1}$ in bands of $1$\,m\,s$^{-1}$; black lines, streamlines; only nodes where the time-mean Liutex field marks a detectable vortex are shown. Physical-coordinate version: Fig.~\ref{fig:liutex-xy}.}
\alttext{Nine-grid of the optical-flow velocity of the Liutex field in the natural coordinates of the trajectories, with streamlines, showing the vortex structures convected with the local flow.}
\label{fig:liutex-field}
\end{figure}

Figure~\ref{fig:liutex-field} gives the velocity of the Liutex field itself---the speed at which the vortex structures are convected---in the natural coordinates of the two trajectories, evaluated with dense optical flow.

The streamlines follow the shear layers, and the front-jet motion fades in the near field while the rear-jet motion persists downstream. The optical-flow speed matches the local mean speed to within a few percent (median ratio $0.95$--$1.06$ over the nine conditions), so the shear-layer vortices are convected with the local flow (Fig.~\ref{fig:liutex-field}). The core tracking and the cross-spectral phase complete the speed hierarchy, summarized in the bottom row of Fig.~\ref{fig:uc-profiles} and in Fig.~S19 of the supplementary material. Figure~\ref{fig:liutex-xy} gives the physical-coordinate counterpart of Fig.~\ref{fig:liutex-field}, comparing the dense optical flow with the OpenPIV cross-correlation of the Liutex pattern and with the mean-flow streamlines.

That the vortices convect at the material speed---rather than as a propagating wave---identifies them as the advective carriers of the momentum whose redistribution the flux budget of Section~\ref{sec:flux} reads. This closes the exchange picture of Sections~\ref{sec:ccsl}--\ref{sec:flux} by identifying its dynamical carrier.

The shedding frequency is governed by the near-field shear-layer instability rather than by the local velocity magnitude. Pooled over the cases it is $1612\pm243$\,Hz for the front jet ($n=67$) and $1534\pm249$\,Hz for the rear jet ($n=67$), a difference below the $\approx\pm10\%$ normalization uncertainty. In the pooled along-trajectory sense the frequency contrast is therefore not a sensitive discriminator of the coupling, unlike the trajectory and decay asymmetries of the companion paper.

\begin{figure}
\centerline{\includegraphics[width=\textwidth]{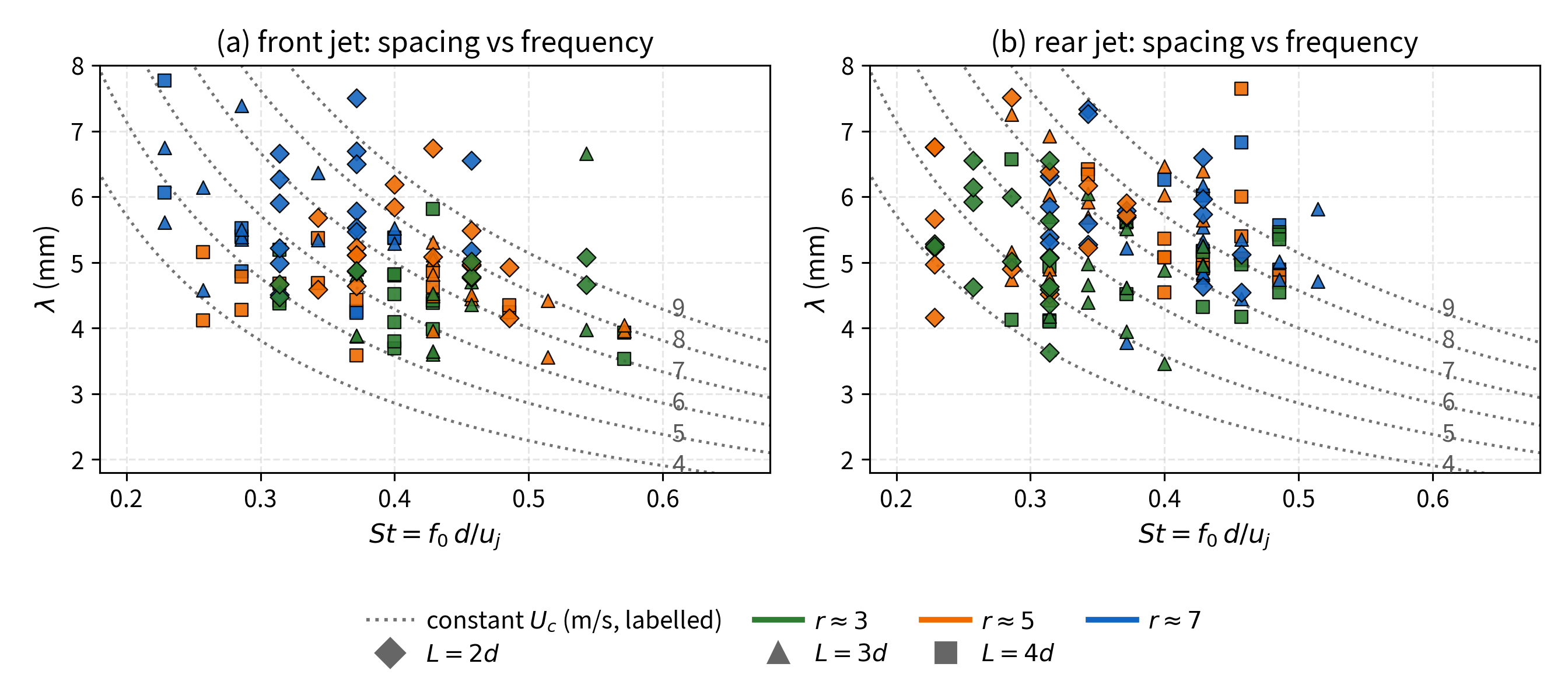}}
\caption{Spacing--frequency relation of the two jets: the directly measured inter-vortex spacing $\lambda$ against $St=f_0\,d/u_j$ for (a) the front and (b) the rear jet, with the constant-$U_c$ hyperbolas labeled in m/s (symbol convention of Section~\ref{sec:expt}). Both quantities are independent measurements---the Liutex time-series peak frequency and the spacing between consecutive detected vortex cores. Core speed: bottom row of Fig.~\ref{fig:uc-profiles}.}
\alttext{Two-panel spacing-frequency relation of the front and rear jets, with the directly measured inter-vortex spacing against the Strouhal number and the constant-convection-speed hyperbolas.}
\label{fig:uc-params}
\end{figure}

The spacing--frequency relation of the two jets is consistent with a common set of constant-$U_c$ hyperbolas (Fig.~\ref{fig:uc-params}). The side-resolved near-field values at the smallest spacings differ by 25--30\%, so the near-equality should not be read as a local one; Table~\ref{tab:metrics} collects these frequency indicators, resolved by side and spacing.

The shedding frequency of the gap-side layers is given in Fig.~\ref{fig:promo-s29} and Table ST5 of the supplementary material.

\begin{figure}
\centerline{\includegraphics[width=0.985\textwidth]{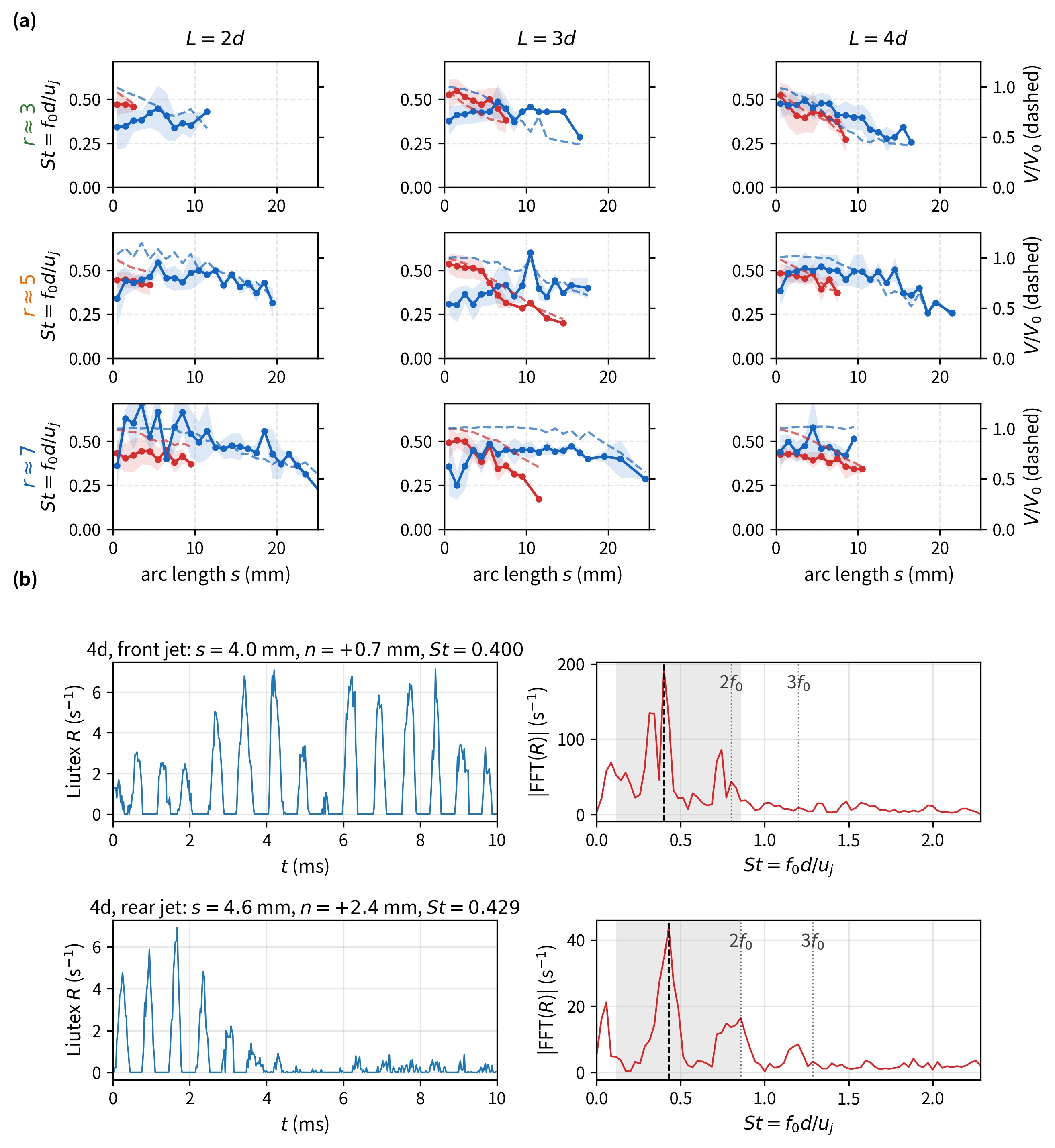}}
\caption{Shedding frequency of the gap-side layers. (a) Dominant Liutex frequency over the trajectory zone ($\pm2\sigma(s)$) with the normalized centerline velocity decay (dashed), for the nine conditions (rows $r\approx3$, 5, 7; columns $L=2d$, $3d$, $4d$); red: front jet, blue: rear jet, and the shaded band the case-to-case spread of the frequency. (b) Time series and FFT spectra of $R(t)$ at the point with the clearest shedding ($L=4d$, $r\approx7$): front jet ($s=4.0$\,mm, $St=0.400$) and rear jet ($s=4.6$\,mm, $St=0.429$).}
\alttext{Dominant Liutex frequency along the trajectory and example time series and spectra.}
\label{fig:promo-s29}
\end{figure}

The equality of the two frequencies is nonetheless consistent with what the enclosed exchange of Section~\ref{sec:ccsl} predicts for two layers coupled through one shared recirculation. The frequency equality alone cannot distinguish a shared vortex train from two independent oscillators tuned to the same Strouhal number, so the coupling statement is read in the weaker sense. The four-part near/far decomposition of the vortex dynamics, and the R-peak station plotted against the gap minimum, the transition height and the row separation, are shown in Fig. S10 of the supplementary material.

The amplitude statistics behind these stations---peak amplitude, peak station and rise rate against the far-field decay rate, over the 136 case/jets---are collected in Fig. S7 of the supplementary material: the front-jet peak occurs systematically earlier than the rear-jet one. The last element of the chain of Section~\ref{sec:ccsl}---the modal character of the unsteady dynamics---is examined in Section~\ref{sec:modal}.

\begin{figure}
\centerline{\includegraphics[width=\textwidth]{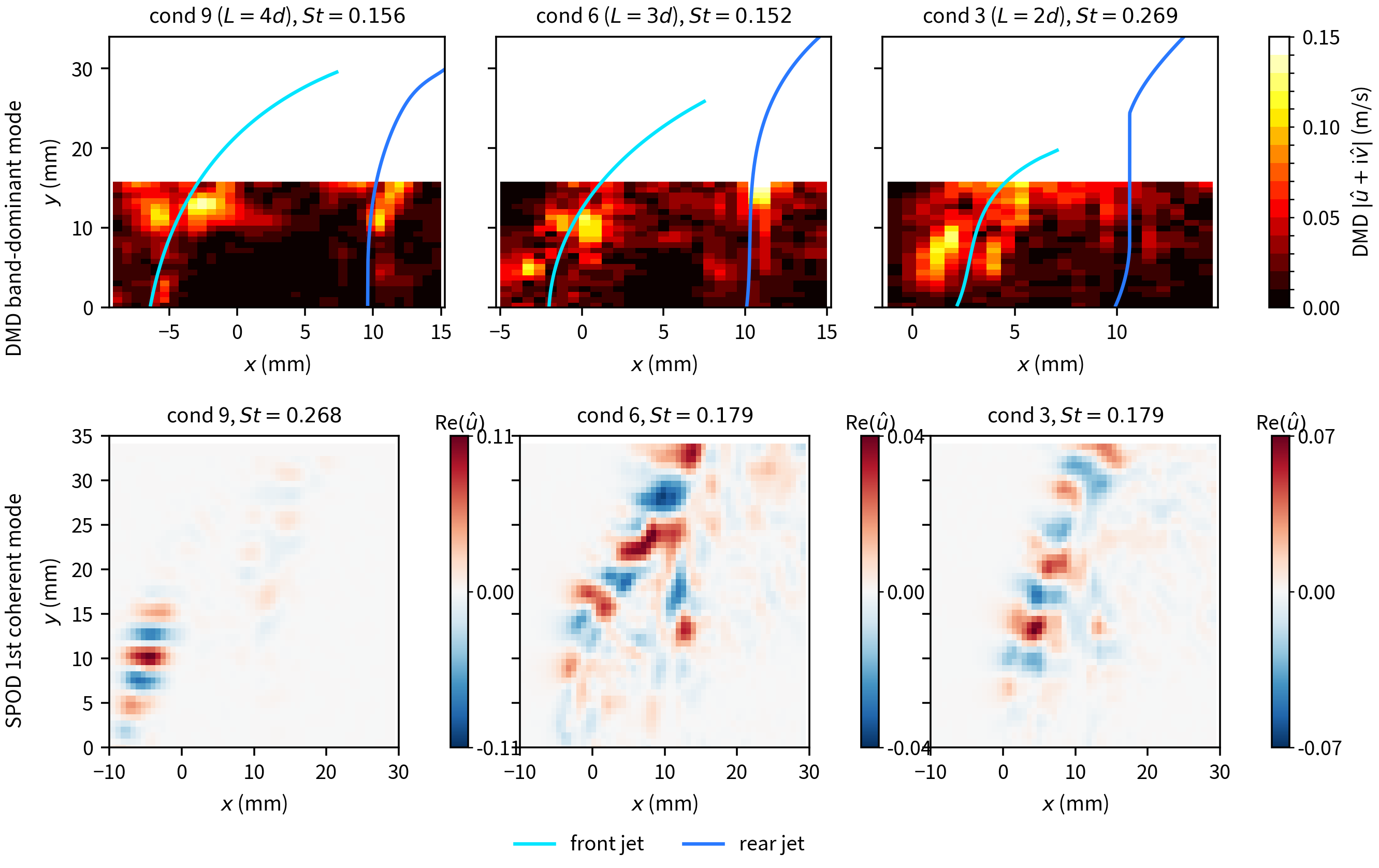}}
\caption{Band-dominant DMD modes of conditions 9, 6 and 3 ($L=4d$, $3d$, $2d$ at $r\approx7$) with the front (cyan) and rear (blue) trajectories (top), and the 1st coherent SPOD modes of the same three conditions at their peak frequencies (bottom), localized along the two shear layers; the DMD modes (top) share the fixed absolute scale $0$--$0.15$\,m\,s$^{-1}$ in bands of $0.01$\,m\,s$^{-1}$, while the SPOD modes (bottom) show $\mathrm{Re}(\hat u)$ on per-panel scales.}
\alttext{Band-dominant DMD modes of three conditions with the front and rear trajectories, and the leading SPOD modes of condition 3 localized along the two shear layers.}
\label{fig:modal-modes}
\end{figure}

The vortex kinematics are therefore those of a convected vortex train, consistent with the exchange mechanism of Section~\ref{sec:ccsl}; the modal character of the instability is examined next.

\subsection{\label{sec:modal}The excluded alternative: no self-excited global mode}

DMD and SPOD are applied to the mean-subtracted 400-frame series of representative cases and, pooled, to all 68 cases (SPOD with $n_{\mathrm{fft}}=128$, 75\% overlap and nine blocks; supplementary material). They locate the dominant modes along the two shear layers within the $0.5$--$1.1$\,kHz zone; no single mode locks the whole field. This modal band lies below the $1.5$--$1.6$\,kHz shedding peak of the Liutex time series (Table~\ref{tab:metrics}): the velocity-field modes resolve the largest, slowly evolving structures of the layer, whose band includes the subharmonic of the shedding cycle, whereas the Liutex amplitude peaks at the fundamental. The band-dominant modes of conditions 9, 6 and 3, and the leading SPOD modes of the same conditions, are shown in Fig.~\ref{fig:modal-modes}.

Extending the decomposition to one representative case of every condition (Fig. S14 of the supplementary material) gives an energy fraction of $64$--$83$\% within $\pm2$\,mm of the two trajectories (SPOD $78$--$90\%$; the single jet gives $85\%$).

The largest eigenvalue modulus corresponds to the zero-frequency mean mode, not to a growing oscillation: in the four records that store the full series the oscillatory modes have moduli $0.991$--$0.999$ while the zero-frequency mode reaches $1.0007$; over all 68 cases the exported statistics give a largest modulus between $0.999$ and $1.013$. The single value above unity corresponds to an e-folding time of about $2$\,ms ($\ln 1.013$ divided by the $25\,\mu$s frame separation). The eigenvalue estimate carries an estimation scatter of about $\pm0.01$ over the resolved modes, so a modulus of $1.013$ cannot be distinguished from unity at this record length; no surrogate test was run, so that scatter is the uncertainty of this statement (Section~\ref{sec:limitations}). The mode is in any case localized to one shear layer and does not span the field (Fig.~\ref{fig:modal-modes}), so it fails the field-spanning criterion of a global mode.

The gap-side layers are therefore convectively unstable, with no evidence within the 10\,ms records and the resolvable $0.4$--$3$\,kHz band of a self-excited global mode---the standard modal criterion for its absence~\citep{taira_modal_2017}. Downstream, the two vortex sheets merge into a single mixing layer. A global mode at frequencies below the inverse record length ($\lesssim100$\,Hz) lies outside the resolvable band of the present records (Section~\ref{sec:limitations}). The exchange that carries the bending is accordingly a broadband turbulent transport rather than a coherent modal oscillation, consistent with the directly measured Reynolds stress of Section~\ref{sec:ccsl} and with the dilution that the flux budget of Section~\ref{sec:flux} reads as a redistribution.

\subsection{\label{sec:discussion}The mechanism in one chain}
The evidence assembled so far can be stated as one chain. The chain begins with the rear-jet blockage: the blocked gap hosts a recirculation with reversed streamwise velocity ($u<0$) and an elevated inter-jet pressure. This reversed gap flow sweeps the front-jet leeward layer while the rear jet encloses and continuously replenishes the recirculation---something the single jet cannot do. The front jet therefore transfers streamwise momentum to the gap far more strongly than any single-jet layer does. The enclosed gap recirculation, together with the reversed flow it drives, thus draws the two trajectories together and produces the front-low/rear-high asymmetry of the companion paper.

The measured mixing length shows the transfer to be carried by the mixing scale rather than by a steeper shear: on the front-jet gap side it exceeds the single-jet leeward value in five of the six conditions at $r\approx5$ and 7, while the mean shear across the layer does not increase (Fig.~\ref{fig:promo-s7}). The near-field turbulent transport is $2.3$--$3.8$ versus $0.47$\,m/s$^2$, and the absolute gap-side exchange is nearly $r$-independent while exceeding the single-jet leeward zone like-for-like by up to $1.5\times$ (zone peak, across conditions) and $2.8\times$ (area-integrated) at $r\approx7$---a momentum dilution equivalent to $r_{\mathrm{eff}}<r$. Consistently with this dilution, the front jet bends earlier (near-field segment up to 40\% shorter in seven of the nine conditions), decays at 1.3--1.8 times the single-jet rate and stays lower, while the shielded rear jet rises higher.

The bending scales with the absolute near-field exchange ($\rho=0.68$, $p=0.042$ at the condition level; Spearman $\rho=0.42$, $p=0.0014$ per case), not with the twin/single ratio (condition-level $\rho=+0.13$; per-case $+0.23$; neither significant).

This chain, whose steps are detailed in Section~\ref{sec:ccsl}, accounts for the trajectory asymmetry, the reversed gap flow, the elevated inter-jet pressure, the gap turbulent kinetic energy (TKE) and the vortex dynamics through a single primary carrier: the recirculation-driven momentum exchange, with the blockage pressure as its accompanying signature. The vortices of Section~\ref{sec:liutex} carry that exchange as turbulent transport, not as a coherent oscillator---the absence of a growing global mode (Section~\ref{sec:modal}) excludes the latter---so the mean exchange and its unsteady carrier are two faces of one mechanism. The pressure is a diagnostic field of the same measured velocity and stress fields (the companion paper), so its correlation with the bending is a common-origin response rather than an independent drive.

The resolved turbulent term exceeds the resolved pressure term in eight of the nine conditions, and the pressure-gradient term becomes comparable only at $L=4d$, where the bending nonetheless scales with the exchange rather than with the pressure (Section~\ref{sec:dose}). The chain disfavors two further competing explanations: the classical Bernoulli low-pressure attraction of twin parallel jets (the gap pressure is elevated, not depressed; the companion paper) and a self-excited global mode (Section~\ref{sec:modal}). The flux budget of Section~\ref{sec:flux} updates the physical reading of the companion-paper law: the mass-addition bookkeeping that motivates the hyperbolic form holds only as the integrated description of the post-merging evolution, so $c$ is an effective capacity constant of the trajectory family, and the companion paper states it in the same terms.

The evidence behind the chain is observational---an inference to the best explanation rather than a proof of causation: the present measurements establish a consistent account of the measured sequence. The absolute near-field transport of the exchange and the near-segment shortening (up to 40\%) vary together across the conditions ($\rho=0.68$--$0.87$), and the exchange is large in absolute terms exactly where the shortening is largest; the two classical alternatives are excluded independently (the companion paper and Section~\ref{sec:dose}). In an experiment without controlled intervention this is the strongest causal statement available, and the reverse pathway---the early bending strengthening the exchange---cannot be excluded from the present records.

The quantitative content of the companion paper is the velocity-dilution scaling and its verified consequences, which hold irrespective of the causal interpretation. Independently of the transition detection, the front-jet trajectory turning angle falls below the single-jet value in seven of the nine conditions (the companion paper); the turning-angle change is carried by the twin/single enhancement ratio (condition-level $\rho=+0.77$, $p<0.05$; per-case $\rho=+0.69$, $p<0.001$; Fig.~S16d of the supplementary material), consistent with the same dilution picture. 

The single-jet leeward shear layer at $r\approx3$ sits at the literature global-instability boundary ($r\approx3.0$--$3.3$~\citep{iyer_numerical_2016,shoji_origins_2020,souza_criterion_2021}). The convective classification of the twin gap-side layers at that ratio therefore rests on the DMD/SPOD evidence of Section~\ref{sec:modal}---which finds no growing, field-spanning mode---rather than on the velocity ratio alone.
\subsection{Comparison with previous studies and testable predictions}

The reversed-gap configuration is the CCSL configuration of the single transverse jet~\citep{huerre_absolute_1985,strykowski_stability_1991,karagozian_transverse_2010}, whose velocity ratio $R_1$ (Eq.~\eqref{eq:r1}) characterizes the counter-current strength and rationalizes the convective-to-global transition of the single-jet shear layer~\citep{iyer_numerical_2016,shoji_origins_2020,souza_criterion_2021}. The present measurements extend this analogy from a single leeward layer to the two enclosed gap-side layers and add the force-level evidence---the resolved exchange and the leading-term momentum balance of Section~\ref{sec:dose}---that the single-jet studies infer only indirectly. The twin geometry also separates the counter-current character from its causal ingredient: the single jet satisfies the permissive condition ($R_1=1.06$--$1.22$ over its leeward layer; Fig.~\ref{fig:promo-s3}a) yet exchanges momentum far more weakly. Since $V_1$ decays with height while $|V_2|$ is largest low in the layer (Fig.~\ref{fig:promo-s3}b), the height dependence of $R_1$ partly follows the jet-side weakening; the enhancement is thus carried by the enclosed recirculation, not by the counter-current configuration itself.

The mechanism is geometric in origin. A tandem arrangement in which the rear jet closes a confined recirculation against the front jet should exhibit the same front-low/rear-high asymmetry within the measured regime, provided the spacing is small enough for the recirculation to close and the velocity ratio high enough for the blockage to sustain the exchange; invariance to nozzle geometry and Reynolds number remains to be tested. The like-for-like exchange ratio grows with the velocity ratio, partly because the absolute exchange is nearly $r$-independent while the single-jet reference weakens with $r$; at $r\approx3$ the single-jet wake itself reverses strongly and the enhancement falls to or below the single-jet value, so the enhancement is the dominant mechanism at $r\gtrsim5$ rather than the exclusive cause at every ratio. This regime view reconciles the earlier observations: single-spacing tandem studies reported a front jet similar to the single jet or only the rear-jet shielding, whereas the present multi-spacing data place the deviation in the closed-recirculation regime; the three-jet tandem dynamics of \citet{kristo_time-resolved_2021} complement the present $r\approx3$--$7$ range at low ratio, and the differing instability boundary of staged jets~\citep{chandh_shear_2023} is traced here to the gap-side layers. The residual lowering at $r\approx3$---with the exit-velocity deficit, the gap pressure and the early merging---is decomposed in the companion paper and is not pursued further here.

The mechanism makes a short list of testable predictions: the vortex kinematics must satisfy $U_c=\lambda f_0$ and the two jets should shed at comparable frequencies, as observed; the enhancement should be carried by an enlarged mixing length rather than a steeper mean shear; and the effective capacity $c$ should vary monotonically with the gap geometry while remaining invariant to the velocity ratio, with the whole-trajectory estimator transferring to other tandem configurations. A further prediction follows directly from the mechanism and is testable without repeating the present experiment: the dilution must weaken or disappear where the rear jet can no longer close the gap recirculation---at large spacing, at low velocity ratio, or when the rear jet is inclined away from the front jet. The present data locate the velocity-ratio side of that boundary at $r\approx3$, where the enhancement is marginal (Table~\ref{tab:enhance-verify}) and the residual lowering can no longer be attributed to the exchange alone.

\subsection{\label{sec:limitations}Limitations}

Several limitations qualify these conclusions, which are an inference to the best explanation rather than a proof of causation. The condition-level correlations of Table~\ref{tab:degree-matrix} are exploratory: 36 metric--quantity combinations are tested without multiplicity correction, the nine condition means are not independent of the 56 cases, and the $p_{\rm int}$ entry is not significant at the condition level; the $c=3.31$ interval is a condition-mean rather than a case-level one.

The enhancement metric depends on the normalization: the like-for-like ratios are $0.58$--$1.46$ (zone peak) and $0.59$--$2.83$ (area-integrated), and the $27\times$ contrast reflects the weak pooled single-jet reference. The Liutex amplitude and the Strouhal number carry the $\pm10\%$ uncertainty of the fixed $u_j=14$\,m/s normalization; $R^*=R\,d/u_j$ sets unity at the jet-exit scale, the normalized amplitude is not bounded, and the fields are shown only above $R^*=0.09$.

The measurement is two-dimensional: $w'$, the spanwise stress gradients and the three-dimensional recirculation are unresolved, and the momentum balance of Section~\ref{sec:dose} is a resolved-term comparison. The 10\,ms records resolve mode localization but not growth rates, and modes below the inverse record length ($\lesssim100$\,Hz) lie outside the band; the convective classification is therefore an inference from localization, downstream travel and the absence of growing modes.

The front and rear jets exit at $12.8$ versus $15.4$\,m/s at $\approx0.5$\,mm, an asymmetry that enters every front--rear comparison; part of it is the near-exit interaction signature, and a residual supply-side contribution cannot be excluded because the upstream lines were not instrumented separately. The 400 frames per record resolve the large structures but are not independent samples, so the stress statistics carry a sampling uncertainty; two conditions show no near-field shortening ($L=2d$, $r\approx5$ and $L=3d$, $r\approx3$) and one is scattered ($L=3d$, $r\approx7$), and these cases are retained, making the correlations conservative. The flux-budget truncation at the gap-side valley biases the excess-mass growth downward; the redistribution statement therefore leans on the centerline deficit and the momentum-flux loss.

\begin{table}
\centering
\caption{Dominant shear-layer frequencies of the twin-nozzle JICF (mean $\pm$ SD over the valid cases of each ratio group at the given spacing): the side-resolved FFT peak of the time-resolved Liutex magnitude in the near field ($St=f\,d/u_j$), an auxiliary metric distinct from the primary along-trajectory pooled frequency of Section~\ref{sec:liutex}.}
\alttext{Table of the dominant shear-layer frequencies and Strouhal numbers of the front and rear jets on their windward and leeward sides for the three spacings.}
\label{tab:metrics}
\footnotesize
\begin{tabular}{llccc}
\toprule
\textbf{Side} & \textbf{Jet} & \textbf{$L=4d$} & \textbf{$L=3d$} & \textbf{$L=2d$} \\
\midrule
Dominant shear-layer  & front windward & \shortstack{$1505\pm371$\\$(0.43\pm0.11)$} & \shortstack{$1758\pm270$\\$(0.50\pm0.08)$} & \shortstack{$1454\pm459$\\$(0.42\pm0.13)$} \\
frequency (Hz)        & front leeward  & \shortstack{$1585\pm256$\\$(0.45\pm0.07)$} & \shortstack{$1708\pm421$\\$(0.49\pm0.12)$} & \shortstack{$1546\pm278$\\$(0.44\pm0.08)$} \\
($St=f\,d/u_j$)       & rear windward  & \shortstack{$1610\pm348$\\$(0.46\pm0.10)$} & \shortstack{$1208\pm434$\\$(0.35\pm0.12)$} & \shortstack{$1067\pm431$\\$(0.30\pm0.12)$} \\
                      & rear leeward   & \shortstack{$1350\pm487$\\$(0.39\pm0.14)$} & \shortstack{$1146\pm413$\\$(0.33\pm0.12)$} & \shortstack{$1448\pm608$\\$(0.41\pm0.17)$} \\
\bottomrule
\end{tabular}
\end{table}

\section{Conclusions}\label{sec:concl}
Time-resolved PIV at 40\,kHz examined twin tandem jets in crossflow over nine spacing--velocity-ratio conditions ($L=2d$--$4d$, $r\approx3$--7; 68 twin cases and 11 single-jet records), the same dataset as the companion paper~\citep{companion_part1}. The measurements point to the mechanism that carries the front-jet modification established in the companion paper and identify the unsteady dynamics that sustains it.

\textbf{The front-jet modification is carried by the enclosed-recirculation momentum exchange.}
The rear-jet blockage closes a gap recirculation whose reversed flow hugs the front-jet leeward layer and places the two gap-side layers in the counter-current shear-layer configuration. The directly measured Reynolds stress shows the absolute gap-side exchange to be nearly independent of $r$ and to exceed the single-jet leeward zone like-for-like by up to $1.5\times$ (zone peak, across conditions) and $2.8\times$ (area-integrated) at $r\approx7$; the peak-to-zone-mean contrast reaches $27\times$ because the pooled single-jet reference is weak. The exchange enhancement is carried by an enlarged Prandtl mixing length rather than by a larger velocity difference across the layer (Fig.~\ref{fig:promo-s7}). The like-for-like exchange enhancement grows with the velocity ratio and nearly vanishes at $r\approx3$, where the single-jet wake itself reverses strongly.

The near-field shortening of the front jet scales with the absolute near-field transport of the exchange ($\rho=0.68$, $p=0.042$ at the condition level, $n=9$; $\rho=0.42$ per case) rather than with the twin/single enhancement ratio (condition-level $\rho=+0.13$; per-case $+0.23$; neither significant). The resolved turbulent term exceeds the resolved pressure term in eight of the nine conditions, and the pressure force and the elevated gap pressure are common-origin responses of the same measured fields rather than an independent drive (the companion paper). The evidence is observational---an inference to the best explanation, not a proof of causation.

\textbf{Within the resolvable window the dilution is a redistribution rather than a net mass addition.}
The flux-budget diagnostics read the dilution of the companion paper as a centerline deficit, a widened profile and a momentum loss to the crossflow within the resolvable pre-merging window. The centerline excess ratio confirms the dilution ranking independently of the trajectory fit, while the direct entrainment readings fall below the fitted capacity $c=3.31$ of the companion paper on average. The mass-addition bookkeeping of the hyperbolic form remains valid only as an integrated description of the post-merging evolution, and $r_{\mathrm{eff}}$ and $c$ are accordingly effective parameters of the trajectory family.

\textbf{The gap-side layers remain convectively unstable within the resolvable window.}
Both jets roll up into quasi-periodic vortex trains that convect with the local flow, the vortex-motion estimators fanning out into a material--pattern--phase hierarchy for the twin jets while nearly coinciding for the single jet. The two jets shed at comparable frequencies ($1612\pm243$\,Hz front, $1534\pm249$\,Hz rear). DMD and SPOD find no growing mode within the resolvable band and record length, and the resolved modes are localized along the two shear layers; there is therefore no evidence of a self-excited global mode.

Together with the companion paper, this gives a working model of a twin-nozzle injector: the front jet is diluted by an enclosed, recirculation-driven momentum exchange carried by a convecting vortex train, and the rear jet is shielded, both effects being geometric in origin and expected in any tandem arrangement in which the rear jet closes the gap recirculation. For the modeling of such injectors, the requirement this places on a closure is a single quantity---the mixing length of the gap-side layer---rather than a resolved instability.

\subsection*{Supplementary Material}

The supplementary material documents the measurement and processing protocols, the uncertainty and robustness checks, and the supporting statistics, organized by the section of the main text to which each item belongs: Figs. S1--S22 and Tables ST1--ST6.

\subsection*{Acknowledgments}

The author thanks the National Natural Science Foundation of China (NSFC) (52076137); the Taishan Scholar Program of Shandong Province (No. tsqn202306209); and the Natural Science Foundation of Shanghai (21ZR1431300).

\subsection*{Conflict of Interest}

The author has no conflicts to disclose.

\subsection*{Author Contributions}

Xunchen Liu: Conceptualization (lead); Methodology (lead); Investigation (lead); Formal analysis (lead); Software (lead); Validation (lead); Visualization (lead); Writing -- original draft (lead); Writing -- review and editing (lead); Funding acquisition (lead); Supervision (lead).

\subsection*{Data Availability Statement}

The data that support the findings of this study are openly available in the Science Data Bank at https://doi.org/\allowbreak 10.57760/\allowbreak sciencedb.014ij, the same release as the companion paper~\citep{companion_part1}. The release contains the per-case time-averaged velocity fields, the extracted trajectories and fitted parameters, and the Reynolds-stress, mixing-length and pressure statistics of the 68 twin cases and the 11 single-jet records, together with the vortex-motion statistics and the DMD/SPOD spectra plotted in the figures; the 400 instantaneous fields per record are not included because of their size and are available from the corresponding author on request.

\subsection*{References}

\bibliographystyle{aipnum4-2-titles}
\bibliography{twin-jet-piv,liutex}
\end{document}


\maketitle

The figures and tables below accompany the manuscript ``Tandem jets in crossflow: the enclosed-recirculation momentum exchange and the unsteady shear-layer dynamics''; they are grouped by the section of the main text to which they belong. Cross-references to the main text are given by the printed numbers of that document (for example ``Eq.~(3) of the main text''), since the two documents are compiled separately. The companion paper \citep{companion_part1} reports the mean-field measurements and the scaling law on which the present analysis builds.

\noindent\textbf{Organization.} This supplementary material follows the order of the argument of the main text: each section below opens with the main-text section it supports and then collects the figures (S1--S22), the tables (ST1--ST6) and the methodological detail belonging to that step. Every caption is written so that the figure can be read without the main text, and the numbers quoted here agree with those of the main text. Table ST3 and Figs. S11--S15 collect supporting statistics and field diagnostics that are not cited individually in the main text.

\noindent\textbf{Road map.} Section~1 supplies the case accounting behind the measurement and data reduction of Sec.~II. Section~2 gives the field-level detail of the momentum exchange of Sec.~III A (Figs. S1--S2 and Table ST2; the gap-side statistics and entrainment-angle figures moved from the main text are Figs. S20--S22). Section~3 treats the dose--response of the bending against its candidate causes (Sec.~III B, Fig. S3 of this supplement, with the partial-correlation carriers in Fig.~S16 of this supplement). Section~4 gives the flux-budget detail of the dilution as a redistribution (Sec.~III C, Fig. S4). Section~5 documents the vortex-dynamics protocols---tracking, Liutex-PIV, cross-spectral phase and optical flow---and their validation (Sec.~III D, Figs. S5--S12 and Tables ST3--ST6). Section~6 collects the modal-decomposition statistics (Sec.~III E, Figs. S13--S15). Section~7 closes the supplement with the supporting figures for the discussion (Secs.~III F--H, Figs. S16--S22).

\section{Measurement and data reduction (Sec.~II of the main text)}

This section documents the measurement and data-reduction bookkeeping behind Sec.~II of the main text; no figure of this supplement belongs to this step. Every record is a 40\,kHz burst of 400 image pairs at $25\,\mu$s separation (10\,ms), processed with DaVis 8.4 multi-pass cross-correlation (24$\times$24 pixel interrogation windows, 75\% overlap, 0.35\,mm vector spacing), with rejected vectors flagged and never interpolated; the validation and uncertainty budget (a few percent random uncertainty per validated vector, below one percent for a record mean, and a systematic calibration limit that largely cancels in the twin-to-single comparisons) is given in Sec.~II of the main text. Table ST1 gives the case accounting of the samples entering the various statistics (the full acquisition accounting of the dataset is given in the companion paper~\citep{companion_part1}).

\begin{table}
\centering
\caption{\textbf{Table ST1.} Case accounting of the samples entering the statistics of this paper (the full acquisition accounting of the dataset is given in the companion paper~\citep{companion_part1}). Of the 68 twin cases entering the statistics, 67 contribute the Reynolds-stress terms (the mean-only record 4d-5r-8 has no instantaneous export), 56 have a detectable front-jet near-field transition and thus a defined near-field shortening $E_{ty}$ with the force-level terms $X_{\rm turb}$ and $X_{\rm pres}$, and all 56 also carry the closest-approach gap pressure $p_{\rm int}$.}
\alttext{Table of the case accounting of the samples entering the statistics of this paper.}
\begin{tabular}{lll}
\toprule
Quantity & Sample & Count \\
\midrule
Twin cases entering the statistics & all conditions & 68 \\
Reynolds-stress terms available & 68 minus 4d-5r-8 & 67 \\
Near-field transition detected (dose--response sample) & 68 minus 12 without transition & 56 \\
Gap pressure $p_{\rm int}$ on the dose--response sample & all 56 cases & 56 \\
\bottomrule
\end{tabular}
\end{table}
\section{The exchange: field-level detail (Sec.~III A of the main text)}

This section provides the field-level detail of the momentum-exchange measurement of Sec.~III A of the main text that is not reproduced in the paper. Figure S1 compares the single-jet leeward recirculation zone with the twin gap-side zones like-for-like, and Fig. S2 resolves the turbulent kinetic energy (TKE) of the shear-layer band in the near and far fields. The plain field figures---the counter-current shear-layer (CCSL) extraction and the velocity-ratio profiles (Figs.~4 and 5), the gap-side exchange fields and their difference from the single jet (Figs.~7 and 8), and the mixing-length fields (Figs.~9 and 10)---are in the main text; the gap-side exchange statistics and the entrainment-angle geometry, moved here from the main text, are given in Figs. S20--S22 of this supplement; the leading-term momentum balance and the mixing-length difference field are given in Fig. S17 and Fig. S4 of this supplement, respectively, and the condition-mean Liutex fields of the nine conditions are given in Fig.~15 of the main text. Table ST2 compares the four exchange metrics used in the main text.

\begin{figure}
\centerline{\includegraphics[width=\textwidth]{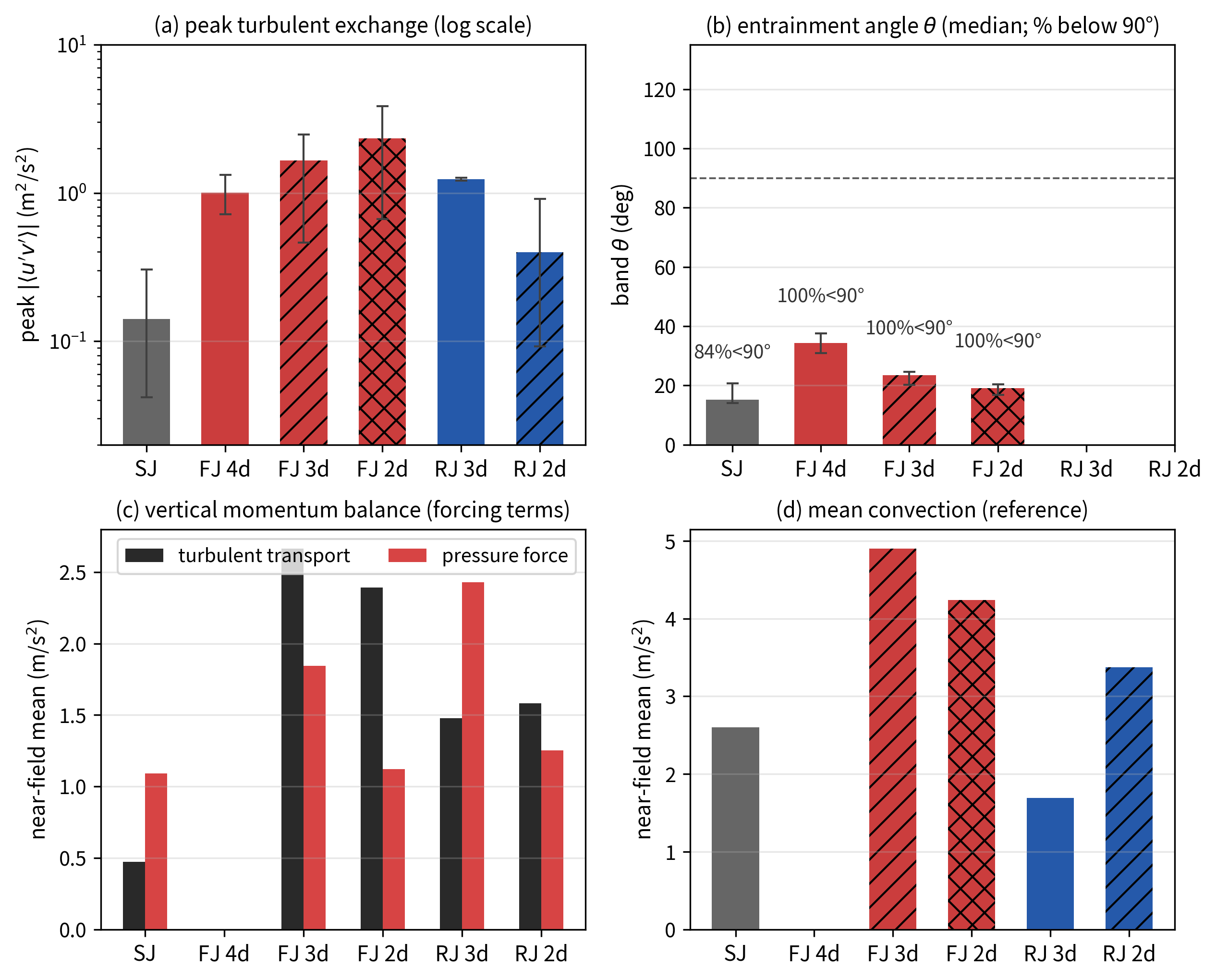}}
\caption{\textbf{Figure S1.} Quantitative comparison of the single-jet leeward recirculation zone (SJ, gray) with the twin-jet gap-side zones (front-jet leeward FJ at $L=4d$, $3d$, $2d$, red; rear-jet windward RJ at $L=3d$, $2d$, blue; the RJ-4d layer is co-flowing), sampled identically (gap-side zone $0\le n\le4$\,mm, $u<0$ points, near field $s=2$--$12$\,mm): (a) the peak turbulent momentum exchange $|\langle u'v'\rangle|$ (log scale), (b) the entrainment angle $\theta$ with the fraction of the zone points below $90^\circ$ (dashed line: $\theta=90^\circ$; only the layers with a defined entrainment zone enter this panel), (c) the near-field turbulent transport (black) and pressure force (red) (bars: group means, error bars: station ranges), and (d) the mean convection term (reference, same bar convention). The geometry of the two recirculations is similar, but the turbulent exchange is stronger on the twin gap side. Panel (a) plots the near-field zone-mean peak of each layer (the SJ value is 0.14\,m$^2$/s$^2$, a zone mean, not the zone peak of 2.4\,m$^2$/s$^2$ in Table~II of the main text); the per-height window-peak ratio to this pooled single-jet reference reaches 27 times, which is a peak-to-zone-mean contrast. The like-for-like ratios (each layer against the single jet of the same velocity ratio) are 0.58--1.46 in zone peak and 0.59--2.83 in the area integral (main text, Sec.~III A); the absolute exchange is nearly independent of the velocity ratio. This is the force-level signature of the early front-jet bending.}
\alttext{Comparison of the single-jet leeward recirculation zone with the twin gap-side zones: peak turbulent exchange, entrainment angle, transport and pressure force, and mean convection term.}
\end{figure}

\begin{figure}
\centerline{\includegraphics[width=\textwidth,height=0.82\textheight,keepaspectratio]{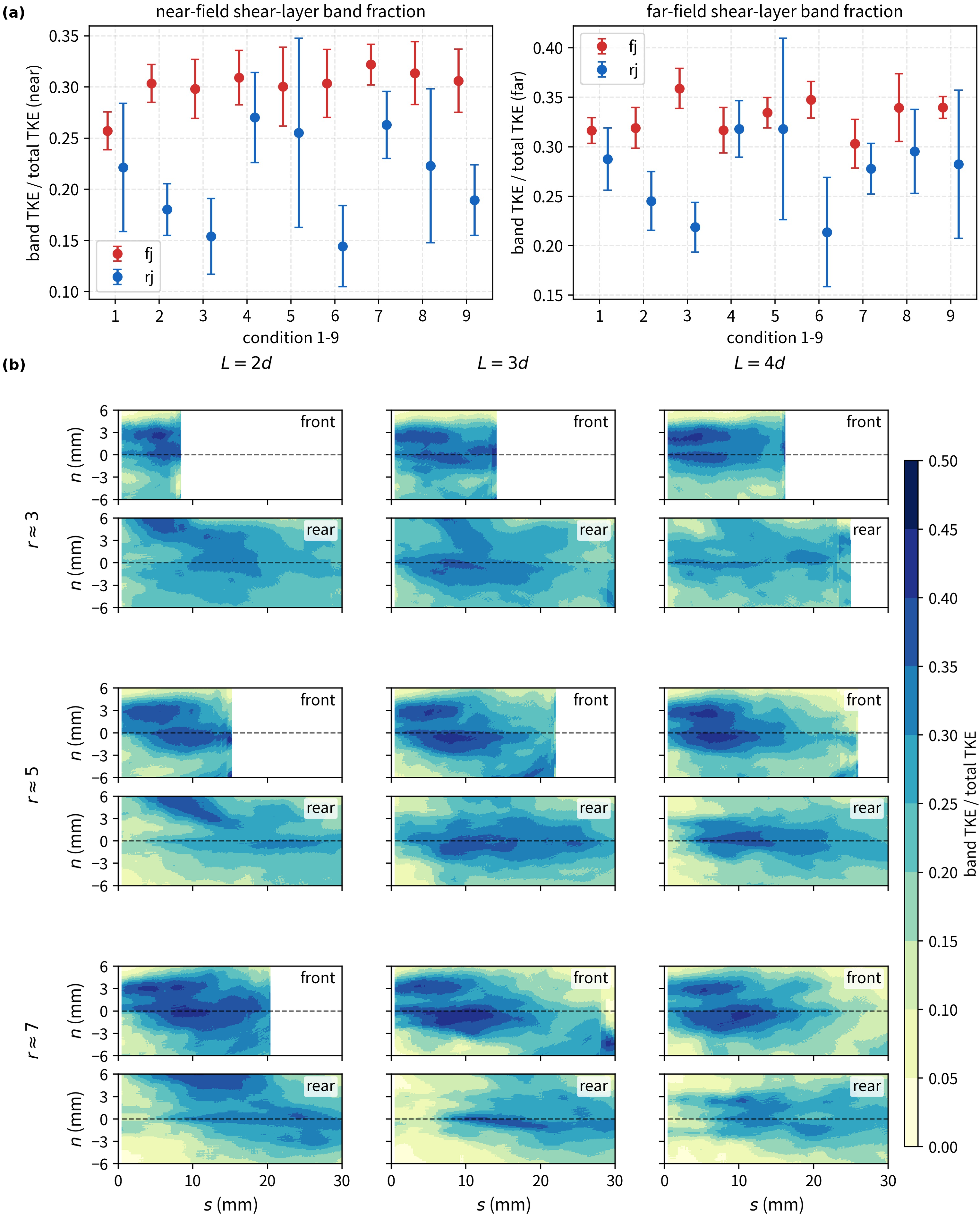}}
\caption{\textbf{Figure S2.} Mean shear-layer-zone fraction of the total TKE. (a) The zone fraction of the total TKE versus the nine conditions (with case-to-case SD error bars) for the near and far fields of the front (red) and rear (blue) jets. (b) Condition-mean band-fraction maps (zone TKE over total TKE, 0.4--3\,kHz; fixed $0$--$0.5$ scale in bands of $0.05$) of the nine conditions in the natural coordinates (front jet in the upper and rear jet in the lower panel). The far-field zone fraction exceeds the near-field value for both jets in nearly every condition, and the coherent shear-layer region carries a larger fraction of the TKE in the far field than in the near field.}
\alttext{Mean shear-layer-zone fraction of the total turbulent kinetic energy and the corresponding band-fraction maps of the nine conditions.}
\end{figure}

\begin{table}
\centering
\caption{\textbf{Table ST2.} The four exchange metrics used in Sec.~III A of the main text, with their definitions, ranges and roles. The window-peak $P_{uv}$ (Eq.~(3) of the main text) is the per-height peak within $\pm2$\,mm windows across each layer; the zone peak $P_{uv}^{\rm zone}$ and the area-integrated $S_{uv}$ are the statistics of Table~II of the main text over the gap-side zone; the per-height window-peak twin/single ratio is the primary exchange-enhancement metric of Sec.~III A; the pooled per-station ratio of Table~III is its robustness form, which pools over stations and heights and therefore lowers the ratio relative to the per-height comparison. Ranges are over the nine conditions unless noted. The last two rows are contrast-type metrics that normalize by the pooled single-jet leeward zone mean; the like-for-like ratios against the single jet of the same velocity ratio are 0.58--1.46 (zone peak) and 0.59--2.83 (area-integrated), and are quoted in Sec.~III A of the main text.}
\alttext{Table comparing the four exchange metrics of the main text with their definitions, ranges and roles.}
\begin{tabular}{>{\raggedright\arraybackslash}p{2.9cm}>{\raggedright\arraybackslash}p{6.9cm}>{\raggedright\arraybackslash}p{3.1cm}>{\raggedright\arraybackslash}p{2.3cm}}
\toprule
Metric & Definition & Range & Role \\
\midrule
Window peak $P_{uv}$ & $\max_{|x-x_e|\le2\,\mathrm{mm}}|\langle u'v'\rangle|$ per height (Eq.~(3)) & $0.46$--$3.85$\,m$^2$/s$^2$ (front leeward) & local exchange level \\
Zone peak $P_{uv}^{\rm zone}$ & peak over the gap-side zone (Table~II) & $1.34$--$6.00$\,m$^2$/s$^2$ & zone exchange level \\
Area-integrated $S_{uv}$ & $\iint|\langle u'v'\rangle|\,ds\,dn$ over the gap-side zone (Table~II) & $13.99$--$88.89$\,m$^2$/s$^2$\,mm$^2$ & integrated exchange intensity \\
Per-height twin/single ratio & $P_{uv}/P_{uv,\rm single}$, window-peak, per case & $7$--$27\times$ & primary enhancement metric \\
Pooled per-station ratio & per-station-mean peak ratio, pooled (Table~III) & $1.7$--$11.2\times$ & robustness metric of the enhancement \\
\bottomrule
\end{tabular}
\end{table}

\section{The dose--response of the bending (Sec.~III B of the main text)}

This section supplies the per-case dose--response behind the smoothed analysis of Fig.~11 of the main text. There the response is a local-linear (LOWESS-type) fit to the case means (0.20 grid of $E_{ty}$, tricube kernel, bandwidth $0.30$) with a 95\% bootstrap confidence band (1000 resamples of the 56 cases) and no binned symbols; Fig. S3 of this supplement plots the 56 individual cases with their continuous values of the near-field shortening, so that the scatter behind the fit can be inspected. The partial-correlation carriers that sharpen the causal assignments are collected, with their mediation statistics, in Fig. S16 of this supplement.

\begin{figure}
\centering
\centerline{\includegraphics[width=\textwidth,height=0.67\textheight,keepaspectratio]{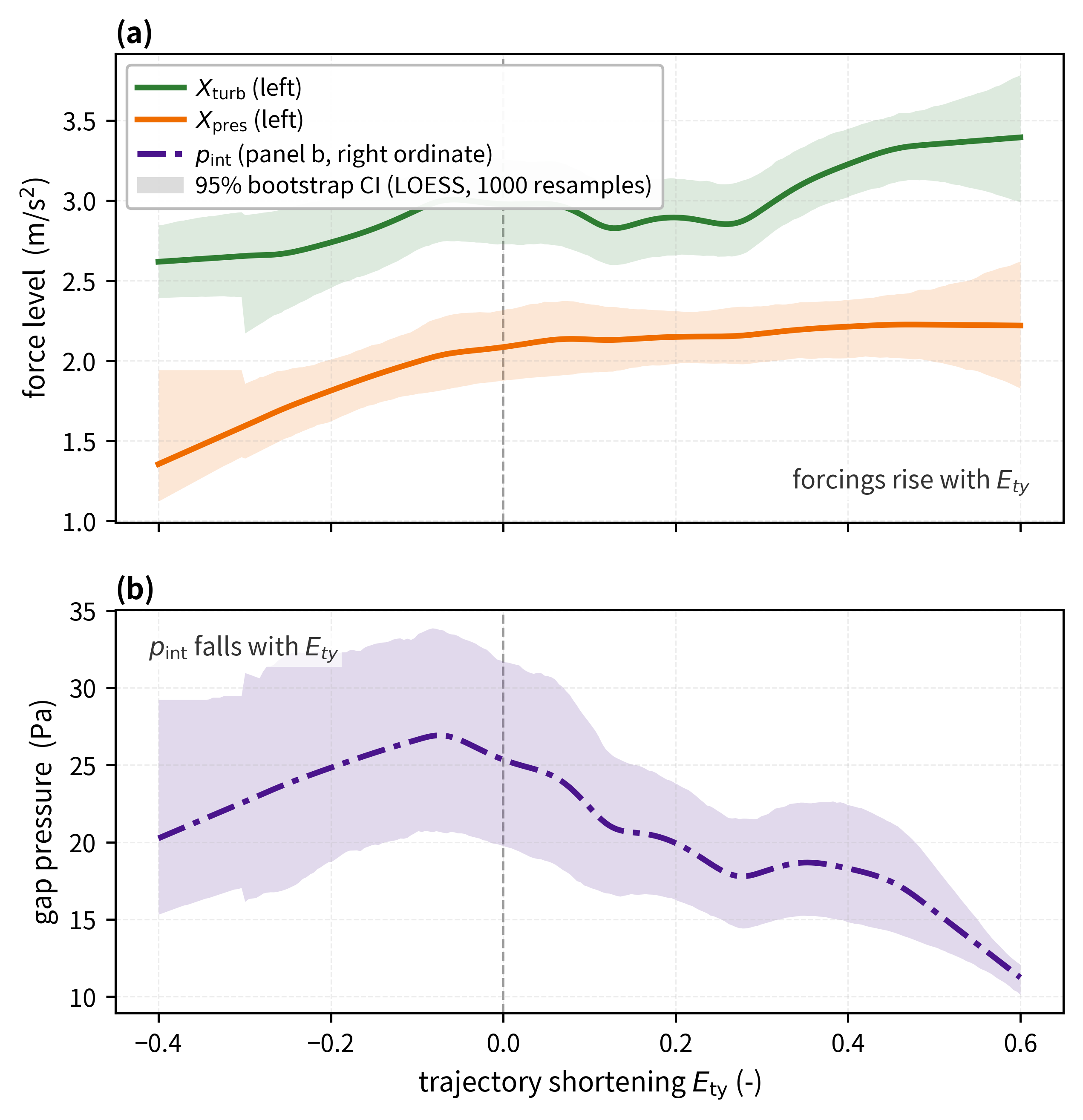}}
\caption{\textbf{Figure S3.} The per-case dose--response of the front-jet bending against its candidate causes: (a) the near-field turbulent transport $X_{\rm turb}$ (green) and the pressure force $X_{\rm pres}$ (orange) on the left ordinate (force level, m\,s$^{-2}$), and (b) the closest-approach gap pressure $p_{\rm int}$ (purple, Pa, own ordinate), each against the near-field shortening $E_{ty}$ (Eq.~(6) of the main text) over the 56 cases with a detectable front-jet near-field transition. The solid curves are local-linear (LOESS-type) fits through the individual cases and the shaded bands their $95\%$ bootstrap confidence intervals (1000 resamples); no bin statistics are drawn. The vertical dashed line marks $E_{ty}=0$ (no shortening). The two forcing terms rise with the shortening while the gap pressure falls: the pressure level is lowest where the bending is strongest, and it never supplies the suction an attraction mechanism would require (main text, Sec.~III B).}
\alttext{Per-case dose-response of the front-jet bending: the near-field turbulent transport, the pressure force and the closest-approach gap pressure against the near-field shortening, with local fits and bootstrap confidence bands.}
\end{figure}

\section{The dilution as a redistribution: flux-budget detail (Sec.~III C of the main text)}

The flux-budget derivation behind Sec.~III C of the main text is documented here; the symbols follow the main text and are re-defined only where their operational definition is needed. Natural-coordinate profiles $(s,n)$ are built from the instantaneous fields (400 frames per record; the mean-only record 4d-5r-8 contributes no stress terms). At each station the excess velocity $\Delta u=u_s-U_\infty \hat t_x(s)$, with $\hat t_x$ the streamwise component of the unit tangent to the trajectory, is integrated over the contiguous jet region ($\Delta u>\max(0.08\,\Delta u_{\rm peak},\,0.2\ {\rm m/s})$, truncated at the gap-side valley, at the windward blockage minimum and at the merging station). The analysis window is the first contiguous pre-merging segment ($\tilde u=s/(rd)\le0.35$ unless merging intervenes earlier). The volume flux is $Q=\int\Delta u\,dn$ and the advective momentum flux $J_{\rm adv}=\int\Delta u^2\,dn$; the bulk readout is the momentum-to-mass ratio $u_{\rm eff}=J_{\rm adv}/Q$. The entrainment readings are $c_{\rm meas}=(Q(\tilde u^*)-Q(\tilde u_0))/(U_\infty d)$ per condition and $c_{\rm excess}$, the excess of $c_{\rm meas}$ over the matched single-jet growth. The bookkeeping is validated on the single-jet records: the near-exit volume flux $Q(s_0)$ averages $41.3$, compared with the nominal $V_j d=40$ (mm\,m/s per unit span), and the leave-one-out momentum-to-mass ratio recovers the single-jet family at $u_{\rm eff}/V_j=1.018$ (range $0.71$--$1.19$). Here $\tilde u_0$ and $\tilde u^*$ are the start and the end of the analysis window. The condition averages exclude the two off-specification records (nozzle positions $x\approx9.4$--$9.6$\,mm instead of 10\,mm).

\begin{figure}
\centerline{\includegraphics[width=\textwidth]{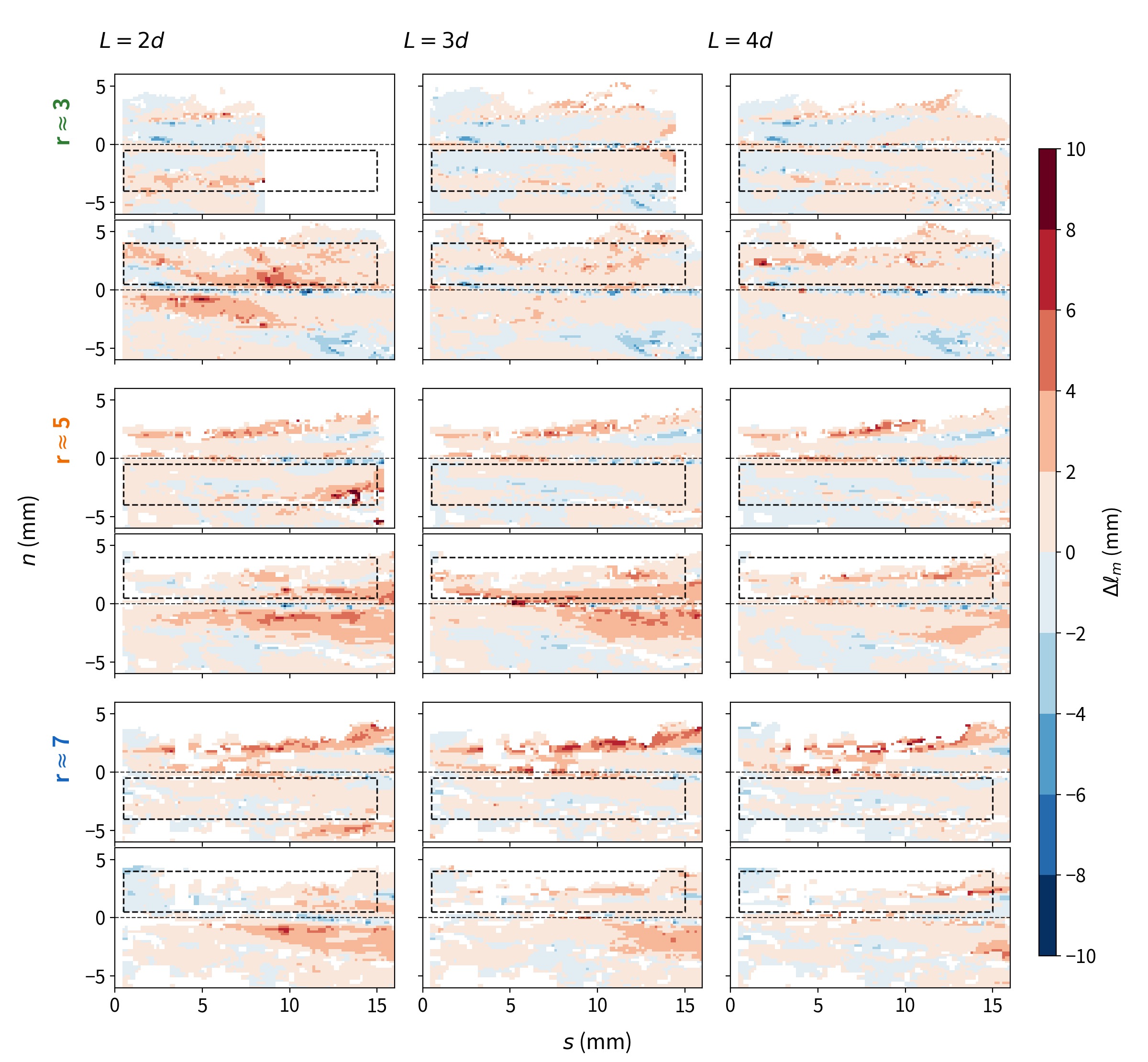}}
\caption{\textbf{Figure S4.} Natural-coordinate mixing-length difference $l_{\rm mix,twin}-l_{\rm mix,single}$ of the condition-averaged fields, referenced to the single jet of the same velocity-ratio group: eighteen panels (rows $r\approx3$, 5, 7; columns $L=2d$, $3d$, $4d$; front jet in the upper and rear jet in the lower panel of every cell), with $s$ the arc length and $n$ the perpendicular offset ($n=0$ dashed black); red: twin exceeds single, blue: twin below single, on the fixed $\pm10$\,mm scale in bands of $2$\,mm (colour bar); black dashed rectangles: the gap-side zone over which the exchange integrals of Table~II of the main text are evaluated; the two off-specification records are excluded from the condition averages. The mixing-length enhancement occupies the same gap-side zone as the exchange enhancement of Fig.~8 of the main text at $r\approx5$ and 7; at $r\approx3$ the difference becomes small and mixed in sign.}
\alttext{Natural-coordinate difference fields of the Prandtl mixing length between the twin jets and the single-jet reference over the nine conditions.}
\end{figure}

\section{Vortex-dynamics protocols (Sec.~III D of the main text)}

The four vortex-motion estimators of Sec.~III D of the main text---Liutex-PIV (the windowed-deformation fit to the measured Liutex field), core tracking, the cross-spectral phase along the layer and the dense optical flow---are documented and cross-validated here. Figure S5 gives the single-jet near/far decomposition and Fig. S6 the kymograph and spectrum on which the estimators operate; Figs. S7 and S8 collect the Liutex-amplitude statistics and the time-mean fields; Tables ST3--ST6 give the single-jet and twin-jet near/far medians and the frequency comparison. The corresponding nine-condition statistics are given in this supplement by the tracked vortex dynamics (Fig. S19) and the Liutex-amplitude profile (Fig. S7), and in the main text by the shedding-frequency figure (Fig.~23), the Liutex-amplitude fields (Fig.~15) and the estimator comparison (Fig.~21). The estimators nearly coincide for the single jet and fan out into the material--pattern--phase hierarchy of Sec.~III D for the twin jets.

The protocols behind these estimators are as follows. All field quantities use the Liutex magnitude $R$ of the rigid-rotation part of the velocity gradient, normalized as $R^*=R\,d/u_j$ with $d=4$\,mm and $u_j=14$\,m/s. The near/far split of every table and figure uses the near-field transition station $s_{tn}$, the arc length at which the near-field power-law fit of the trajectory gives way to the far-field fit (Sec.~II of the main text). Vortex cores are the local maxima of $R$ inside the trajectory band $|n|\le3$\,mm (arc length $s\le30$\,mm for the twin jets and $s\le20$\,mm for the single jet), thresholded at half the in-band maximum, and are matched between every fourth frame (100\,$\mu$s) by nearest neighbour within 1.5\,mm; tracks with fewer than four points are discarded. The tracked core speed is the arc-length rate $ds/dt$ of the matched cores; the spacing $\lambda$ is the separation of adjacent simultaneous cores and the passage rate is one crossing per track and station, with the windward and leeward rows separated by the sign of $n$ ($n>0$: windward, upstream, convex side). The cross-spectral phase speed follows from the phase slope between adjacent stations inside the vortex band in a 6\,mm sliding window (0.4--3\,kHz). The Liutex-PIV estimate applies the OpenPIV windef chain---a 32-pixel first pass refined on a 16-pixel deformed window, Gaussian sub-pixel interpolation and a peak-to-mean signal-to-noise criterion---to the Liutex field on the windows given in Tables ST4 and ST6. The dense optical flow uses the Farneb\'ack method (pyramid level 3, 15-pixel interrogation with Gaussian weighting) on frame pairs four frames apart (100\,$\mu$s). Frequencies are the peaks of the side-resolved FFT of the 400-frame (10\,ms, 100\,Hz resolution) Liutex time series within 0.4--3\,kHz, quoted as $St=f_0d/u_j$.

\begin{figure}
\centerline{\includegraphics[width=\textwidth]{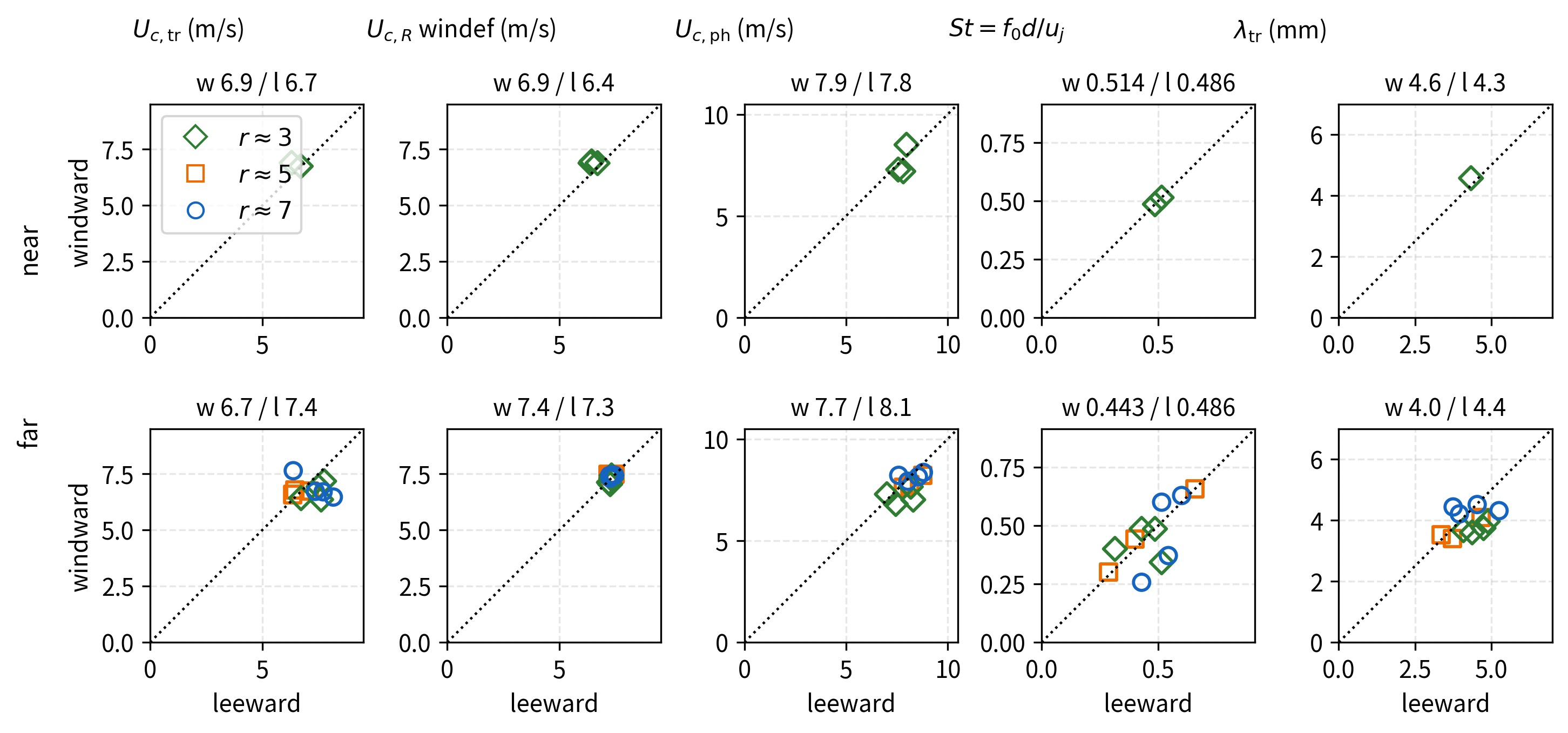}}
\caption{\textbf{Figure S5.} Near/far decomposition of the single-jet vortex motion, windward versus leeward: rows near ($s\le s_{tn}$) and far ($s>s_{tn}$); columns, from left to right, tracked core speed $U_{c,\rm tr}$, Liutex-PIV speed $U_{c,R}$, cross-spectral phase speed $U_{c,\rm ph}$, FFT frequency as $St$ and tracked spacing $\lambda_{\rm tr}$. The windward value is the ordinate and the leeward value the abscissa, the dotted line is the 1:1 line, and the medians over the 11 records are printed above each panel (w/l: windward/leeward); symbols: diamond $r\approx3$, square $r\approx5$, circle $r\approx7$. The near/far split is at $s_{tn}$ except for the Liutex-PIV windows $U_{c,R}$, which are $s\in[0.5,\,5]$ and $[8,\,20]$\,mm.}
\alttext{Near/far decomposition of the single-jet vortex motion on the windward and leeward sides.}
\end{figure}

\begin{figure}
\centerline{\includegraphics[width=0.92\textwidth]{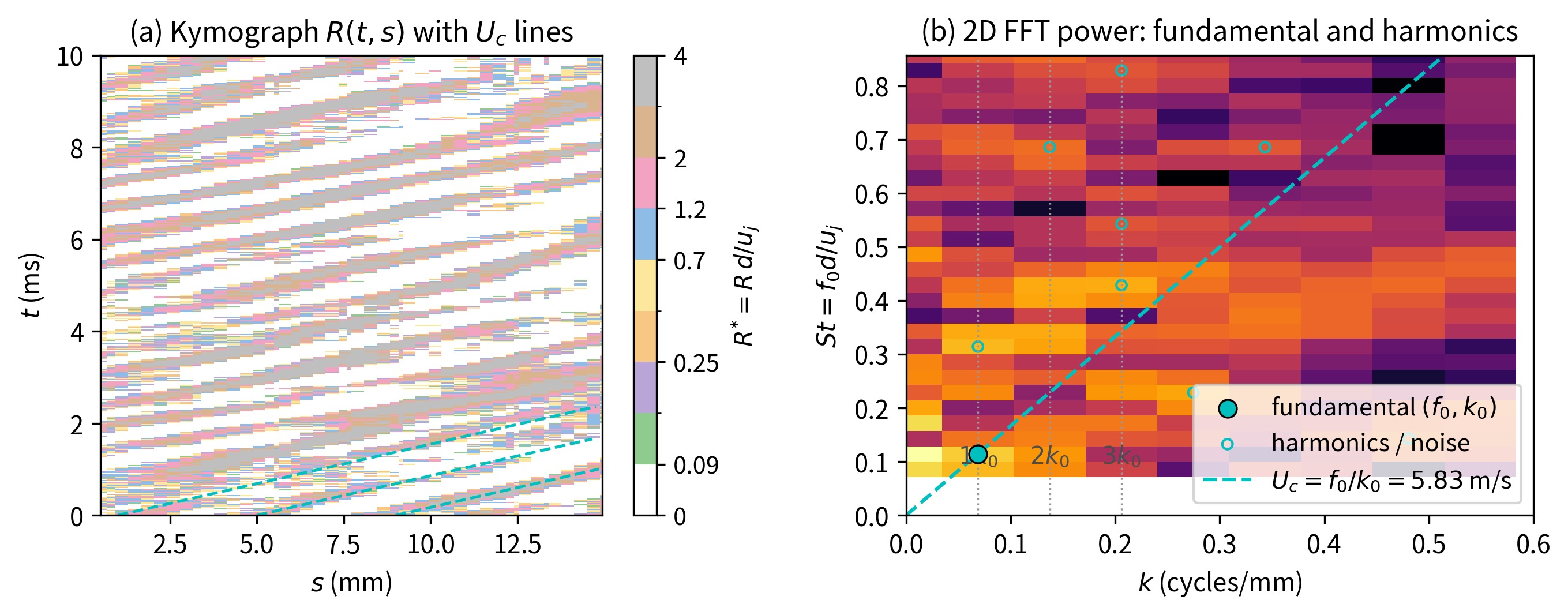}}
\caption{\textbf{Figure S6.} Convection speed of a representative case (condition 9, front jet): (a) kymograph $R(t,s)$ of the Liutex magnitude on the fixed discrete $R^*=R\,d/u_j$ scale (colour bar) versus time $t$ and arc length $s$, with the convection lines of slope $U_c$ (cyan dashed); (b) power spectrum $P(k,f)$ of its two-dimensional FFT versus the wavenumber $k$ and the Strouhal number $St=f_0\,d/u_j$. A train convecting at $U_c$ concentrates power along $f=U_c k$ (cyan dashed line); the strongest peak (filled circle) is the fundamental $(f_0,k_0)$, giving $U_c=f_0/k_0$; open circles: harmonics and noise of the same spectrum.}
\alttext{Kymograph and two-dimensional spectrum of a representative case with the convection-speed line.}
\end{figure}

\begin{figure}
\centerline{\includegraphics[width=\textwidth,height=0.82\textheight,keepaspectratio]{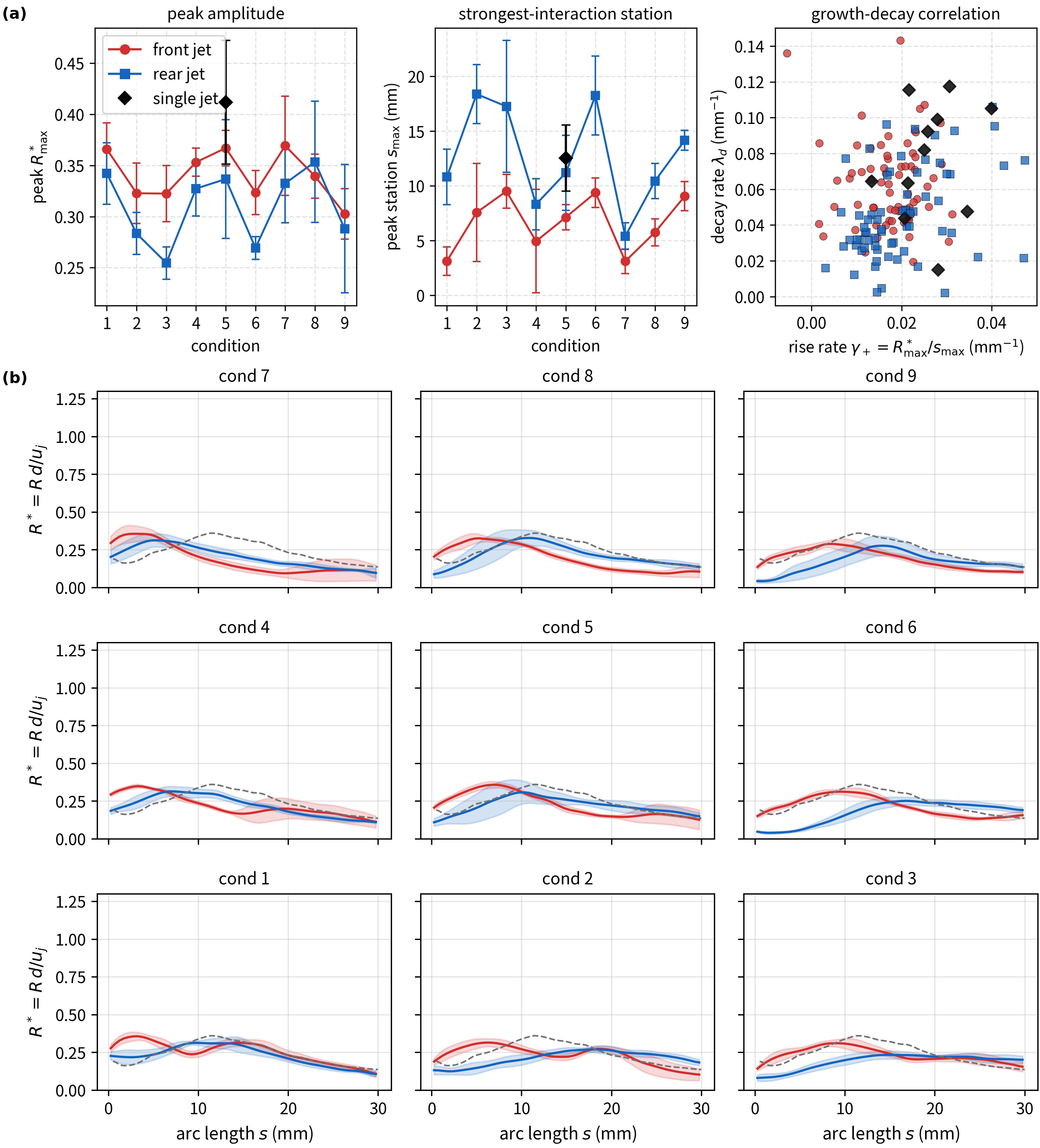}}
\caption{\textbf{Figure S7.} Liutex-amplitude statistics and profiles. (a) Amplitude statistics over the 136 case-jets (68 cases $\times$ 2 jets): peak amplitude, peak station and rise rate versus the far-field decay rate; the peak occurs systematically earlier for the front jet. (b) Condition-mean normalized Liutex amplitude $\bar R^*(s)=R\,d/u_j$ along the trajectory (zone mean over $|n|\le3$\,mm) for the nine conditions (red front, blue rear; bands $\pm1$ SD; gray dashed: single-jet mean).}
\alttext{Liutex-amplitude statistics over the 136 case-jets and the condition-mean normalized amplitude profiles along the trajectories.}
\end{figure}

\begin{figure}
\centerline{\includegraphics[width=\textwidth]{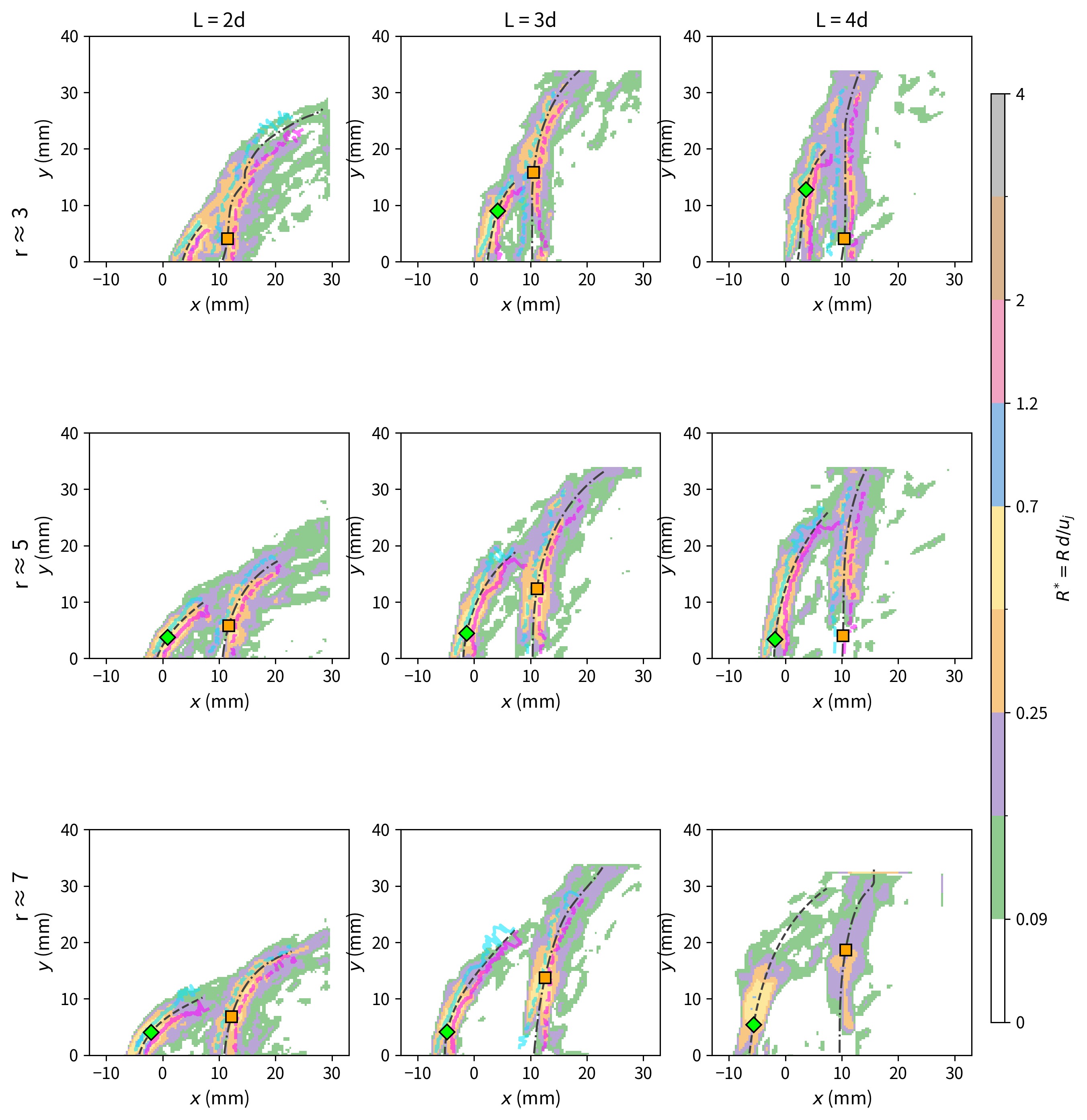}}
\caption{\textbf{Figure S8.} Time-mean Liutex fields $R^*=R\,d/u_j$ of the nine conditions in the physical $(x,y)$ coordinates (rows $r\approx3$, 5, 7; columns $L=2d$, $3d$, $4d$, the arrangement of Table~I), on the fixed discrete absolute scale of the main text (colour bar; white below the lowest band), with the two vortex rows of each jet (cyan windward, magenta leeward; dashed curves: the mean trajectories) and the near/far transition markers (diamond and square). The vortex rows are the two shear-layer branches whose cores are tracked in Fig.~14 of the main text; the space--time vortex tubes of the same cores are shown in Fig.~18 of the main text; the mean field shows how the rows spread and fade downstream.}
\alttext{Time-mean Liutex fields of the nine conditions with the two vortex rows of each jet and the near/far transition diamonds.}
\label{si-sup-liutex-mean}
\end{figure}

\begin{figure}
\centerline{\includegraphics[width=0.92\textwidth]{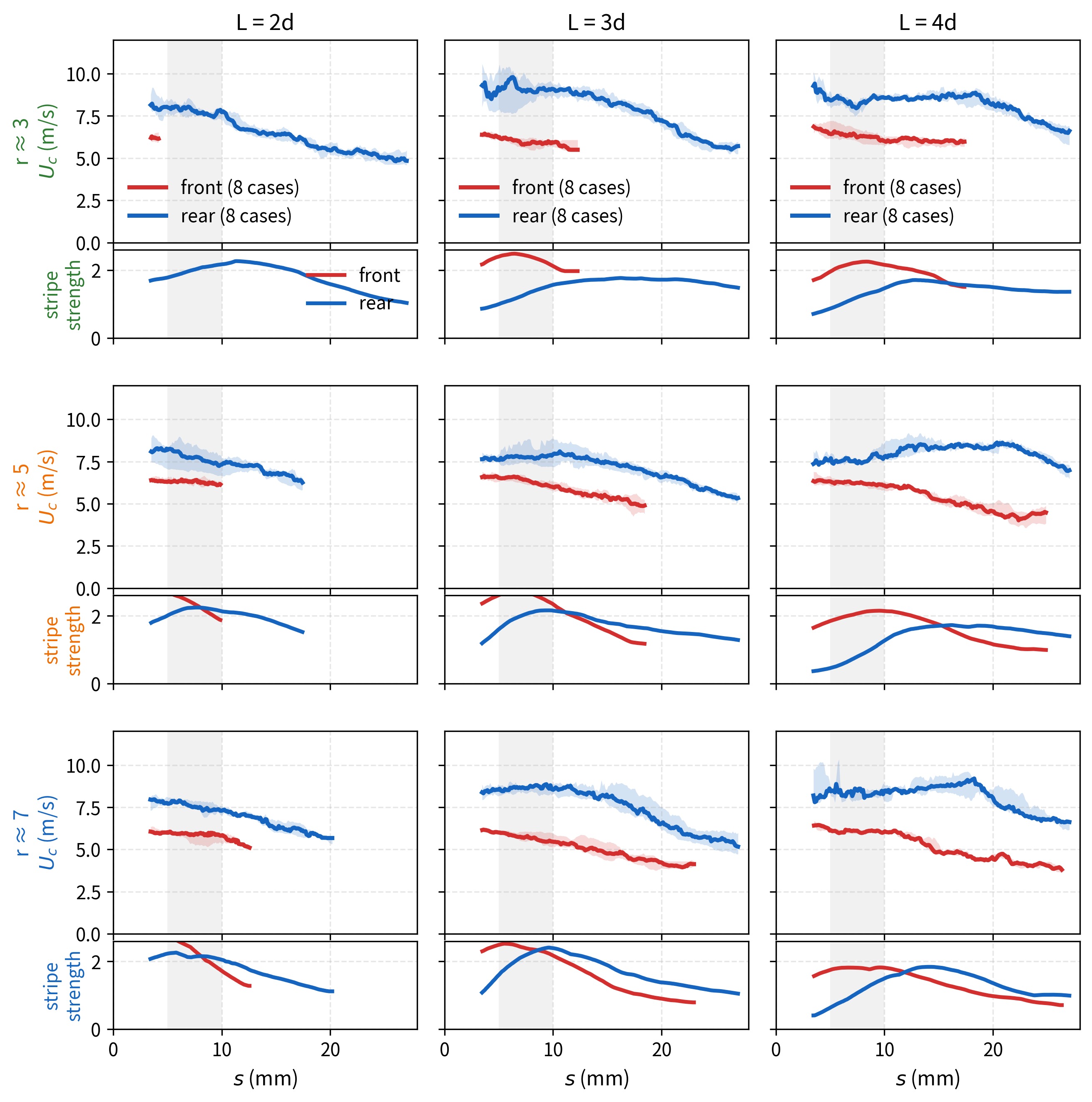}}
\caption{\textbf{Figure S9.} Vortex convection speed resolved in position: the vortex convection speed $U_c(s)$, measured station by station in the space--time diagrams of Figs.~16 and~17 of the main text (6\,mm sliding window, cross-spectral phase speed inside the vortex band), is shown in the upper sub-panel of every cell, with the interquartile band over the cases of each condition (front red, rear blue; the case count is given in each legend); the lower sub-panel gives the stripe strength (the space--time contrast of the vortex train); rows $r\approx3$, 5, 7, columns $L=2d$, $3d$, $4d$. The grey vertical band marks the fixed near-field window common to all panels, and the abscissa is the arc length $s$.}
\alttext{Vortex convection speed resolved in position with the interquartile band over the cases of each condition and the space--time stripe strength.}
\label{si-sup-uc-pos}
\end{figure}

\begin{figure}
\centerline{\includegraphics[width=\textwidth]{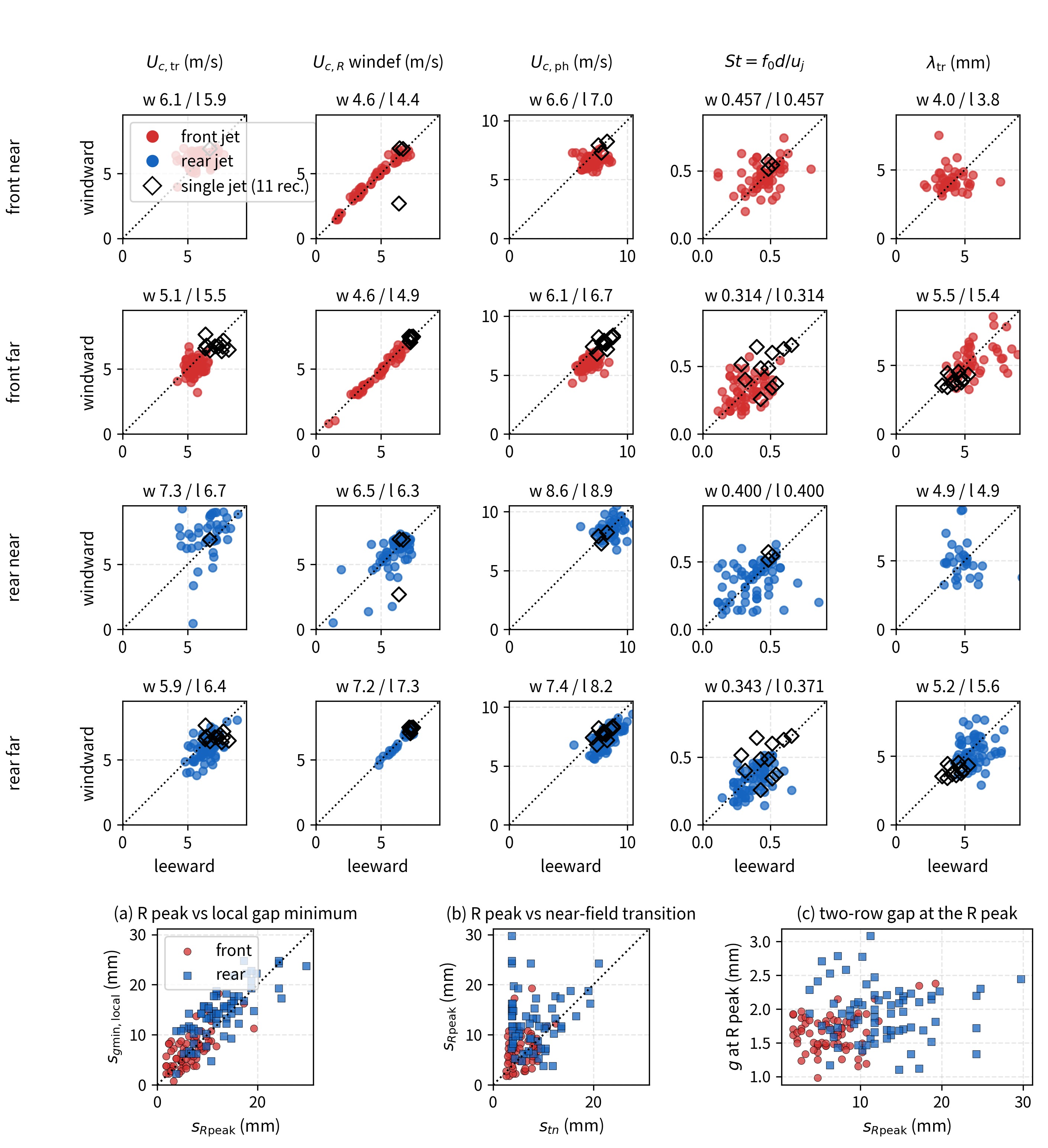}}
\caption{\textbf{Figure S10.} Four-part near/far decomposition of the twin-jet vortex dynamics and R-peak station diagnostics. Upper block, for the four quadrants (rows: front and rear jet; columns: near and far field): the vortex-motion estimators plotted against each other, windward (ordinate) versus leeward (abscissa), with the dotted 1:1 line and the medians over the cases printed above each panel (w/l); columns, from left to right: tracked core speed, Liutex-PIV speed, cross-spectral phase speed, frequency as $St$ and tracked spacing. Symbols: red front jet, blue rear jet, open diamonds the 11 single-jet records. Lower block: (a) R-peak station versus the local gap minimum, (b) R-peak station versus the near-field transition and (c) the two-row gap at the R peak versus the R-peak station, over the 136 case-jets (red front, blue rear).}
\alttext{Four-part near/far decomposition of the twin-jet vortex dynamics and the R-peak station diagnostics against the gap minimum, the near-field transition and the row separation.}
\label{si-promo-s11}
\end{figure}

\begin{figure}
\centerline{\includegraphics[width=\textwidth]{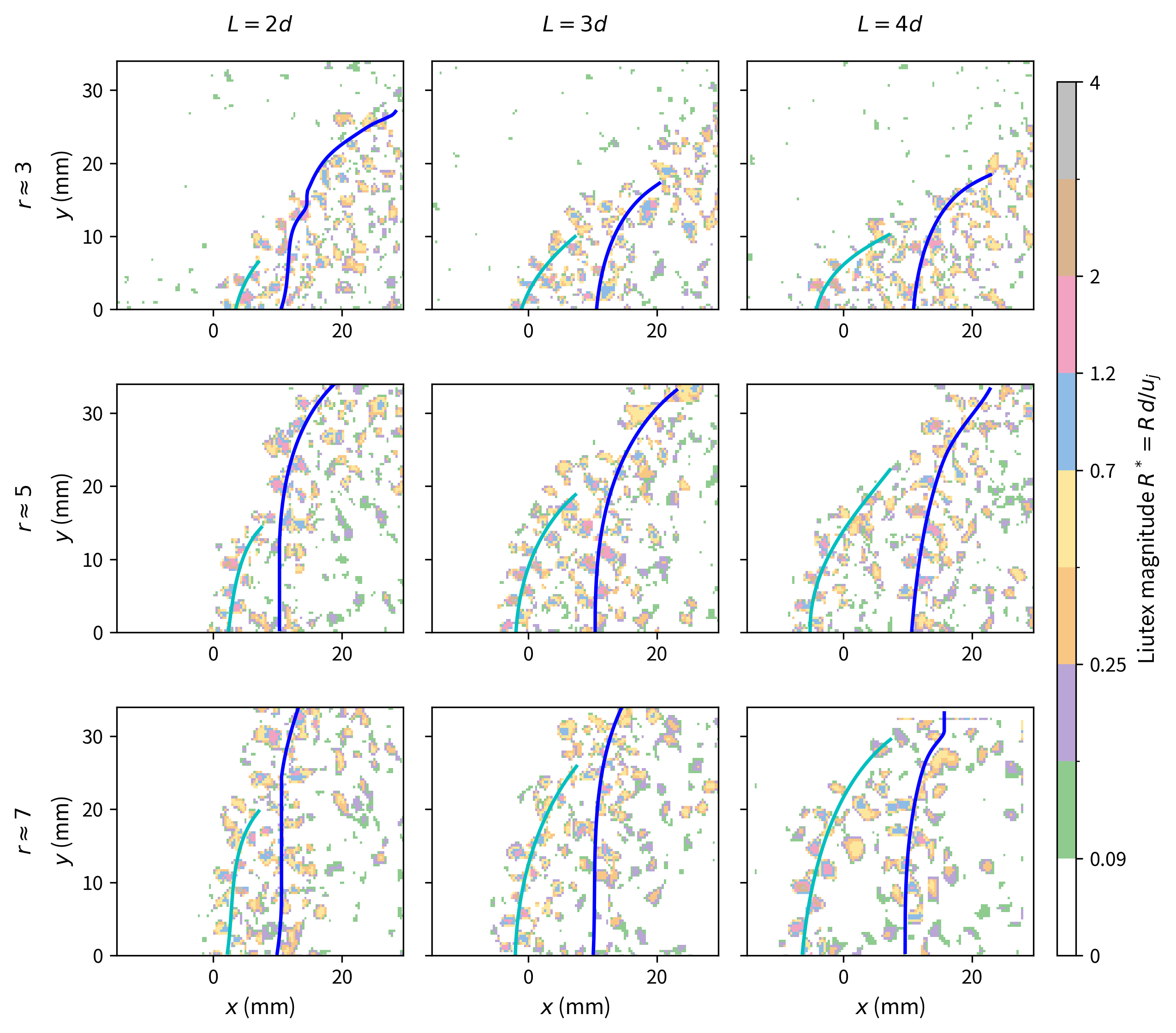}}
\caption{\textbf{Figure S11.} Instantaneous normalized Liutex fields $R^*=R\,d/u_j$ (one frame) of representative cases of the nine conditions, in the format of the nine-condition mean-field figure of the main text (Fig.~6; rows $r\approx3$, 5, 7; columns $L=2d$, $3d$, $4d$, the arrangement of Table~I), on the fixed discrete absolute scale of the main text (colour bar; white below the lowest band), with the extracted front (cyan) and rear (blue) trajectories. The shear-layer vortex sheets are visible on both sides of each jet, and at small spacings the gap-side vortex activity is enhanced and the sheets merge closer to the exit.}
\alttext{Instantaneous normalized Liutex fields of representative cases of the nine conditions with the extracted front and rear trajectories.}
\label{si-promo-s12}
\end{figure}

\begin{figure}
\centerline{\includegraphics[width=\textwidth]{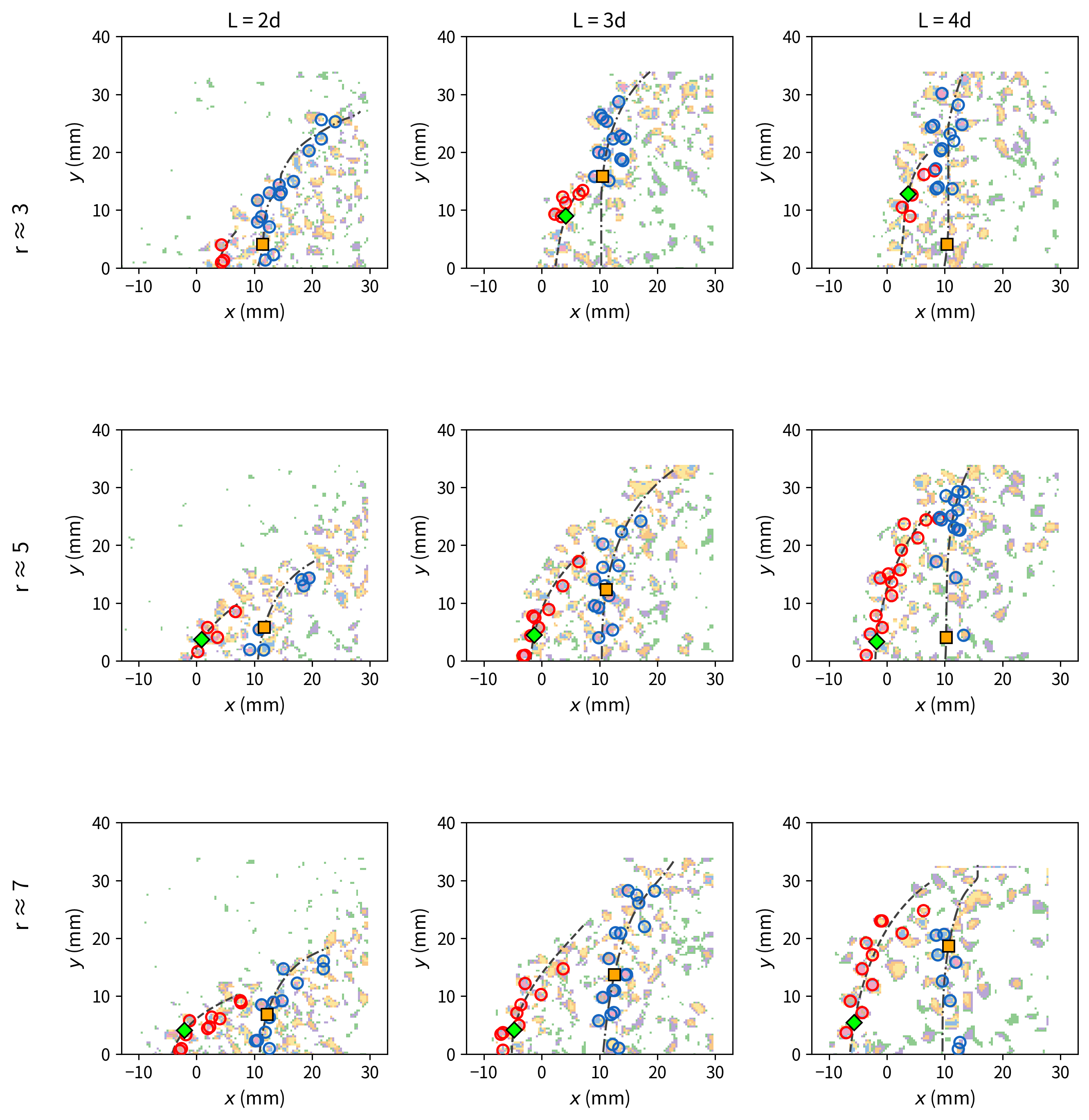}}
\caption{\textbf{Figure S12.} Instantaneous Liutex fields of the nine conditions in the physical $(x,y)$ coordinates, the coordinate counterpart of Fig.~14 of the main text (rows $r\approx3$, 5, 7; columns $L=2d$, $3d$, $4d$), with the mean trajectories (dashed), the vortex cores detected in the front-jet (red circles) and rear-jet (blue circles) shear layers, and the front-jet (green diamonds) and rear-jet (orange squares) near-field/far-field transitions; the colour scale is the fixed absolute one of the main text (white below $R^{*}=0.09$, then eight non-uniform bands up to $R^{*}=4$). The field of view spans the full physical plane of both jets.}
\alttext{Instantaneous Liutex fields of the nine conditions in the physical coordinates, with the mean trajectories, the detected vortex cores and the near/far transitions.}
\end{figure}

\begin{table}
\centering
\begin{tabular}{cccc}
\hline
record & $s_{R,\mathrm{peak}}$ (mm) & $g(s_{R,\mathrm{peak}})$ (mm) & $s_{g,\min}$ (mm) \\
\hline
$r\approx7$ ($s_{g,\min}=15.5$\,mm) & $12.5$--$17.0$ & $1.4$--$2.4$ & $15.5$--$16.0$ \\
$r\approx5$ ($s_{g,\min}=15$--$16.5$\,mm) & $13.7$--$14.9$ & $1.3$--$1.5$ & $15.0$--$16.5$ \\
$r\approx3$ ($s_{g,\min}=9$--$18.5$\,mm) & $7.0$--$13.2$ & $1.2$--$1.6$ & $9.0$--$18.5$ \\
\hline
\end{tabular}
\caption{\textbf{Table ST3.} Streamwise stations of the peak of the time-mean Liutex strength and of the minimum separation $g$ of the two vortex rows for the 11 independent single-jet records (ranges).}
\alttext{Table of the streamwise stations of the Liutex peak and of the minimum row separation for the single jet.}
\label{tab:app-single-R-merge}
\end{table}

\begin{table}
\centering
\caption{\textbf{Table ST4.} Single-jet near/far decomposition: medians over the 11 independent records of the windward and leeward sides in each stage (near $s\le s_{tn}$, far $s>s_{tn}$ through the tail; $U_{c,R}$ Liutex-PIV windows $s\in[0.5,5]$ / $[8,20]$\,mm). $U_{c,\mathrm{tr}}$: tracked core speed; $U_{c,R}$: Liutex-PIV speed; $U_{c,\mathrm{ph}}$: cross-spectral phase speed; $f_0$: side-resolved FFT frequency; $\lambda_{\mathrm{tr}}$: tracked spacing.}
\alttext{Table of the single-jet near/far decomposition medians of the vortex-motion estimators.}\label{tab:single-fourpart}
\scriptsize
\begin{tabular}{llccccc}
\toprule
stage & side & $U_{c,\mathrm{tr}}$ (m/s) & $U_{c,R}$ Liutex-PIV (m/s) & $U_{c,\mathrm{ph}}$ (m/s) & $f_0$ (kHz, $St=f_0d/u_j$) & $\lambda_{\mathrm{tr}}$ (mm) \\
\midrule
near & windward & 6.89 & 6.88 & 7.79 & 1.80 (0.51) & 4.59 \\
near & leeward & 6.71 & 6.42 & 7.82 & 1.70 (0.49) & 4.34 \\
far & windward & 6.72 & 7.42 & 7.80 & 1.70 (0.49) & 3.97 \\
far & leeward & 7.36 & 7.26 & 8.15 & 1.70 (0.49) & 4.38 \\
\bottomrule
\end{tabular}
\end{table}

\begin{table}
\centering
\caption{\textbf{Table ST5.} Frequency metrics of the twin-jet shear layer compared (front/rear): the primary along-trajectory pooled dominant frequency of the main text (dominant frequency sampled over the trajectory zone $\pm2\sigma(s)$ and pooled over the cases with a detectable dominant frequency, 67 for each jet), the side-resolved medians over $s\le20$\,mm (amplitude above the 70th percentile; Fig.~23 of the main text), and the near-field side-resolved values of Table~VI of the main text ($s\le s_{tn}$, windward/leeward). An earlier pooling that mixed both sides and the near and far fields is not used for the frequency comparison; the near-field sampling of the last row resolves the shielded, not yet fully established rear-jet stage at the smaller spacings.}
\alttext{Table comparing the front and rear shear-layer frequency metrics.}\label{tab:st5-freq-metrics}
\begin{tabular}{lcc}
\toprule
metric & front & rear \\
\midrule
along-trajectory pooled (primary) & $1612\pm243$\,Hz ($St=0.46$) & $1534\pm249$\,Hz ($St=0.44$) \\
side-resolved, $s\le20$\,mm & $1.4$--$1.5$\,kHz ($St=0.40$--$0.43$) & $1.5$--$1.6$\,kHz ($St=0.43$--$0.46$) \\
near-field side-resolved (Table~VI) & $1454$--$1758$\,Hz & $1067$--$1610$\,Hz \\
\bottomrule
\end{tabular}
\end{table}

\begin{table*}
\centering
\caption{\textbf{Table ST6.} Twin-jet near/far decomposition: medians over the 68 cases of the windward and leeward sides in each quadrant (near $s\le s_{tn}$, far $s>s_{tn}$ through the tail; $U_{c,R}$ Liutex-PIV windows $s\in[0.5,5]$ / $[8,20]$\,mm); the last four rows repeat the single-jet reference of Table ST4 for comparison.}
\alttext{Table of the twin-jet near/far decomposition medians of the vortex-motion estimators.}\label{tab:twin-fourpart}
\scriptsize
\begin{tabular}{llccccc}
\toprule
quadrant & side & $U_{c,\mathrm{tr}}$ (m/s) & $U_{c,R}$ Liutex-PIV (m/s) & $U_{c,\mathrm{ph}}$ (m/s) & $f_0$ (kHz, $St=f_0d/u_j$) & $\lambda_{\mathrm{tr}}$ (mm) \\
\midrule
front near & windward & 6.06 & 4.65 & 6.62 & 1.60 (0.46) & 3.97 \\
front near & leeward & 5.95 & 4.43 & 7.04 & 1.60 (0.46) & 3.81 \\
front far & windward & 5.12 & 4.65 & 6.07 & 1.10 (0.31) & 5.49 \\
front far & leeward & 5.51 & 4.89 & 6.72 & 1.10 (0.31) & 5.41 \\
rear near & windward & 7.26 & 6.47 & 8.58 & 1.40 (0.40) & 4.92 \\
rear near & leeward & 6.67 & 6.28 & 8.91 & 1.40 (0.40) & 4.87 \\
rear far & windward & 5.92 & 7.19 & 7.42 & 1.20 (0.34) & 5.18 \\
rear far & leeward & 6.37 & 7.30 & 8.22 & 1.30 (0.37) & 5.58 \\
\midrule
single near & windward & 6.89 & 6.88 & 7.79 & 1.80 (0.51) & 4.59 \\
single near & leeward & 6.71 & 6.42 & 7.82 & 1.70 (0.49) & 4.34 \\
single far & windward & 6.72 & 7.42 & 7.80 & 1.70 (0.49) & 3.97 \\
single far & leeward & 7.36 & 7.26 & 8.15 & 1.70 (0.49) & 4.38 \\
\bottomrule
\end{tabular}
\end{table*}

\section{Modal-decomposition statistics (Sec.~III E of the main text)}

This section collects the DMD/SPOD mode statistics behind Sec.~III E of the main text. Both decompositions use the mean-subtracted 400-frame velocity series of each record; SPOD uses $n_{\mathrm{fft}}=128$ with 75\% overlap and nine blocks, while the DMD is computed on the full 400-frame series. The largest admissible frequency is set by the $25\,\mu$s frame interval, and the eigenvalue modulus carries an estimation scatter of about $\pm0.01$ over the resolved modes; a modulus of $1.013$ therefore cannot be distinguished from unity at the 10\,ms record length ($\ln 1.013/\Delta t$ corresponds to an e-folding time of about 2\,ms). Extending the decomposition to one representative case of every condition (Fig. S14) gives an energy fraction of $64$--$83$\% within $\pm2$\,mm of the two trajectories in every condition (SPOD $78$--$90\%$; the single jet gives $85\%$), and the largest DMD eigenvalue magnitude stays between $0.999$ and $1.013$. The band-dominant mode of one representative case per condition is shown in Fig. S15; the modes are localized along the two shear layers in every condition, with no single mode locking the whole field.

\begin{figure}
\centerline{\includegraphics[width=\textwidth]{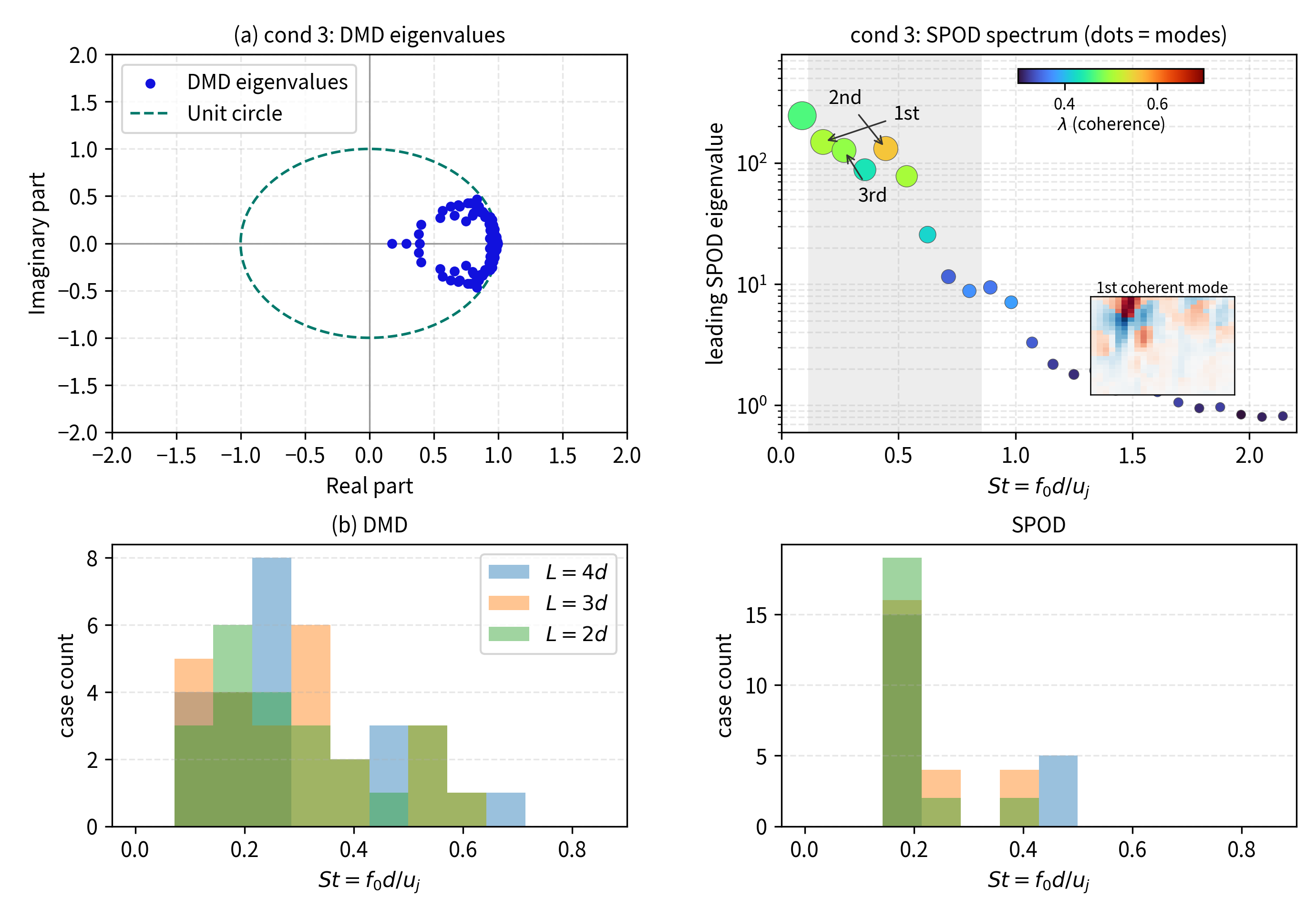}}
\caption{\textbf{Figure S13.} Modal-decomposition diagnostics of condition 3. (a) Left: DMD eigenvalues in the complex plane with the unit circle (dashed); the clustering near the circle within the shear-layer band is visible, but near-unit-circle eigenvalues alone do not discriminate a global mode. Right: leading SPOD eigenvalues (dots) versus the Strouhal number $St=f_0\,d/u_j$ ($n_{\mathrm{fft}}=128$, 75\% overlap, 9 blocks), with the 1st, 2nd and 3rd coherent modes marked and an inset showing the field of the first coherent mode on its own colour scale; the grey band marks the range in which the spectrum peaks. (b) Histograms of the per-case dominant DMD (left) and SPOD (right) frequencies, stacked by spacing ($L=4d$, $3d$, $2d$); they are broad and overlapping.}
\alttext{Modal diagnostics: DMD eigenvalues with the unit circle, the leading SPOD eigenvalue spectrum versus Strouhal number with the first modes and the first coherent-mode field inset, and histograms of the per-case dominant DMD and SPOD frequencies.}
\end{figure}

\begin{figure}
\centerline{\includegraphics[width=\textwidth]{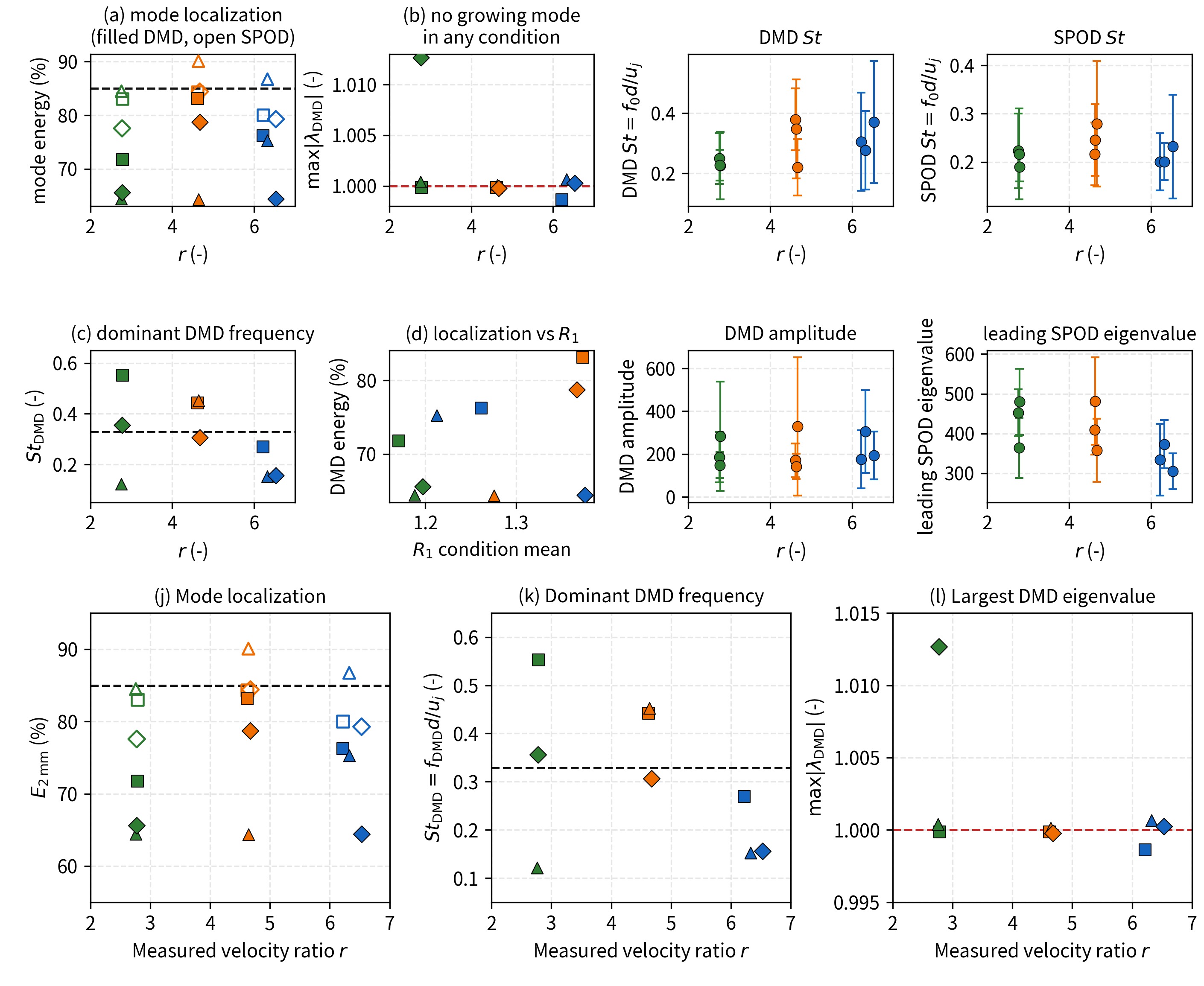}}
\caption{\textbf{Figure S14.} Modal statistics. Upper block: (a) mode energy fraction within $\pm2$\,mm of the trajectories (filled DMD, open SPOD; dashed: single jet), (b) largest DMD eigenvalue magnitude (dashed line: unity, the no-growth reference), (c) dominant DMD Strouhal number and (d) localization fraction versus the front-jet exchange strength; right: per-case DMD/SPOD frequencies, DMD amplitude and leading SPOD eigenvalue (mean $\pm$ SD over the cases of each velocity-ratio group; colours: $r\approx3$, 5, 7). Lower block: the same statistics of the nine conditions versus the measured velocity ratio: (j) the mode energy fraction within $\pm2$\,mm of the trajectories, (k) the dominant DMD Strouhal number and (l) the largest DMD eigenvalue magnitude (filled: DMD, open: SPOD; dashed: single jet). The two blocks deliberately show the same statistics twice---the upper block resolved per case and per jet, the lower block as condition means against the measured velocity ratio---so that the absence of a growing mode can be read both case by case and condition by condition.}
\alttext{Modal statistics: mode energy fractions, DMD eigenvalue magnitudes and dominant Strouhal numbers.}
\end{figure}

\begin{figure}
\centerline{\includegraphics[width=\textwidth]{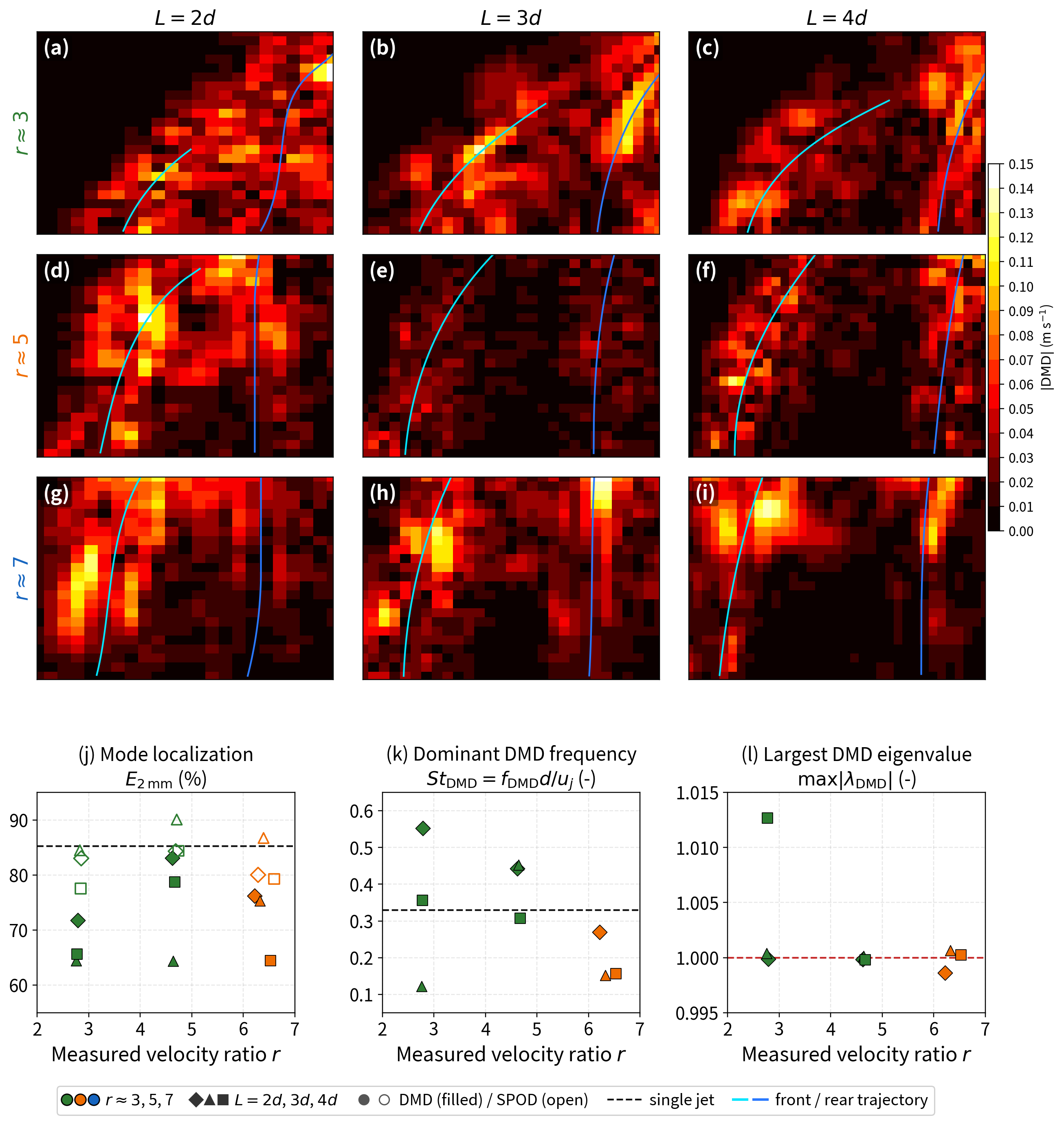}}
\caption{\textbf{Figure S15.} Band-dominant DMD mode magnitude of one representative case per condition in the nine-grid of Table~I of the main text (rows $r\approx3$, 5, 7; columns $L=2d$, $3d$, $4d$; panels (a)--(i)), with the front (cyan) and rear (blue) trajectories; the modes are localized along the two shear layers in every condition and are drawn on the fixed absolute scale $0$--$0.15$\,m\,s$^{-1}$ in bands of $0.01$\,m\,s$^{-1}$ (colour bar). (j)--(l) Statistics against the measured velocity ratio: (j) mode localization $E_{2\,{\rm mm}}$, (k) dominant DMD Strouhal number and (l) largest DMD eigenvalue magnitude; the symbol convention is given in the legend below the panels (colour: velocity-ratio group; shape: spacing; filled/open: DMD/SPOD; dashed: single jet).}
\alttext{Band-dominant DMD mode magnitude for one representative case per condition, with the trajectories, and the statistics of the mode localization, dominant Strouhal number and largest eigenvalue against the measured velocity ratio.}
\end{figure}

\section{Supporting figures for the discussion (Secs.~III F--H of the main text)}

Eight supporting figures carry the quantitative detail behind the discussion (Secs.~III F--H). Five of them document the present analysis: the condition-mean Liutex amplitude profiles along the trajectory (Fig. S7), the partial-correlation carriers and mediation statistics behind the causal assignments (Fig. S16), the nine-condition leading terms of the vertical momentum balance (Fig. S17), the example flux evolutions behind the closure test of Fig.~12 of the main text (Fig. S18) and the tracked vortex dynamics of the two jets (Fig. S19). The remaining three were moved here from the main text and give the gap-side exchange statistics (Fig. S20) and the entrainment-angle geometry, at $r\approx7$ and over the nine conditions (Figs. S21 and S22).

\begin{figure}
\centerline{\includegraphics[width=\textwidth,height=0.70\textheight,keepaspectratio]{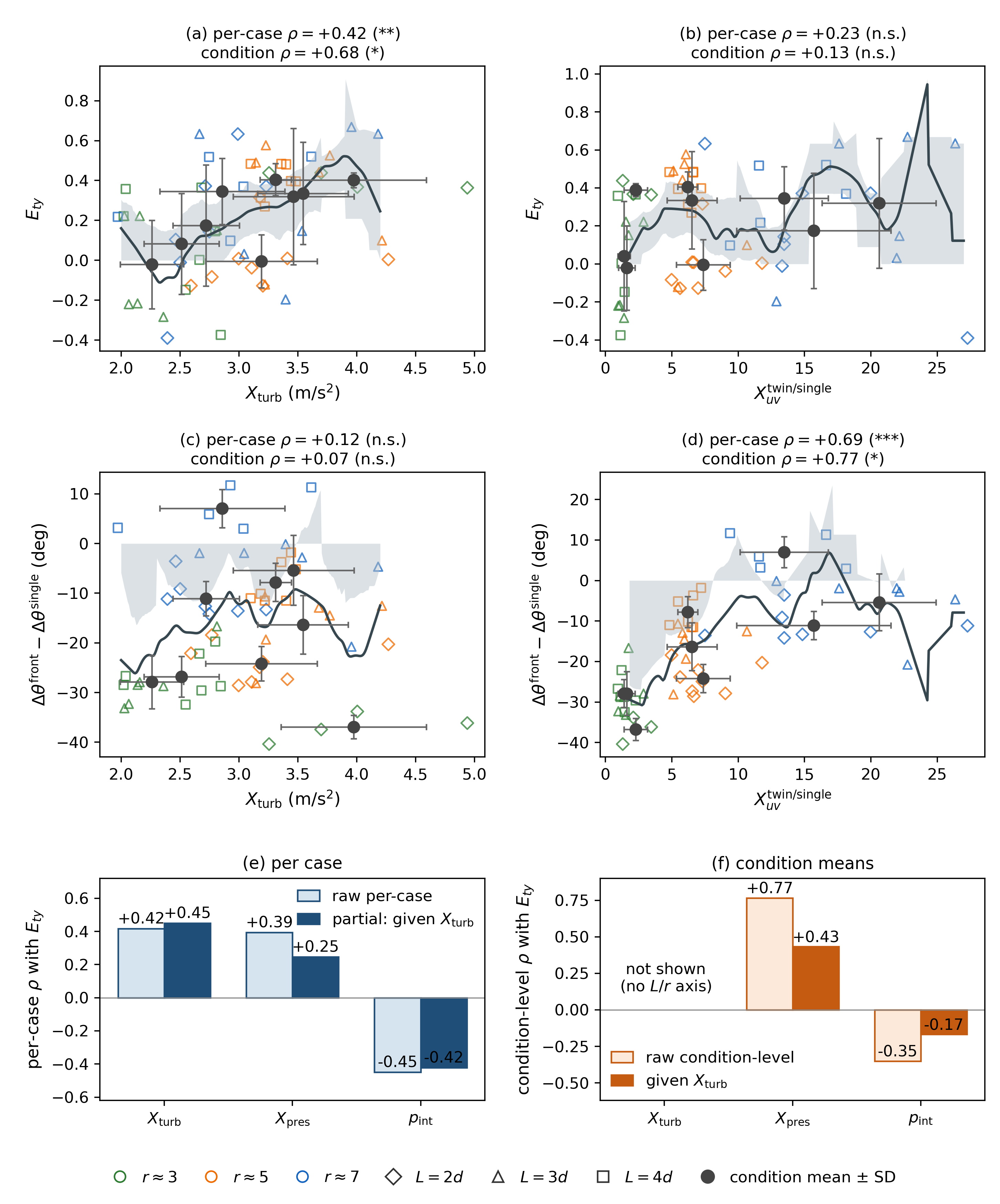}}
\caption{\textbf{Figure S16.} Partial-correlation carriers and mediation statistics behind the dose--response of Fig.~11 of the main text. (a)--(d) The near-field shortening $E_{ty}$ (a, b) and the turning-angle change (c, d) against the absolute turbulent transport $X_{\rm turb}$ (a, c) and the twin/single enhancement ratio $X_{uv}^{\rm twin/single}$ (b, d), with the nine-condition symbol convention, condition means ($\pm1$ SD), the local (LOESS-type) fit (black curve) with its $95\%$ bootstrap confidence band (grey) and the Spearman coefficients; light gray bands mark the fit confidence interval. (e)--(f) Spearman $\rho$ of $E_{ty}$ against $X_{\rm turb}$, the pressure force $X_{\rm pres}$ and the gap pressure $p_{\rm int}$, raw (light bars) and partialled given $X_{\rm turb}$ (dark bars), per case (e) and per condition (f); in (f) the condition-level raw coefficient for $X_{\rm turb}$ is not defined (no $L/r$ axis) and is marked as not shown. The pressure level anti-correlates with the shortening, while both forcing terms correlate positively with it; partialling the transport reduces the pressure coefficient, so the pressure is a common-origin response rather than an independent driver.}
\alttext{Partial-correlation carriers of the modification metrics with local fits and confidence bands, and the raw and partialled Spearman coefficients of the shortening against the forcing terms and the gap pressure.}
\end{figure}

\begin{figure}
\centerline{\includegraphics[width=\textwidth,height=0.70\textheight,keepaspectratio]{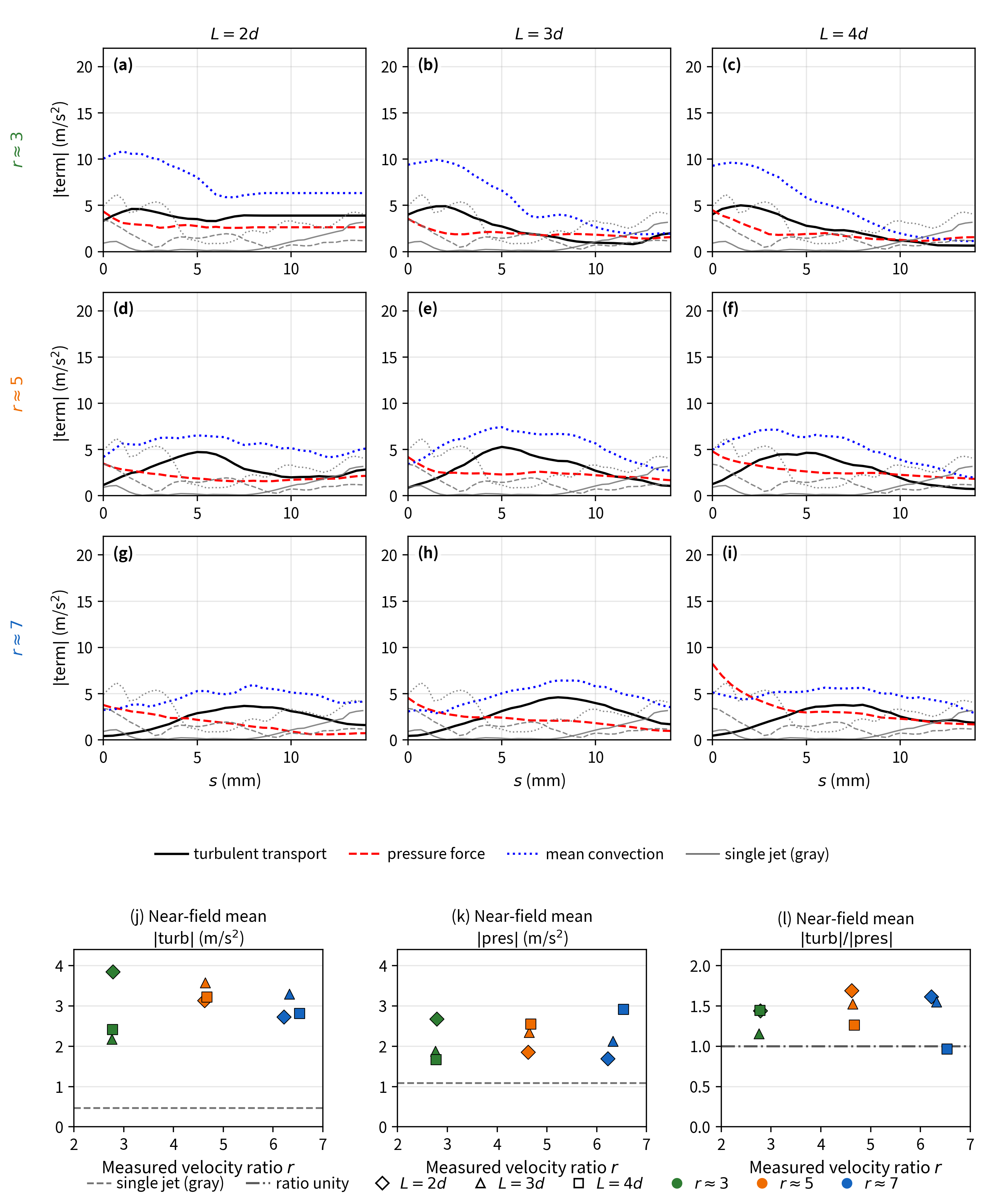}}
\caption{\textbf{Figure S17.} Condition-mean leading-term comparison of the vertical momentum balance for the front jet on the gap (leeward) side. (a)--(i) Zone maxima over $0\le n\le4$\,mm of the turbulent transport (solid black), pressure force (dashed red) and mean convection (dotted blue) terms versus the arc length $s$, over the nine conditions (rows $r\approx3$, 5, 7; columns $L=2d$, $3d$, $4d$; ordinate: $|$term$|$ in m\,s$^{-2}$; thin gray: single-jet reference; legend below the grid). (j)--(l) The near-field ($s=2$--$12$\,mm) zone means of the two forcing terms and their ratio against the measured velocity ratio $r$, with the single-jet reference (gray dashed) and the ratio unity (dash-dotted); symbol colour: velocity-ratio group, shape: spacing (legend below).}
\alttext{Condition-mean leading-term vertical momentum balance of the front jet on the gap side over the nine conditions, with the near-field zone means of the forcing terms and their ratio.}
\end{figure}

\begin{figure}
\centerline{\includegraphics[width=0.82\textwidth]{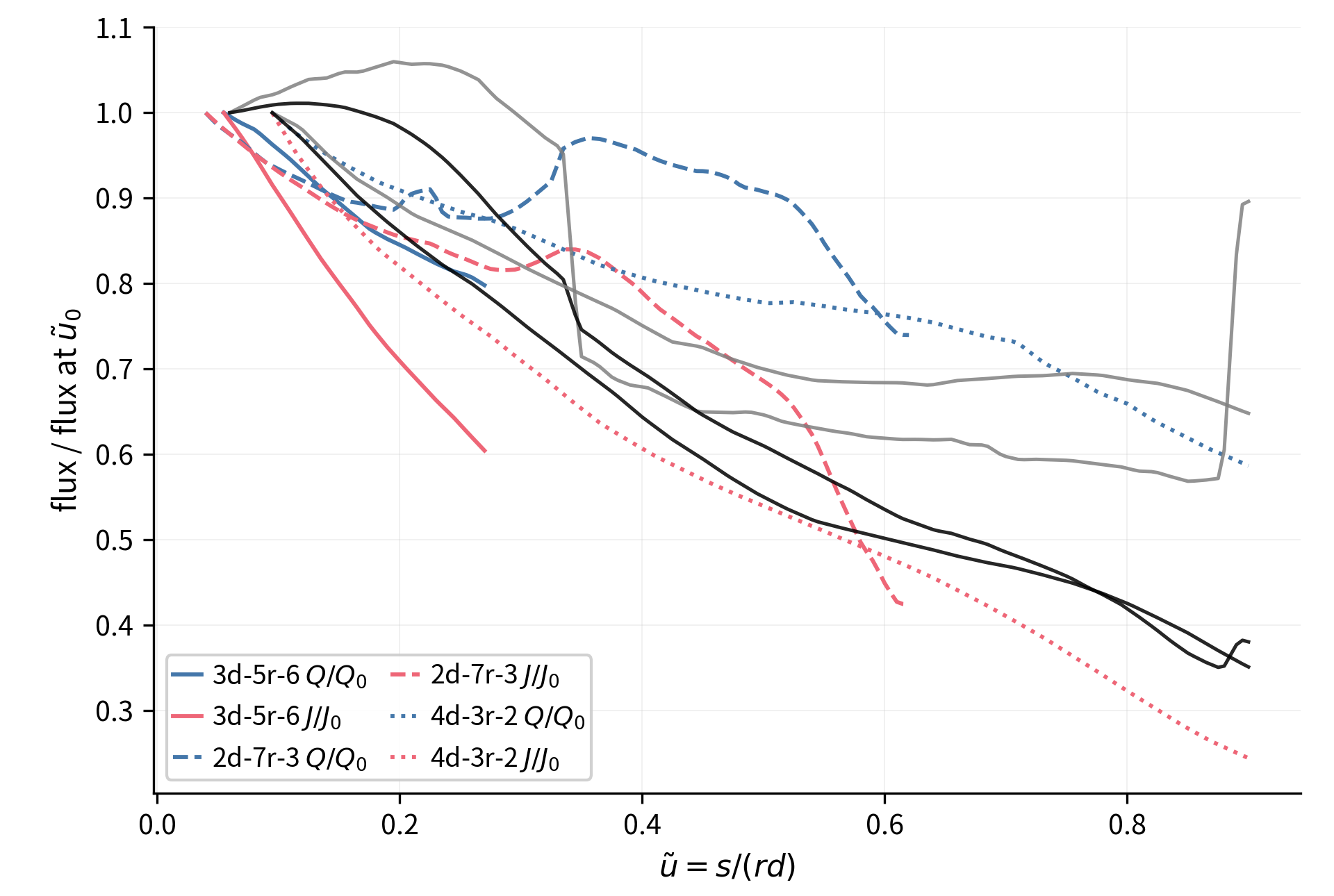}}
\caption{\textbf{Figure S18.} Example flux evolutions $Q/Q_0$ and $J/J_0$ of three twin and two single jets (grey and black: single jets) against $\tilde u=s/(rd)$ (abscissa: normalized arc length; ordinate: flux normalized by its value at $\tilde u_0$), i.e.\ the sample behind the closure test of Fig.~12 of the main text. For the twin jets the volume flux $Q/Q_0$ and the momentum flux $J/J_0$ of a condition share the same line style (solid, dashed and dotted for the three example conditions, as labelled in the legend).}
\alttext{Example flux evolutions of three twin and two single jets against the normalized arc length.}
\end{figure}

\begin{figure}
\centerline{\includegraphics[width=\textwidth]{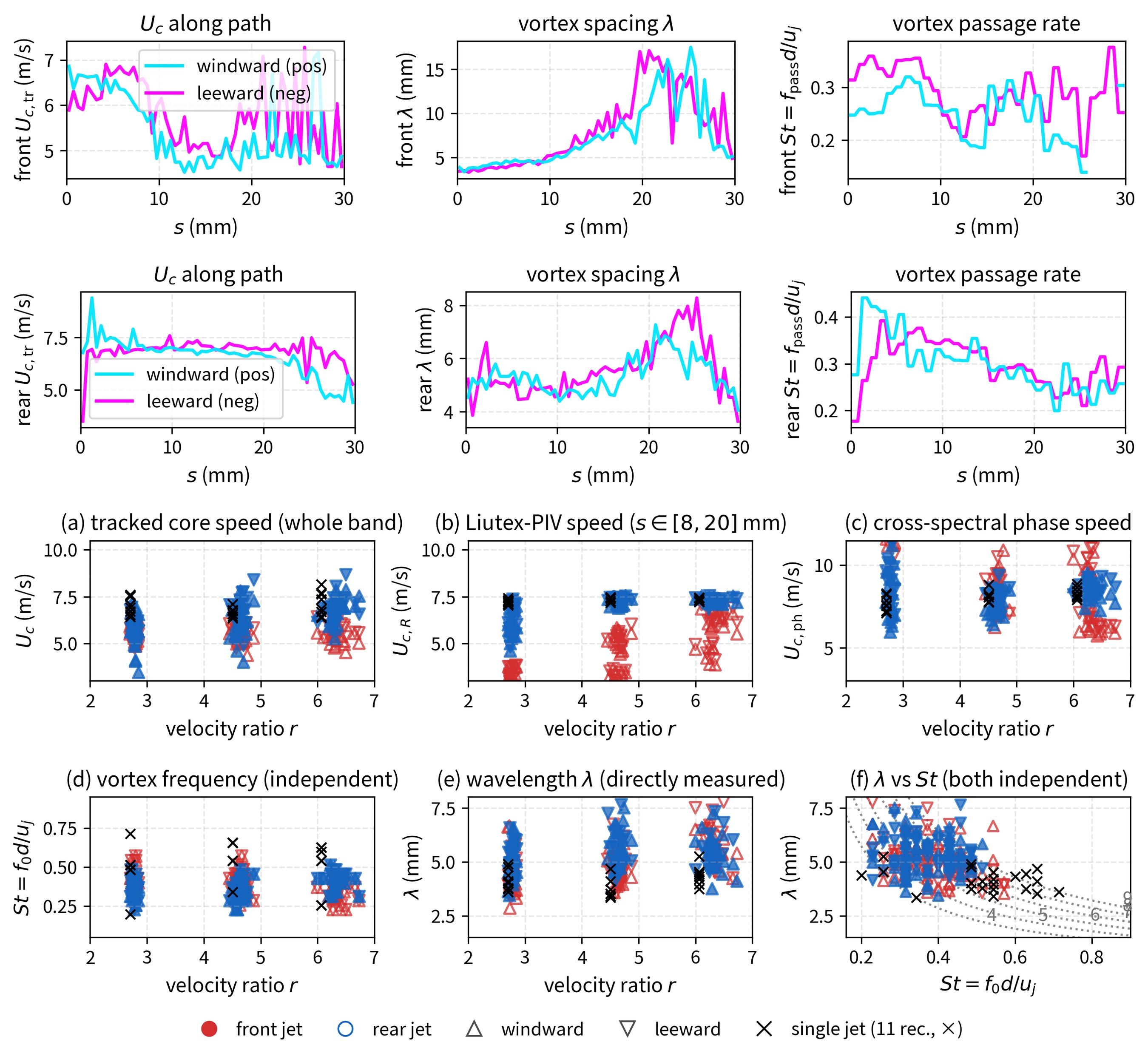}}
\caption{\textbf{Figure S19.} Tracked vortex dynamics of the two jets. Upper block: (a) convection speed $U_{c,\rm tr}$ along the path, (b) vortex spacing $\lambda$ and (c) vortex passage rate $St=f_{\rm pass}d/u_j$ versus the arc length $s$, for the front (upper row) and rear (lower row) jets; cyan: windward (positive $n$), magenta: leeward (negative $n$), as pooled segment medians over the 68 twin cases. Lower block: single-jet baseline versus twin jets (front red, rear blue; windward $\triangle$, leeward $\nabla$; crosses: the 11 single-jet records): (a) tracked core speed (whole band), (b) Liutex-PIV speed ($s\in[8,20]$\,mm), (c) cross-spectral phase speed, (d) frequency as $St$, (e) spacing $\lambda$ and (f) $\lambda$ versus $St$, with the dotted curves the constant-$U_c$ hyperbolas (the counterpart of Fig.~22 of the main text).}
\alttext{Tracked vortex dynamics: convection speed, spacing, passage rate and the method-by-method single-jet versus twin comparison.}
\end{figure}

\begin{figure}
\centerline{\includegraphics[width=\textwidth]{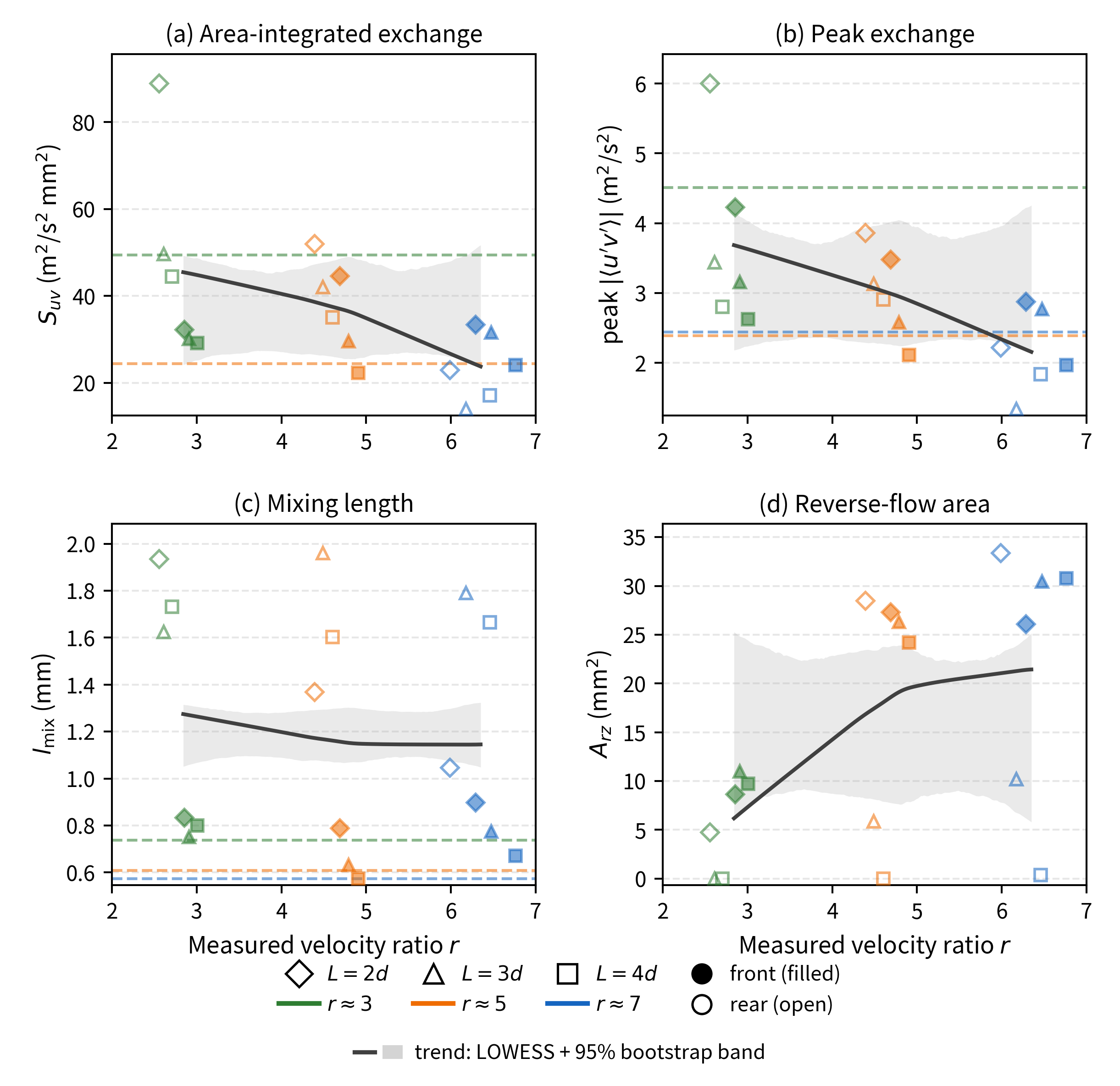}}
\caption{\textbf{Figure S20.} Gap-side statistics behind the fields of Fig.~7 of the main text (formerly panels (j)--(m) of the main text), against the measured velocity ratio: (a) area-integrated exchange $S_{uv}$, (b) peak exchange $|\langle u'v'\rangle|$, (c) zone-median mixing length $l_{\rm mix}$, (d) reverse-flow area $A_{rz}$ (symbol convention of Sec.~II of the main text; markers offset horizontally within each ratio group; dashed: single-jet reference of the same group, which has no reverse-flow area); the weighted local-linear (LOWESS-type, tricube kernel, bandwidth $2.2$ in measured $r$) trend with a $95\%$ bootstrap band ($1000$ block resamples of the nine conditions, both jet sides per condition) is drawn over the measured $r$ range.}
\alttext{Four-panel figure of the gap-side statistics, the area-integrated exchange, the peak exchange, the mixing length and the reverse-flow area, against the measured velocity ratio, with the single-jet reference lines.}
\label{fig:exchange-stats}
\end{figure}

\begin{figure}
\centerline{\includegraphics[width=\textwidth]{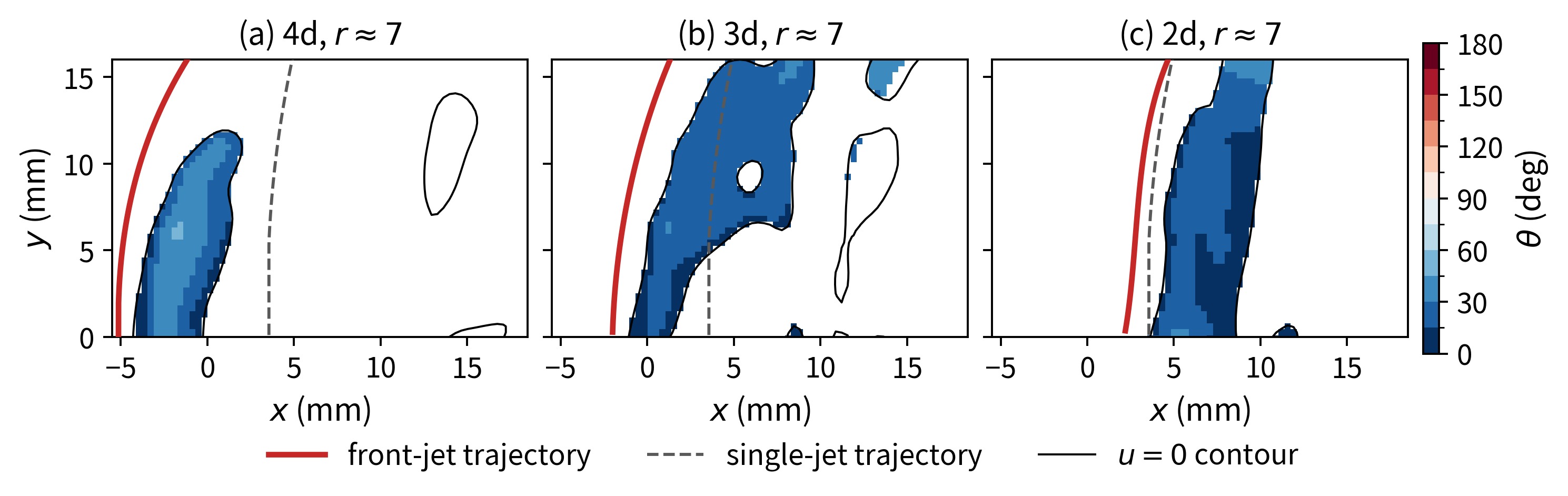}}
\caption{\textbf{Figure S21.} Geometric condition of the vertical-momentum entrainment at $r\approx7$ (columns: $L=4d$, $3d$, $2d$; moved here from the main text): maps of the angle $\theta$ between the gap-zone velocity ($u<0$) and the front-jet tangent (blue: $\theta<90^\circ$, red: $\theta>90^\circ$; fixed $0$--$180^\circ$ scale in bands of $15^\circ$, white at $90^\circ$; no $\theta>90^\circ$ region occurs in these three conditions), with the $u=0$ contour (black), the front-jet (red) and single-jet (gray dashed) trajectories. The $\theta(s)$ profiles of these three conditions, and of the whole nine-condition grid, are collected in Fig. S22.}
\alttext{Maps and profiles of the entrainment angle between the gap-zone velocity and the front-jet tangent at r about 7.}
\label{fig:promo-s8}
\end{figure}

\begin{figure}
\centerline{\includegraphics[width=\textwidth]{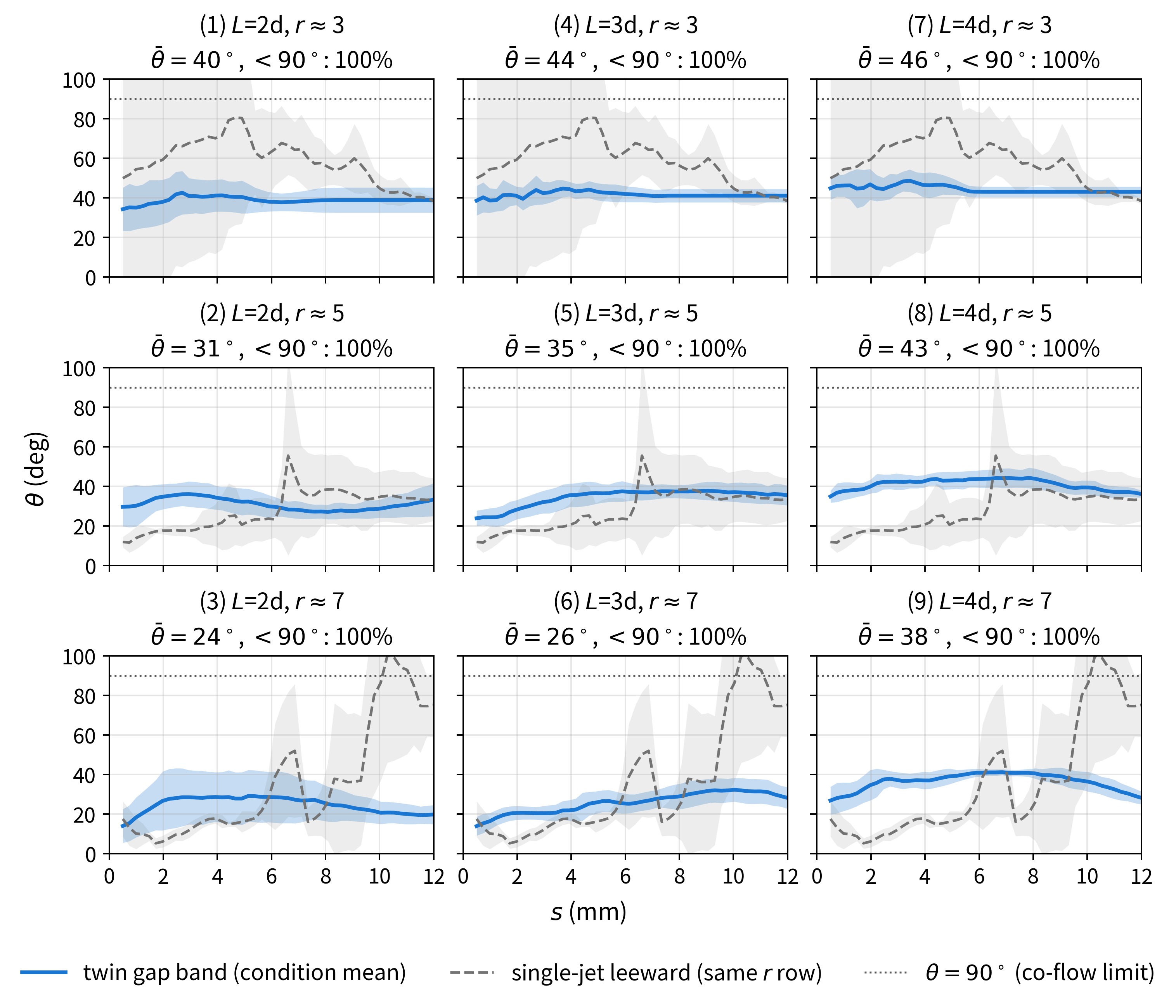}}
\caption{\textbf{Figure S22.} Entrainment angle over the nine conditions (rows $r\approx3$, 5, 7; columns $L=2d$, $3d$, $4d$, as in Table~I of the main text; moved here from the main text; the condition number of Table~I is given in each panel): zone-median $\theta$ between the reversed gap velocity ($u<0$) and the front-jet tangent versus the arc length $s$ (gray dashed: single-jet leeward zone of the same ratio row; dotted: $\theta=90^\circ$), with the near-field ($s=2$--$12$\,mm) median $\bar\theta$ and the per-cell station fraction below $90^\circ$ annotated in every panel. The three conditions of Fig. S21 are the panels numbered 3, 6 and 9.}
\alttext{Nine-condition statistics of the entrainment angle along the arc length with the near-field medians.}
\label{fig:promo-s9}
\end{figure}

\bibliographystyle{aipauth4-1-titles}
\bibliography{twin-jet-piv,liutex}